\documentclass[twocolumn, times, tighten, appendixfloats]{aastex631}

\usepackage{natbib}
\usepackage{enumitem}
\usepackage{amsmath}
\usepackage{soul}
\usepackage{longtable}

\usepackage{graphicx}		
\usepackage{latexsym}		
\usepackage{bm}			
\usepackage[english]{babel}
\usepackage{color}		
\newcommand{\etal}	{\hbox{et~al.}}%

\newcommand{\DEL}[1]{}
\newlength{\txw}
\newlength{\txh}

{\relax}%

\begin{document}

\title{Candidate Galaxies at $\mathbf{z\approx 11.3}$--21.8 and beyond: results from JWST's public data taken in its first year}

\author[0000-0001-7592-7714]{Haojing Yan} 
\affiliation{Department of Physics and Astronomy, University of Missouri, Columbia, MO 65211, USA}

\author[0000-0001-7957-6202]{Bangzheng Sun} 
\affiliation{Department of Physics and Astronomy, University of Missouri, Columbia, MO 65211, USA}

\author[0000-0003-3270-6844]{Zhiyuan Ma}
\affiliation{Department of Astronomy, University of Massachusetts, Amherst, MA 01003, USA}

\author[0000-0003-4952-3008]{Chenxiaoji Ling}
\affiliation{National Astronomical Observatories, Chinese Academy of Sciences, Beijing 100101, China}


\correspondingauthor{Haojing Yan}
\email{yanha@missouri.edu}

\setwatermarkfontsize{1in}

\shortauthors{Yan \etal}
\shorttitle{Candidate Galaxies at $z>11.3$}

\begin{abstract}

   We present a systematic search of candidate galaxies at $z\gtrsim 11.3$
using the public Near Infrared Camera data taken by the James Webb Space 
Telescope (JWST) in its Cycle 1, which include six blank fields totalling 
386~arcmin$^2$ and two lensing cluster fields totalling 48~arcmin$^2$. 
The candidates are selected as F150W, F200W and F277W dropouts, which 
correspond to $z\approx 12.7$ ($11.3\lesssim z\lesssim 15.4$),
17.3 ($15.4\lesssim z\lesssim 21.8$) and 24.7 ($21.8\lesssim z\lesssim 28.3$),
respectively. Our sample consists of 123 F150W dropouts, 52 F200W dropouts and
32 F277W dropouts, which is the largest candidate galaxy sample probing the 
highest redshift range to date. The F150W and F200W dropouts have sufficient 
photometric information that allows contaminant rejection, which we do by 
fitting to their spectrum energy distributions. Based on the purified samples of
F150W and F200W dropouts, we derive galaxy luminosity functions at
$z\approx 12.7$ and 17.3, respectively. We find that both are better described
by power law than Schechter function and that there is only a marginal
evolution (a factor of $\lesssim 2$) between the two epochs. The emergence of
galaxy population at $z\approx 17.3$ or earlier is consistent with the 
suggestion of an early cosmic hydrogen reionization and is not necessarily a 
crisis of the $\Lambda$CDM paradigm. To establish a new picture of galaxy
formation in the early universe, we will need both JWST spectroscopic 
confirmation of bright candidates such as those in our sample and deeper 
surveys to further constrain the faint-end of the luminosity function at 
$M\gtrsim -18$~mag.

\end{abstract}

\keywords{dark ages, reionization, first stars --- galaxies: evolution
--- galaxies: high-redshift} 


\section{Introduction}

   In merely one year, James Webb Space Telescope (JWST) has greatly changed
our view of galaxy formation in the early universe. From a few medium-deep 
fields observed by its Near Infrared Camera \citep[NIRCam;][]{Rieke2023a}, 
several tens of candidate galaxies at 
$z>11$ have been claimed, 
\citep[e.g.,][]{Adams2023a,Adams2023b, Atek2023, Castellano2022, Donnan2023, 
Finkelstein2022b, Finkelstein2023, Harikane2023, Naidu2022a, Rodighiero2023, 
Yan2023a, Yan2023b}, with some potentially being at $z\approx 20$ 
\citep[][]{Yan2023a, Yan2023b}. This is in stark contrast to the widely 
accepted, pre-JWST picture that there should be very few galaxies at $z>10$
\citep[e.g.][]{Oesch2018}, and the inferred number density is at least
an order of magnitude higher than any previous model predictions
\citep[e.g.][]{Behroozi2020, Vogelsberger2020, Haslbauer2022, Kannan2023, Yung2023}.
With the spectroscopic confirmations of a few $z>11$ galaxies up to $z=13.2$
\citep[][]{Robertson2023a, CurtisLake2023, WangUncover2023, AH2023b}, it is now
a consensus that galaxies at $z>11$ are indeed much more abundant than 
previously thought.

   A few new models have been proposed to resolve the tension between the 
JWST observations and the theories of galaxy formation in the early universe 
within the standard $\Lambda$CDM paradigm
\citep[e.g.,][]{Inayoshi2022, Ferrara2023, Mason2023, Yung2023, Mirocha2023, Shen2023}. 
However, a new picture of early galaxy formation is still far away from being
established. On the observational side, we will need
spectroscopic confirmation of more galaxies at $z>11$ and a large, reliable
photometric candidate sample probing the highest redshift possible. 

   JWST has gathered sufficient NIRCam data to allow for the latter. We have 
finished a systematic search of candidate galaxies at $z\gtrsim 11.3$ using the
public data taken during the first year (Cycle 1) of JWST operation. The widely
separated fields minimize the impact of cosmic variance, and the different 
combinations of field coverage and depth naturally form a multi-tiered dataset
sampling a large range in luminosity. We present our results in this paper,
which is organized as follows. Sections 2 and 3 describe the data and the 
reduction processes, respectively. Photometry is presented in Section 4. The 
candidate selection using the dropout technique is detailed in Section 5, along 
with the sample. Using the purified sample, we derive galaxy luminosity 
functions at $z\approx 12.7$ and 17.3 in Section 6 and discuss the implications 
in Section 7. We conclude with a summary in Section 8. Throughout the paper,
the quoted magnitudes are in the AB system, and all coordinates are in the ICRS 
frame (equinox 2000). We adopt a flat $\Lambda$CDM cosmology with parameters 
$H_0=71$~km~s$^{-1}$~Mpc$^{-1}$, $\Omega_M=0.27$ and $\Omega_\Lambda=0.73$.


\section{Data Overview}

   In this study, we make use of six ``blank'' fields and two galaxies cluster 
fields, which all have data in at least six NIRCam broad bands. These fields 
are summarized in Table~\ref{tab:allfields} and are detailed below.

\begin{table*}[hbt!]
    \centering
    \caption{Summary of data}
    \resizebox{\textwidth}{!}{
    \begin{tabular}{cccccccc} \hline
         Field & Ctr. R.A. & Ctr. Decl. &  Pipeline & Context & Area (arcmin$^2$) & Exposure (ks) & Astrometry \\ \hline
         PRIMER/COSMOS & 150.12299 & 2.34764 & 1.10.2 & 1089 & 137.13 & $\sim$2.5 & CANDELS \\
         PRIMER/UDS1 & 34.35004 & -5.20001 & 1.10.2 & 1089 & 114.45 & $\sim$1-2 & SXDS  \\ 
         CEERS & 215.00545 & 52.93451 & 1.9.4 & 1046 & 86.45 & $\sim$2.5 & CANDELS \\
         GLASS & 3.49130 & -30.33756 & 1.10.2 & 1084 & 12.31 & $\sim$8 & GAIA DR3 \\
         NGDEEP & 53.24923 & -27.84753 & 1.9.4 & 1069 & 9.14 & $\sim$30-40 & HUDF12 \\
         JADES & 53.16444 & -27.78256 & - & - & 25.50 & $\sim$14-60 & GAIA DR3 \\ 
         \hline
         UNCOVER & 3.55273 & -30.38000 & 1.9.4 & 1078 & 37.04 & $\sim$10-20 & GAIA DR3 \\
         SMACS J0723-7327 & 110.75927 & -73.46797 & 1.11.0 & 1094 & 11.08 & $\sim$7 & RELICS \\ \hline
    \end{tabular}
    }
    \begin{flushleft}
    \tablecomments{Basic information of the JWST Cycle 1 public NIRCam data
    used in this study. The blank fields and the cluster fields are separated by the horizontal line. The listed coordinates are for the central 
    positions of the final mosaics. The exposure times (in kilo-seconds) are
    representative estimates. The external catalogs used for astrometric
    calibration and image alignment are listed in the last column.}
    \end{flushleft}
    \label{tab:allfields}
\end{table*}

\subsection{Blank fields}

   $\bullet$ {\it PRIMER COSMOS and UDS1}\,\,\,
    The shallowest but the widest data are from the Public Release IMaging for
Extragalactic Research program \citep[PRIMER, PID 1837;][]{Dunlop2021}
\footnote{\url{https://primer-jwst.github.io/index.html}}.
It observes the COSMOS and the UDS fields in eight NIRCam bands, namely, F090W,
F115W, F150W, and F200W in the short-wavelength (SW) channel and F277W, F356W,
F410M, and F444W in the long-wavelength (LW) channel. As of this writing (July
2023), the observations in the COSMOS field have been finished but the UDS field
is only covered by half. Following the PRIMER team's
designation, hereafter the finished areas are referred to as the ``COSMOS''
and the ``UDS1'' fields, respectively. Our dropout search was done in the areas
that are fully overlapped in all the used NIRCam bands, which amount to 137.13 
and 115.45~arcmin$^2$ in the
COSMOS and the UDS1 fields, respectively. The UDS1 field suffered a small 
amount of bad data due to problems in the observations, and the quoted coverage
above has taken this into account by excluding the affected regions. By design,
the NIRCam observations are done as coordinated parallels to the primary MIRI 
observations; as a result, the NIRCam exposure times are rather non-uniform
over the footprint. Roughly speaking, the exposure
times UDS1 are about 1--2~ksec and those in COSMOS1 are about 2.5~ksec.

   $\bullet$ {\it CEERS in EGS}\,\,\,
    The Cosmic Evolution Early Release Science Survey
\citep[CEERS, PID 1345;][]{Finkelstein2023}
\footnote{\url{https://ceers.github.io}}, one of the Early Release Science (ERS)
programs, contributed another set of shallow, wide-field data. It observed the
EGS field in seven NIRCam bands, namely, F115W, F150W and F200W in the SW
channel and F277W, F356W, F410M, and F444W in the LW channel, and the fully 
overlapped coverage in all bands is 86.45~arcmin$^2$. These observations were 
obtained as coordinated
parallels to either MIRI or NIRISS primaries, and the resultant exposure times
are also non-uniform; on average, these are about 2.5~ksec.

   $\bullet$ {\it GLASS near Abell 2744}\,\,\,
   ``Through the Looking GLASS'' program \citep[GLASS, PID 1324;][]{Treu2022}
\footnote{\url{https://glass.astro.ucla.edu/ers}},
also one of the ERS programs, was another public survey. Its main target was 
the galaxy cluster Abell 2744. Its
NIRCam data, however, are outside of the cluster region because they were taken 
as coordinate parallels to the primary spectroscopic observations on the
cluster. These data are in seven bands: F090W, F115W, F150W, and F200W in SW,
and F277W, F356W and F444W in LW. They cover $\sim$12.31~arcmin$^2$ of 
fully overlapped area, and the exposure times are around $\sim$8~ksec.

   $\bullet$ {\it NGDEEP Ep1 in HUDF-Par2}\,\,\,
   The Next Generation Deep Extragalactic Exploratory Public Survey
\citep[NGDEEP, PID 2079;][]{Bagley2023} provided $\sim$10$\times$ deeper NIRCam
data than the three wide-fields mentioned above. The primary observation of
this program is the NIRISS spectroscopy in the Hubble Ultra Deep Field (HUDF),
and the NIRCam observations are taken as the coordinated parallels. The NIRCam
field largely overlaps the footprint of the HUDF-Par2, which is one of the two
very deep HST parallel fields to the HUDF. Six NIRCam bands are used: F115W,
F150W and F200W in SW and F277W, F356W and F444W in LW.
Currently, the program has finished only half of its observations (Ep1); 
the fully overlapped coverage by all bands is 9.14~arcmin$^2$, and 
the average exposure times are roughly $\sim$30--40 ksec. 

   $\bullet$ {\it JADES DR1 in GOODS-S}\,\,\,
   The deepest data used in this work are from the first data release (DR1) of
the JWST Advanced Deep Extragalactic Survey 
\citep[JADES, PID 1180, 1210, 1283, 1286, 1287;][]{Eisenstein2023, Rieke2023b,
Bunker2023b}
\footnote{\url{https://jades-survey.github.io}}, which is the largest JWST
Guaranteed Time Observations (GTO) program and is still ongoing. The JADES DR1 
NIRCam data are in the historic GOODS South (GOODS-S) field and cover the HUDF.
The observations were made in nine
bands: F090W, F115W, F150W, and F200W in SW, and F277W, F355M, F356W, F410M,
and F444W in LW. The fully overlapped area by all bands is 25.5~arcmin$^2$.
The exposure times range from $\sim$14--60~ksec, depending on the bands. 

\subsection{Galaxy cluster fields}

   $\bullet$ {\it UNCOVER and DD2756 in Abell 2744}\,\,\,
   The Ultradeep NIRSpec and NIRCam ObserVations before the Epoch of 
Reionization \citep[UNCOVER, PID 2561;][]{Bezanson2022}
\footnote{\url{https://jwst-uncover.github.io}},
one of the JWST Treasury Programs, provided deep NIRCam data towards the galaxy
cluster Abell 2744 (at $z=0.308$). Currently, these are the deepest NIRCam data
in the public domain on a lensing cluster. It observed in seven bands, namely,
F115W, F150W and F200W in SW, and F277W, F356W, F410M, and F444W in LW. 
The exposure times are roughly 
$\sim$10--20~ksec over most area. 
In addition, there was a Director's Discretionary Time (DDT)
program that carried out NIRSpec spectroscopy in the field (PID 2756), and its
NIRCam parallel observations overlapped the original UNCOVER footprint. Here we
consider these data together. The total coverage (overlapped by all bands)
is $\sim$37.0~arcmin$^2$. For simplicity, hereafter we refer to this field as
``UNCOVER''.

   $\bullet$ {\it ERO SMACS J0723.3-7327}\,\,\,
   Among the first science-grade data taken by the JWST Early Release 
Observations \citep[ERO;][]{Pontoppidan2022}, there were NIRCam images taken 
on the galaxy cluster SMACS J0723.3-7327. These data were in six bands: F090W,
F150W and F200W in SW, and F277W, F356W and F444W in LW. The fully overlapped
coverage by all bands is $\sim$11.08~arcmin$^2$, and the exposure times are 
$\sim$7~ksec. This field is referred to as ``SMACS0723'' hereafter.

\section{Data Reduction}

   We reduced the aforementioned data on our own except those from the JADES 
program in GOODS-S. The data were retrieved from the Mikulski Archive for Space 
Telescopes (MAST). Reduction started from the so-called ``uncal'' 
products, which are the single exposures from the standard JWST data reduction 
pipeline after Level~1b processing. The JWST data reduction pipeline and the
reference files in the processing ``context'' have been evolving, and we list 
their versions in Table~\ref{tab:allfields}. 

   We started from running ``Stage 1'' of the reduction pipeline. This stage 
applies detector-level corrections to the ``uncal.fits" images, and outputs 
count-rate images in the units of counts per second. We adopted the default 
parameters except those involving the detection and removal of the 
``snowballs'' defects (caused by large cosmic ray events), for which we set 
{\tt jump.expane\_large\_events} to ``True'' and set {\tt jump.expand\_factor}
to 1.5 so that the neighboring pixels of the detected snowballs were also 
flagged. We estimated and subtracted the median background count rate for all
individual exposures in the SW channel to remove the differences in the 
baseline bias levels among SW detectors. To estimate this background, each
image was segmented into sections of $128\times 128$ pixels in size (with
sources masked), and a $3\times3$ median filter was applied to obtain the
median values. For all exposures involving detectors A3, A4, B3, and B4 in 
F115W, F150W and F200W, we removed the large-scale defects referred to as
``wisps'' by subtracting the corresponding 
templates\footnote{\href{https://jwst-docs.stsci.edu/jwst-near-infrared-camera/nircam-instrument-features-and-caveats/nircam-claws-and-wisps}{https://jwst-docs.stsci.edu/jwst-near-infrared-camera/nircam-instrument-features-and-caveats/nircam-claws-and-wisps}}. 

    The images were then processed through ``Stage 2'' of the pipeline. 
This stage applies flat-fielding and flux calibration, and the output images are
in the units of MJy~sr$^{-1}$. For this step, we adopted all the default 
parameters. We then corrected for the ``$1/f$ noise'' on each level 2 
calibrated image in the SW channel. This was done following the method 
implimented by the external tool 
``image1overf''\footnote{\href{https://github.com/chriswillott/jwst}{https://github.com/chriswillott/jwst}}, 
and the corrections were applied on a per-amplifier basis along both rows and 
columns of the image. 

   The calibration of astrometry and image alignment was done using an external 
tool, ``JHAT"\footnote{\href{https://github.com/arminrest/jhat}{https://github.com/arminrest/jhat}}, 
to substitute the default ``TweakReg'' step in the standard pipeline. This
method is designed to align each exposure using the external reference 
catalog of user's choice. Table~\ref{tab:allfields} lists the reference 
catalog used for each field.

    The ``Stage 3'' of the pipeline was then run on the calibrated exposures to
create the final mosaics using the drizzle algorithm. 
Before this process, however, another round of 
background subtraction was done for each ``cal.fits" image. This background
was also estimated by segmenting each source-masked image into blocks of 
$128\times 128$ pixels in size and applying a $3\times 3$ median filter. We
found that this extra step was necessary, otherwise the final mosaics would have
significant non-uniformity in the background.


    For all the data that we processed on our own, we produced the final stacks
at the pixel scale of 0\arcsec.06 (``60mas'') following
\citet[][]{Yan2023a, Yan2023b}. This scale critically samples
the LW data but undersamples the SW ones, however this choice
is sufficient for our purpose of dropout selection.
The JADES GOODS-S data, on the other hand, are at the pixel scale of 
0\arcsec.03 (``30mas'').

    We note that some data suffer from random contaminants of various kinds, 
which required specialized treatments. These are: 
\begin{enumerate}
	\item CEERS: the F200W images in Observation 52 have very bright edges.
	We flagged those areas as {\tt DO\_NOT\_USE} in their respective data 
    quality ({\tt DQ}) arrays so that they were not used during the drizzling 
    process. 
	\item NGDEEP: issues with flat-field calibration led to an inconsistent 
    background in F444W images, shown as small-scale fluctuations. 
    We applied a secondary flat-field correction for remedy. 
	This was done by modeling and dividing the background after the 
	original flat-fielding process during Stage 2 of the pipeline processing. 
	\item UDS1: there are broad, line-like structures striking through a
    significant portion of the field. 
	These defects are of unknown origin, and they appeared in the same sky
    areas in different observations over $\sim$6 months. Such regions are
    excluded from our analysis.
\end{enumerate}

\section{Photometry}

   We carried out source extraction and photometry using SExtractor
\citep{Bertin1996} in dual-image mode. Following \citet[][]{Yan2023a, Yan2023b},
we used the F356W mosaics for source detection, because these are the deepest
images as compared to those in other bands. An additional advantage is that the
F356W images, being taken in the LW channel, are also cosmetically cleaner
than the SW images. The convolution filter was a $5\times 5$ 
Gaussian function with the full width at half maximum (FWHM) of 2 pixels, and
the detection threshold was set to 1.0. The error maps produced from the
data reduction pipeline were used as the ``root mean square'' (RMS) maps in
the detection as well as in deriving the photometric errors.
We adopted \texttt{MAG\_ISO} magnitudes, which are robust for color 
measurements. The F356W point spread function (PSF) is comparable to that in 
F444W but is larger than those in the SW bands. The sources of interest are 
small enough that the F356W \texttt{MAG\_ISO} apertures include nearly all the
source flux while minimizing the background noise. 
Hereafter we denote the magnitudes in 
F090W, F115W, F150W, F200W, F277W, F335M, F356W, F410M, and F444W as
$m_{090}$, $m_{115}$, $m_{150}$, $m_{200}$, $m_{277}$, $m_{335}$, $m_{356}$,
$m_{410}$, and $m_{444}$, respectively. We note that F335M and F410M are two
medium bands, while all others are broad bands.
To minimize false detections, we kept only the sources that have
${\rm S/N}\geq 5.0$ and \texttt{ISOAREA\_IMAGE} $\geq 10$~pixels in F356W.

\begin{figure*}[htbp]
    \centering
    \includegraphics[width=\textwidth]{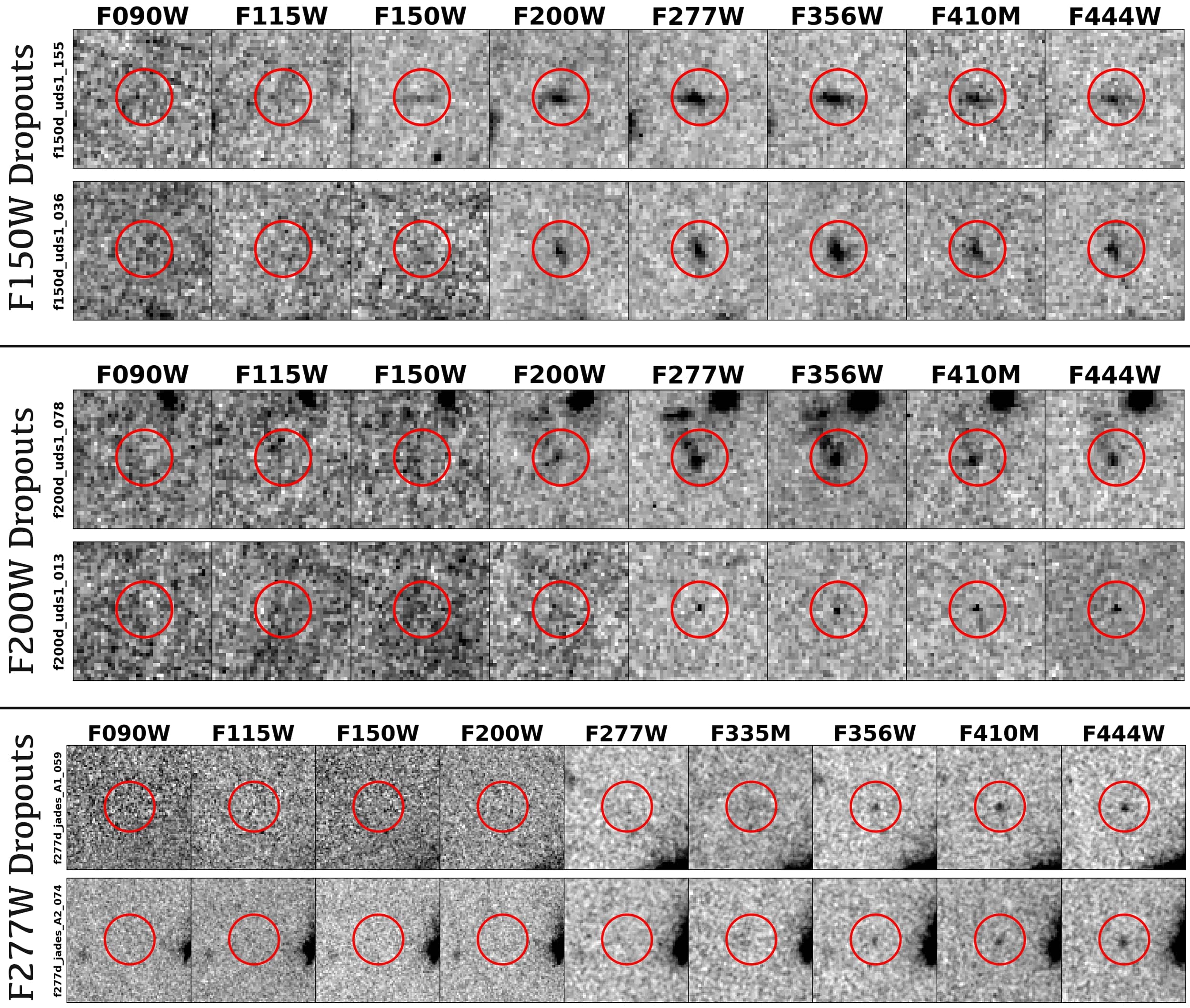}
    \caption{NIRCam image cutouts of dropout examples, two in each category.
    The cutouts are 2.\arcsec4$\times$2.\arcsec4 in size. The red circles are 
    centered on the dropout positions and are 0.\arcsec48 in radius. The image 
    cutouts of all dropouts in our sample are presented in Appendix A.}
    \label{fig:dropoutdemo}
\end{figure*}

\section{Dropout Selection and Purification}

    In this work, we select candidate galaxies at $z\gtrsim 11.3$ using the
standard dropout technique, which does not depend on the intrinsic properties
of galaxies. After the dropouts are selected, we ``purify'' the sample by
keeping those that are consistent with our best assumptions of the
intrinsic properties of galaxies at high redshifts. The purification is done
for the F150W and F200W dropouts. 

\begin{table*}[hbt!]
    \centering
    \caption{Numbers of selected dropouts}
    \resizebox{\textwidth}{!}{
    \begin{tabular}{cccccccccc} \hline
        Field & COSMOS & UDS1 & CEERS & GLASS & NGDEEP & JADES & UNCOVER & SMACS0723 & Total \\ \hline
        F150W dropouts & 23 & 15 & 15 & 7 & 4 & 20 & 18 & 21 & 123 \\ 
        S1/S2 purified & 16/16 & 4/10 & 8/6 & 3/0 & 0/1 & 5/4 & 11/10 & 7/15 & 54/62 \\ \hline
        F200W dropouts & 15 & 15 & 3 & 6 & 1 & 4 & 4 & 4 & 52 \\ 
        S1/S2 purified & 7/14 & 8/13 & 1/3 & 3/4 & 0/0 & 0/1 & 1/4 & 1/4 & 21/43 \\ \hline 
        F277W dropouts & 7 & 6 & 3 & 2 & 2 & 10 & 1 & 1 & 32 \\ \hline
    \end{tabular}
}
    \begin{flushleft}
    \tablecomments{For the F150W and F200W dropouts, the number of objects after
    the purification in Scheme 1 and 2 are also listed separately. The F277W
    dropouts are not purified due to the limited SED information.}
    \end{flushleft}
    \label{tab:counts}
\end{table*}

\subsection{Selection overview}

   The dropout identification of a high-$z$ object utilizes the fact that the
cumulative Ly$\alpha$ and Lyman limit absorptions due to line-of-sight cosmic
neutral hydrogen (H~I) towards this object effectively wipe out its emission 
bluer than the rest-frame 1216\AA, which creates an abrupt break 
- ``Lyman break'' -
in its spectral energy distribution (SED). When observed in two adjacent bands,
one straddling the break and the other sampling the redder part of the SED, the
object appears to ``drop out'' from the bluer band (hereafter the drop-out
band) and ``shift into'' the redder band (hereafter the shift-in band). Using 
the aforementioned data, we searched for candidate high-$z$ galaxies as F150W, 
F200W and F277W dropouts, respectively. Their shift-in bands are F200W, F277W
and F356W, respectively. Following \citet[][]{Yan2023a}, we adopted the 
following criteria.

    $\bullet$ The object should have ${\rm S/N}\geq 5.0$ in the shift-in band.
This is to increase the reliability of the source and to ensure a reliable
measurement of the break amplitude.

    $\bullet$ The break amplitude, i.e., the color between the drop-out band
and the shift-in band, should be $\geq 0.8$~mag. If the object has 
${\rm S/N}< 2.0$ in the drop-out band, the 2~$\sigma$ upper limit within a 
circular aperture 0\arcsec.2 in radius at the source location would be used to 
calculate the lower limit of the color index in between the two bands, and this
lower limit should be $\geq 0.8$~mag. The rationale of this choice of the
break amplitude is as follows. The throughput curves of the NIRCam bands can be
approximated by rectangles, and the truncation of a
flat spectrum (in $f_\nu$), which is characteristic of Lyman-break at high
redshifts, will create a color decrement of $\sim$0.75~mag in between the 
drop-out band and the shift-in band when the break is redshifted out halfway of
the former. For simplicity, we require the break be $\geq 0.8$~mag. 

    $\bullet$ The object should have ${\rm S/N}\geq 5.0$ in at least one other
band redder than the shift-in band. This requirement is to further increase
the reliability of the source. Recall that only objects with 
${\rm S/N}\geq 5.0$ in F356W are retained in our initial catalog; this 
requirement means that our candidates should have ${\rm S/N}\geq 5.0$ in at
least three bands and therefore are highly reliable detections.

    $\bullet$ The source should be a non-detection in all the ``veto'' bands,
i.e., the bands bluer than the drop-out band. This is because the cosmic
H~I absorption should completely wipe out any emission bluer than Lyman
break if the source is at $z>11$. Quantitatively, we require
${\rm S/N}< 2.0$ in the veto bands.

    All the initial candidates were visually examined in all NIRCam bands to
ensure that they were not affected by image defects and that they were indeed
invisible in the veto bands. When available (e.g., in the JADES GOODS-S
region), the HST ACS images were incorporated as part of the veto-band images 
and were also examined.

   A surprising fact about the dropouts that we selected is that there are some
very bright ones, some of which can be as bright as $m_{356}\lesssim 23$~mag.
Similar dropouts have been found previously by \citet[][]{Yan2023a, Yan2023b}.
The most extreme ones are non-detections in F150W (cut-off at 2.0~$\mu$m),
which qualifies them as the so-called ``HST-dark'' galaxies
\citep[e.g.,][]{Huang2011, Wang2016, Barrufet2023}
because they would be completely invisible in the reddest band of the HST
(WFC3 IR F160W; cut-off at 1.7~$\mu$m) even under the deepest exposures. 
The nature of such galaxies is still under debate
\citep[e.g.,][]{McKinney2023, Meyer2023}.
Some of our fields have deep far-IR/sub-mm data (e.g., those from Herschel
SPIRE, JCMT SCUBA2, ALMA etc.), radio data (e.g., those from the VLA) and/or
X-ray data (e.g., those from Chandra). However, the vast majority of these
very bright dropouts do not seem to be related to the sources at these
wavelengths.

   While we cannot rule out the possibility that some of these bright dropouts
could indeed be legitimate high-$z$ candidates, we take a conservative approach
in this study and exclude those with $m_{356}<26.0$~mag from our final sample.
Admittedly, the choice of this brightness threshold is somewhat arbitrary; 
on the other hand, it corresponds to $M_{\rm UV}<-22.0$~mag at $z=14$
(Lyman break moving to the red edge of F150W at this redshift) and approaches
the regime of luminous AGNs. The recently confirmed AGN at $z=10.073$ 
\citep[][]{Goulding2023} shows that AGNs indeed are already in place at 
$z\gtrsim 10$, and therefore it is not impossible that some of our very bright 
dropouts could be high-$z$ AGNs \citep[see also][]{Juodzbalis2023}. However, it 
would be inappropriate to discuss them in the context of ``normal'' galaxies.
Therefore, we believe that it is prudent to exclude the bright dropouts from
the ``normal'' galaxy sample, and we will defer the discussion
of such bright dropouts to a future paper.

    In total, we selected 123 F150W dropouts, 52 F200W dropouts and 32 F277W 
dropouts. Given the rectangle-shaped throughput curves and the adopted break 
amplitude as explained above, the redshift selection windows for F150W, F200W
and F277W dropouts are $11.3\lesssim z\lesssim 15.4$ 
(hereafter $z\approx 12.7$), $15.4\lesssim z\lesssim 21.8$ 
(hereafter $z\approx 17.3$) and $21.8\lesssim z\lesssim 28.3$
(hereafter $z\approx 24.7$), respectively.
Table~\ref{tab:counts} gives the detailed breakdown of the selected dropouts in 
each field. Figure~\ref{fig:dropoutdemo} shows the NIRCam image cutouts of two  
example objects in each category. The full catalogs of these dropouts are given 
in Appendix A, together with their image cutouts.

\begin{figure*}[htbp]
    \centering
    \includegraphics[width=0.8\textwidth]{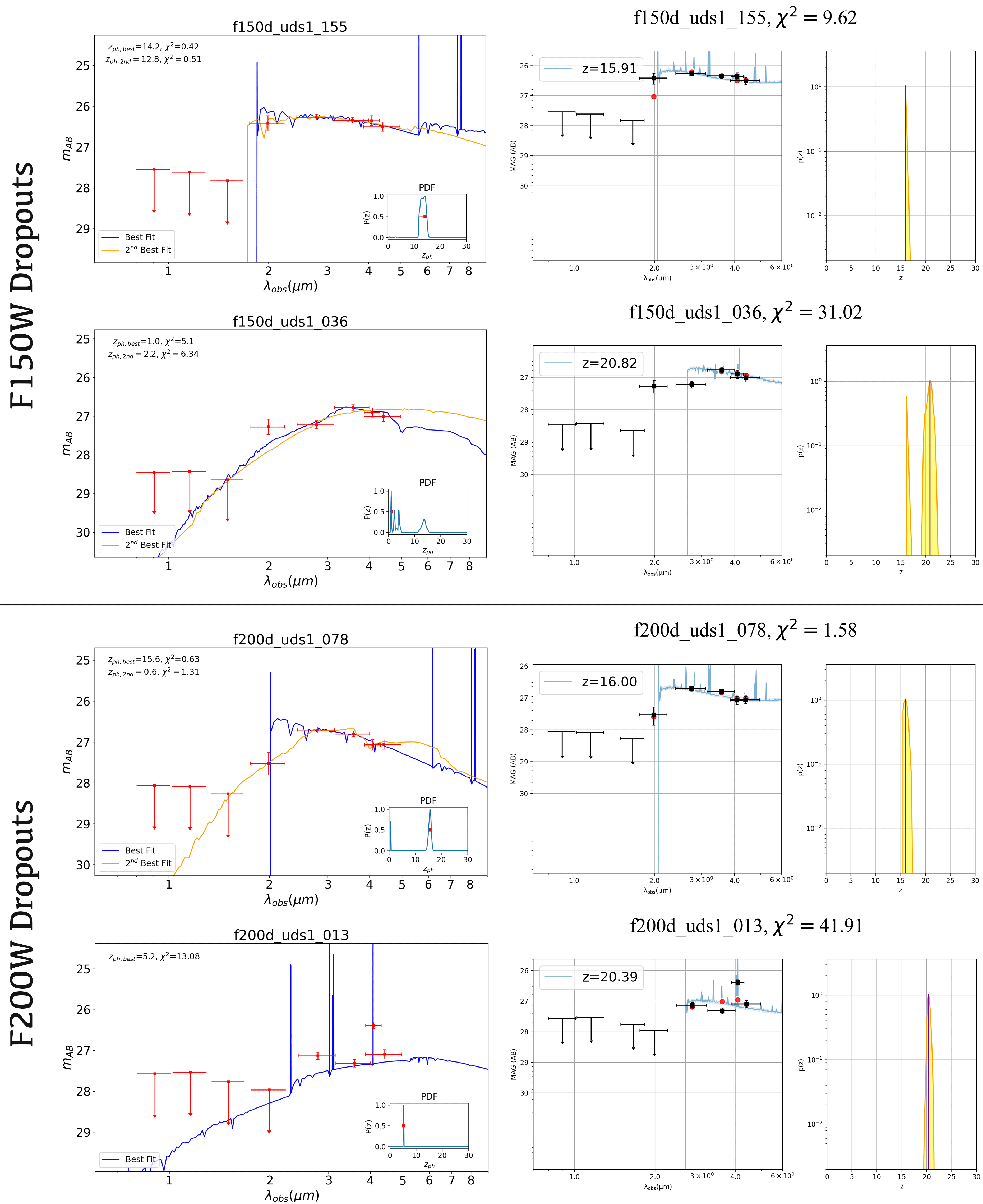}
    \caption{Demonstration of contaminant rejection by SED fitting, using two 
    F150W dropouts (top two rows) and two F200W dropouts (bottom two rows) 
    shown in Figure~\ref{fig:dropoutdemo} as examples. The left panels are
    the SED fitting results in ``Scheme 1'' (S1) using Le Phare and BC03
    models, while the right panels are those in ``Scheme 2'' (S2) using EAZY
    and the ``set 3+4'' templates. In all cases, the data points
    and limits form the SED based on the photometry described in 
    Section 4, and the curve is the best-fit spectrum (bright and dark blue in 
    S1 and S2, respectively) corresponding to the first peak of the probability 
    distribution functions (PDF; displayed in the inset in S1 and in the
    smaller panel to right in S2) that gives $z_{\rm ph}$ of the
    object. The $\chi^2$ value is also labeled. In S1, the model spectrum 
    corresponding to the secondary peak of the PDF is also shown (orange 
    curve), together with the secondary $z_{\rm ph}$ and $\chi^2$. 
    Among these examples, \texttt{f150d\_uds1\_155} and 
    \texttt{f200d\_uds1\_078} pass the purification in both S1 and S2. On the
    hand, \texttt{f150d\_uds1\_036} and \texttt{f200d\_uds1\_013} do not pass 
    either S1 ($z_{\rm ph}$ being too low) or S2 ($\chi^2$ being too high) and
    are rejected as contaminants. The full SED fitting results of all F150W and 
    F200W dropouts in our sample are presented in Appendix A.}
    \label{fig:seddemo}
\end{figure*}

\subsection{Contamination assessment and purification of F150W and F200W dropouts}

   Dropout selections must consider the effect due to possible contaminators. 
Hereafter we confine our analysis to the F150W and F200W dropouts, because 
most of our F277W dropouts are only detected in two reddest broad bands
(F356W and F444W) and their SEDs do not contain sufficient information for
reliable diagnostics.

   In the dropout searches at lower redshifts ($z\approx 7$--9 and lower), a
common type of contaminants are Galactic brown dwarfs because their broad
molecular absorption bands could create features in the SEDs that mimic Lyman
break. For this reason, point sources are usually excluded from a dropout 
sample for galaxies. As pointed out in \citet[][]{Yan2023a}, however, brown 
dwarfs are not likely a major source of contamination because their molecular
absorption bands do not coincide with the dropout bands of interest for the 
search at $z>11$. On the other hand, \citet[][]{Yan2023c} shows that many
point-like dropouts in SMACS0723 have SEDs that could be explained by 
supernovae at much lower redshifts ($z\lesssim 7$), which justifies the
rejection of point-like dropouts. Interestingly, there are only three
such sources among the dropouts that have $m_{356}>26.0$~mag, all of which
are in SMACS0723. These three are already excluded from our final catalog.

   The most severe contamination to our sample could be caused by ordinary
galaxies at low redshifts that have red SEDs due to either their old stellar
populations and/or dust reddening. In the pre-JWST era when the dropout
selections had only one or two bands redder than the shift-in band, assessing
the rate of contamination of this type was usually done by color-color diagram
diagnostics. Such traditional diagnostics using the NIRCam bands were 
demonstrated in \citet[][]{Yan2023a} in SMACS0723. 
\citet[][]{Yan2023a} also took an alternative, more appropriate approach
for their F150W and the F200W dropouts because there are two to three bands
(among the total of six bands)
redder than the shift-in band. This afforded the opportunity to use 
SED fitting to screen out low-$z$ interlopers, which is equivalent to 
using the different projections of the color space simultaneously but has the
advantage of being able to consider a very wide range of contaminant SEDs. 
Here we use an approach similar to that of \citet[][]{Yan2023a} to purify the 
F150W and the F200W dropouts, bearing in mind a possible caveat that this
approach relies on the assumption of the intrinsic SED properties of 
high-$z$ galaxies that are still unknown to us. 
We refrained from applying this purification method to the F277W dropouts, as 
most of them only have detections in two bands (F356W and F444W) redder than
the shift-in band and an SED fitting would be highly uncertain.

    It is well known that SED fitting results depend on the software and
the templates in use. While an exhaustive test is impossible, we carry out
the purification of our F150W and F200W dropouts in two schemes that utilize
vastly different fitting tools and templates. Hereafter these two schemes are
referred to as ``S1'' and ``S2'', respectively. In S1, we use Le Phare 
\citep[][]{Arnouts1999, Ilbert2006} as the fitting tool and the population
synthesis models of 
\citet[][BC03]{Bruzual2003} to construct the templates. We assume 
exponentially declining star formation histories in the form of 
SFR $\propto e^{-t/\tau}$, where $\tau$ ranges from 0 to 13~Gyr
(0 for SSP and 13~Gyr to approximate a constant star formation). These models
adopt the Chabrier initial mass function \cite[][]{Chabrier2003}. We apply the
Calzetti extinction law \citep{Calzetti2001}, with $E(B-V)$ ranging from
0 to 1.0 mag. On top of these continuum models, nebular emission lines are 
added by turning on the {\tt EM\_LINES} option in Le Phare. 
In S2, we use EAZY \citep[][]{Brammer2008} as the fitting tool and adopt two 
sets of new templates from \citet[][]{Larson2022}, which are currently
referred to as the ``set 3 and 4'' (here after ``3+4'') templates when used 
in EAZY. These templates are tuned to derive photometric redshifts for
galaxies at $z>8$. To a certain degree, these choices of templates for S1 and
S2 examine our dropouts from two different angles: 
S1 is more on testing how severely they could be
contaminated by low-$z$ interlopers, while S2 is more on testing how well 
they could be consistent with being at high-$z$ under the templates that are 
optimized for the high-$z$ interpretation. In both schemes, we reject the 
fits that violate the upper limits in the photometry. This is done in S1 by 
setting the magnitude error to $-1.0$ in Le Phare. 
In S2, we modify the EAZY code to implement this functionality. 
As a common practice 
in SED fitting, we add in quadrature 0.05~mag to the measured photometric 
errors in both schemes to account for the possible system effects between 
bands as well as the imperfectness of the templates.
We do not apply any magnitude prior in either scheme.

   In S1, an F150W (F200W) dropout is retained only when it satisfies these 
conditions: ($i$) the primary solution gives $z_{\rm ph} \geq 11$ 
($\geq 15.4$), ($ii$) $\chi^2\leq 10$ for the primary solution, and ($iii$) 
if the probability density function $P(z)$ have multiple peaks, the integral of 
$P(z)$ over $z\geq 10$ ($\geq 14$) should be at least twice as high as the 
integral of $P(z)$ over lower redshifts. In S2, the same requirement ($i$) in
S1 is also used for purification. We find that the 
implementation of using upper limits and the adoption of the 3+4 templates 
largely eliminate secondary peaks in $P(z)$, and therefore the requirement
of $P(z)$ integrals ($iii$) is not necessary. However, the fits tend to have
larger $\chi^2$ as compared to those in S1. Therefore, we relax the
requirement $(ii)$ to $\chi^2\leq 20$ in S2. Figure \ref{fig:seddemo} shows 
some of the retained and the rejected dropouts in both schemes. The full SED 
fitting results of all the F150W and F200W dropouts are presented in Appendix A.


\section{Luminosity Functions at $z\approx 12.7$ and 17.3}

    Assuming that our F150W and F200W dropouts thus purified are indeed at
$z\gtrsim 11.3$ as expected, here we derive the luminosity functions of 
galaxies at the nominal redshifts of $z\approx 12.7$ and $z\approx 17.3$.

\subsection{Effective areas for density calculation}

   A problem with the aforementioned surveys is that most of them have highly
non-uniform exposures across their survey areas, which leads to highly
non-uniform depths within the surveys. The consequence is that the effective
area used for the dropout source density calculation within a given survey is
not a fixed number. For illustration, imagine a faint dropout found in a deep
region of a survey that would not even be detected in its shallower regions.
The effective survey area for this dropout cannot be the full area that this
survey extends to but must be a smaller area within which such a dropout could
be detected. The adopted dropout selection criteria further complicate the
situation. Simply put, the effective area must be derived on a per-source
basis. For a given dropout, the procedures to derive its effective area are 
outlined as follows.

    {\it Step 1:}\  We first determine $A_1$, the region where the 5~$\sigma$ 
depth in F356W is the same or deeper than $m_{356}$ of the dropout. This 
corresponds to the selection criterion that a valid dropout should have
${\rm S/N}\geq 5$ in F356W. 

    {\it Step 2:}\  Within $A_1$, we then determine $A_{12}$, the region where
the 5~$\sigma$ depth in the shift-in band is the same or deeper than the 
magnitude of the dropout in this band. This corresponds to the criterion of 
${\rm S/N}\geq 5$ in the shift-in band. 

    {\it Step 3:}\ Within $A_{12}$, we further determine $A_{123}$, the region
where the color limit as calculated by the 2~$\sigma$ depth in the drop-out
band and the dropout's magnitude in the shift-in band is at least 0.8~mag.

    The value of $A_{123}$ is then adopted as the effective survey area,
$A_{\rm eff}$, of the given dropout. In case when the dropout is within a
cluster field, we need to take into account the fact that it would be 
magnified differently at other locations in the field. For this purpose, we 
adopt the lens model of \citet[][]{Furtak2023} for the UNCOVER field and that
of \citet[][]{Pascale2022} for the SMACS0723 field, respectively.
The magnitudes of the dropout are
replaced by its de-magnified magnitudes while the depths of the images at
different locations are artificially increased by the amounts as predicted by
the lens model. We then perform the aforementioned three steps to derive the
effective survey area.

    In all the above, the depths are obtained by using the ``depth maps'' that
are appropriate for the dropout under question. The depth maps are explained
in Appendix B.

\subsection{Incompleteness correction}

    Before using the density calculated above to derive the luminosity
functions, the incompleteness of the dropout selections must be corrected. This
is done by putting simulated dropouts in the images and check their recovery
rates. One approach to simulate such galaxies is to use analytic functions
(such as the S\'{e}rsic profile). However, this approach has the problem that
analytic functions rarely reproduce the morphologies of real dropouts.
We opt to use the real dropouts in our sample for simulation and to derive the
incompleteness for each object separately. The basic procedure is as follows.

    For a given dropout in a given band, we cut out a square area of
20$\times$20 pixels centered on this object and then put this cut-out at a
large number of random positions on the science image, which is done by
replacing the square area of the same size centered on these random positions 
with the cut-out. This is repeated for all relevant bands. Note that this is 
done only on the science images; the RMS maps are not altered. Similar to the 
argument above for the effective area, we need to consider the exposure 
non-uniformity across the survey field. It is possible that a faint dropout 
found in a deep region would not even be detected in a shallow region; 
therefore, for the simulation of a given dropout we only use the random 
positions in the area that has exposure time within $\pm 20\%$ of the exposure 
time where the dropout was found. We require that there should be $\geq 300$ 
usable random positions so that there are sufficient statistics. As the shape 
of the non-uniformity is highly irregular, it is impossible to generate the 
random positions only in the usable area; instead, the random positions are 
generated over the full field of a given survey, and the usable positions are 
selected based on the exposure time criterion. If one run does not produce
$\geq 300$ usable random positions, the process is repeated until the 
condition is satisfied.

    We then carry out photometry on the simulated images (in the same way as
described in Section 4) and check whether those simulated objects would be
recovered. The recovery rate, $p_{\rm r}$, is the ratio between the number of
simulated objects recovered and the total number of usable simulated objects.
The incompleteness correction factor, $f_{\rm ic}$, is simply $1/p_{\rm r}$.

    To reiterate Section 5, there are five parts in the dropout selection
criteria: it should have (1) the minimum size of $\geq 10$ pixels and
${\rm S/N}\geq 5$ in F356W, (2) break amplitude of $\geq 0.8$~mag, (3)
${\rm S/N}\geq 5$ in the shift-in band, (4) ${\rm S/N}\geq 5$ in at least one
more band other than F356W and the shift-in band, and (5) non-detection in all
the veto bands. In recovering the simulated objects for each real dropout, we
do not need to consider part (5) because the objects in our sample have been
visually vetted and looking at the same ``simulated'' veto-band image for
$\geq 300$ times would still result in the same conclusion because we would
be examining the replacement from the real veto-band image that we already
examined.

    A subtle but important point is that the recovering process should not
include part (2), i.e., we do not need to consider the break amplitude of
$\geq 0.8$~mag. This is because the differences in the break amplitude of the
simulated objects are dictated by photometric errors, which follow a normal
distribution. To understand this point, let us use a dropout that has the
break amplitude of 0.8~mag for illustration. Such a source is at the boundary
of being selected, and a slight error in photometry could scatter it below the
selection threshold. On the other hand, a source with the break amplitude
slightly smaller than 0.8~mag could also be scattered above the selection
threshold due to photometric errors. In other words, as long as the 
photometric errors are not skewed, a source would have equal chance of being 
scattered above or below the break amplitude threshold. Therefore, the break 
amplitude requirement should not be considered in the recovering process.

\subsection{Results}

  After obtaining the effective area and the incompleteness correction factor 
for each dropout, the luminosity functions can be constructed. As mentioned 
above, we use the F150W dropouts and the F200W dropouts retained in the 
purified samples to derive the luminosity functions at $z\approx 12.7$ and 
$17.3$, respectively. As we apply two different contaminant removal schemes, 
Scheme 1 and 2, we derive two versions of luminosity functions accordingly 
for each redshift range.

   As our source detection was based on the F356W images, we first calculate
the dropout surface density as a function of the 
apparent magnitude $m_{356}$ in the step size of $\Delta m = 0.5$~mag.
For the dropouts in the two cluster fields, their $m_{356}$ values have been
corrected for the magnification (see Section 6.1).
Within a given survey $s$, the contribution of the dropout $i$ to the total
surface density is $dc^i=f_{\rm ic}^i/A_{123}^i$, where $A_{123}^i$ is
its effective area and $f_{\rm ic}^i$ is its incompleteness correction factor,
respectively. The error due to the Poissonian noise is $edc^i=dc^i$.
The total surface density within the magnitude bin $k$ inferred from this 
survey, $\sigma_s$, is then the sum of the contributions from all the 
dropouts within this bin, i.e., $\sigma_s=\sum_i dc^i$, and its associated 
error, $\Delta \sigma_s$, is $edc^i$ added in quadrature, i.e.,
$\Delta \sigma_s=\sqrt[]{\sum_i (edc^i)^2}$.
The final surface density in this magnitude bin inferred from all surveys is 
then the weighted average of the densities from all surveys, 
$\sum_s \sigma_s w_s/\sum_s w_s$, where 
$w_s=(\Delta \sigma_s)^{-2}$ is the weight of the survey $s$.

\begin{figure*}[htbp]
    \centering
    \includegraphics[width=0.8\textwidth]{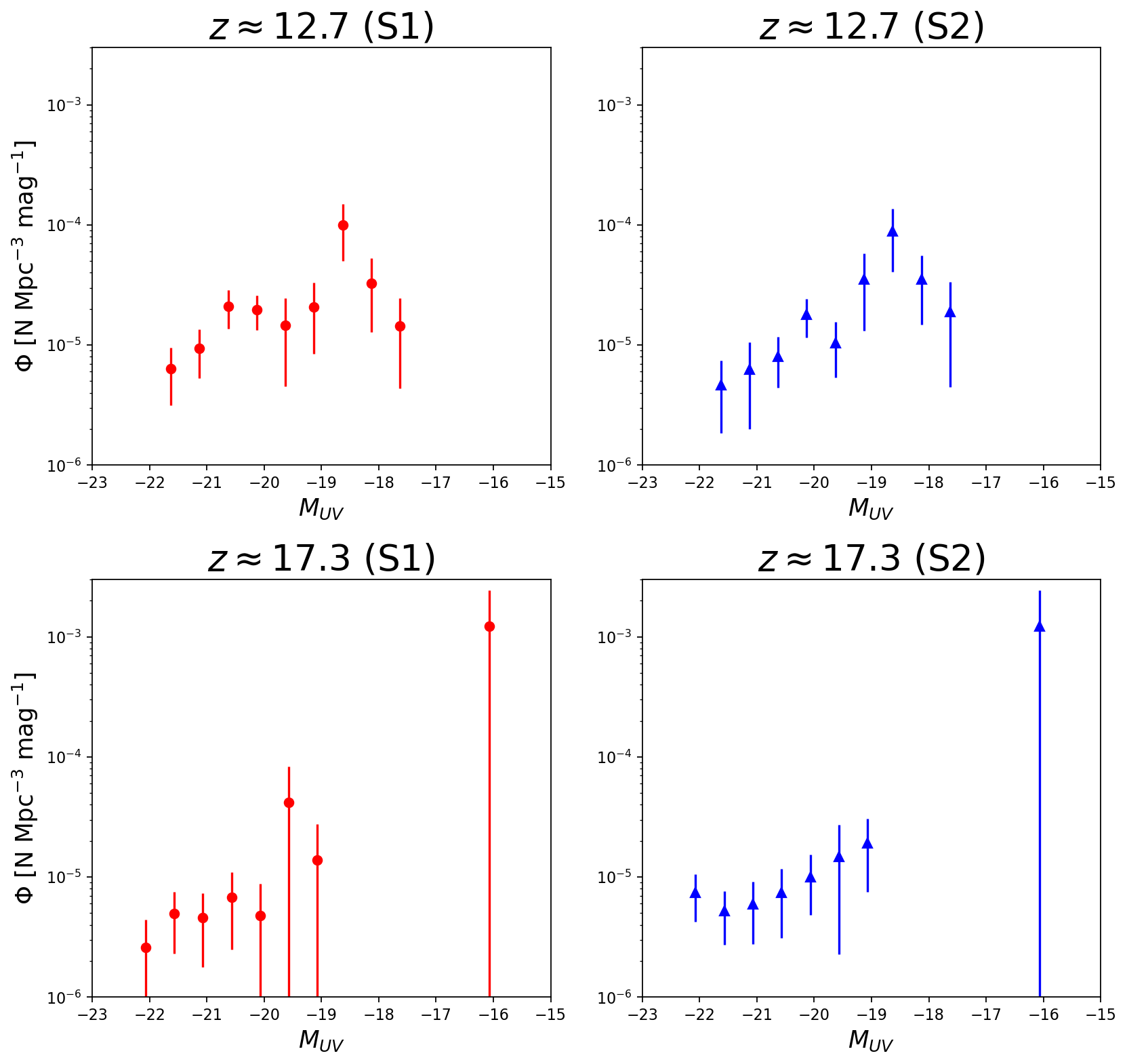}
    \caption{Luminosity functions at $z\approx 12.7$ (top) and 17.3 (bottom)
    based on the purified samples in S1 (left, red symbols) and S2 (right, blue
    symbols).
}
    \label{fig:LFdata}
\end{figure*}

   We express the luminosity functions in terms of number density per unit
magnitude as a function of absolute magnitude. Dividing the surface density 
calculated above by the co-moving volume per unit area, one gets the number 
density. As discussed in Section 5.1, the redshift selection windows for the 
F150W and the F200W dropouts can be approximated by a block function over
$11.3\leq z\leq 15.4$ and $15.4\leq z\leq 21.8$, respectively. The co-moving 
volume per unit area at these two redshift ranges is 
$5.51\times 10^3$~Mpc$^{-3}$~arcmin$^{-2}$ and
$6.08\times 10^3$~Mpc$^{-3}$~arcmin$^{-2}$, respectively. To convert $m_{356}$
to absolute magnitude $M_{\rm UV}$, we adopt $z=12.7$ and 17.3 for the F150W
and the F200W dropouts, respectively.
These luminosity functions are tabulated in Table \ref{tab:LF} and shown in 
Figure \ref{fig:LFdata}. 

\begin{table*}[hbt!]
    \centering
    \caption{Luminosity functions at $z\approx 12.7$ and 17.3}
    \resizebox{\textwidth}{!}{
    \begin{tabular}{cccccccccc} \hline
        $M_{UV}$, $z\approx12.7$ & $-$21.63 & $-$21.13 & $-$20.63 & $-$20.13 & $-$19.63 & $-$19.13 & $-$18.63 & $-$18.13 & $-$17.63 \\ \hline
        $\Phi$ (S1, $10^{-6}$~Mpc$^{-3}$~mag$^{-1}$) & 6.32\textpm3.16 & 9.40\textpm4.10 & 21.14\textpm7.40 & 19.60\textpm6.32 & 14.57\textpm10.05 & 20.73\textpm12.23 & 100.30\textpm49.95 & 32.81\textpm20.06 & 14.45\textpm10.09 \\ 
        $\Phi$ (S2, $10^{-6}$~Mpc$^{-3}$~mag$^{-1}$) & 4.65\textpm2.80 & 6.24\textpm4.25 & 8.06\textpm3.63 & 17.86\textpm6.35 & 10.46\textpm5.12 & 35.51\textpm22.33 & 89.05\textpm48.21 & 35.29\textpm20.55 & 18.95\textpm14.49 \\ 
        $\Phi$ (Avg, $10^{-6}$~Mpc$^{-3}$~mag$^{-1}$) & 5.48\textpm2.98 & 7.82\textpm4.18 & 14.60\textpm5.83 & 18.73\textpm6.33 & 12.51\textpm7.98 & 28.12\textpm18.00 & 94.68\textpm49.09 & 34.05\textpm20.31 & 16.70\textpm12.48 \\
        \hline \hline
        $M_{UV}$, $z\approx17.3$ & $-$22.07 & $-$21.57 & $-$21.07 & $-$20.57 & $-$20.07 & $-$19.57 & $-$19.07 & ... & $-$16.07 \\ \hline 
        $\Phi$ (S1, $10^{-6}$~Mpc$^{-3}$~mag$^{-1}$) & 2.59\textpm1.84 & 4.93\textpm2.63 & 4.57\textpm2.79 & 6.74\textpm4.24 & 4.76\textpm4.04 & 41.62\textpm41.62 & 13.88\textpm13.88 & ... & 1225.50\textpm1225.50 \\
        $\Phi$ (S2, $10^{-6}$~Mpc$^{-3}$~mag$^{-1}$) & 7.40\textpm3.16 & 5.19\textpm2.47 & 5.95\textpm3.19 & 7.40\textpm4.31 & 10.06\textpm5.23 & 14.76\textpm12.50 & 19.07\textpm11.54 & ... & 1225.50\textpm1225.50 \\ 
        $\Phi$ (Avg, $10^{-6}$~Mpc$^{-3}$~mag$^{-1}$) & 5.00\textpm2.58 & 5.06\textpm2.55 & 5.26\textpm3.00 & 7.07\textpm4.27 & 7.41\textpm4.67 & 28.19\textpm30.73 & 16.47\textpm12.76 & ... & 1225.50\textpm1225.50 \\ 
        \hline
    \end{tabular}
    }
    \begin{flushleft}
       \tablecomments{The faint-end values of these luminosity functions are
       not reliable due to the selection limits (see Section 6.4). These happen
       at $M_{UV}>-18.1$~mag and $>-19.1$~mag for $z\approx 12.7$ and 17.3,
       respectively.}
    \end{flushleft}
    \label{tab:LF}
\end{table*}

\subsection{Systematic effects}

   These luminosity functions, while derived based on the largest public 
dataset available in Cycle 1, still suffer from some caveats. To understand
these, we first note that the fields in use contribute to different 
parts of the luminosity functions. The dividing point is roughly at 
$m_{356}\approx 28.5$~mag: the data points brighter than this threshold are
mostly contributed by the three wide fields (COSMOS, UDS1 and CEERS), while
those fainter than this threshold are mostly contributed by the two deep
fields (JADES and NGDEEP) and the lensing fields (UNCOVER and SMACS0723).
The selection limits of these fields, therefore, affect our results at 
different luminosity ranges and create some artificial ``features'' in our 
luminosity functions.

\begin{figure*}[htbp]
    \centering
    \includegraphics[width=0.8\textwidth]{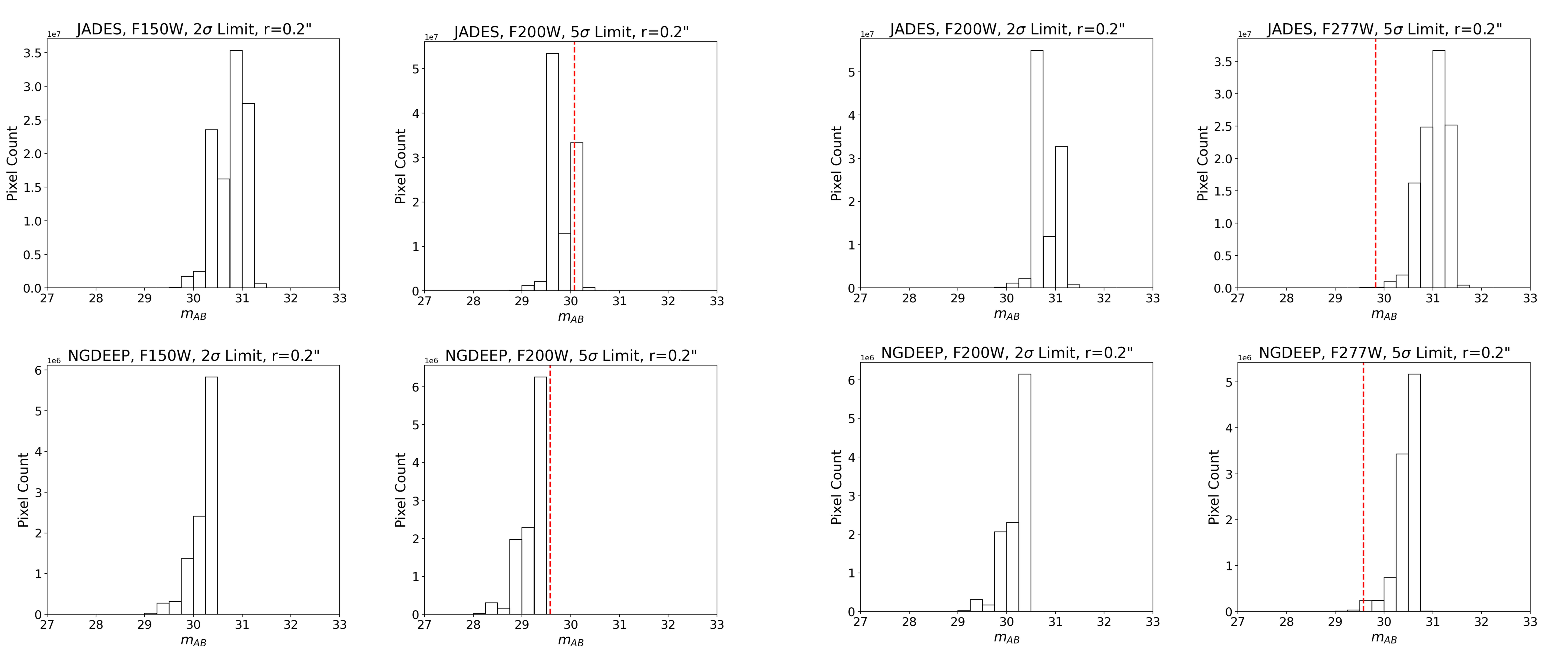}
    \caption{Image detection limits in JADES GOODS-S (top row) and CEERS
    (bottom row) relevant to the selections of F150W dropouts (left panels)
    and F200W dropouts (right panels). The histograms show the 2~$\sigma$ limits
    ($m_{2\sigma}$) measured within a circular aperture of 0.\arcsec2 radius 
    centered on each
    pixel of the drop-out images (the F150W image for F150W dropout 
    selection and the F200W image for F200W dropout selection) and the
    5~$\sigma$ limits measured in the same way for the shift-in images (the
    F200W image for F150W dropout selection and the F277W image for F200W
    dropout selection). Given the adopted dropout amplitude of 0.8~mag, only
    the objects that have $\leq m_{2\sigma}-0.8$~mag in the shift-in image
    (indicated by the vertical dashed red line) could be selected. In the
    meantime, a valid candidate should have ${\rm S/N\geq 5}$ in the shift-in
    band, and therefore the brighter of the two (5~$\sigma$ limit and 
    $\leq m_{2\sigma}-0.8$~mag) is the limit of the dropout selection. These
    limits of these two deepest fields
    explain the faint-end cutoffs of the luminosity functions. See Section 6.4
    for details.
}
    \label{fig:slimitdeep}
\end{figure*}

\begin{figure*}[htbp]
    \centering
    \includegraphics[width=0.8\textwidth]{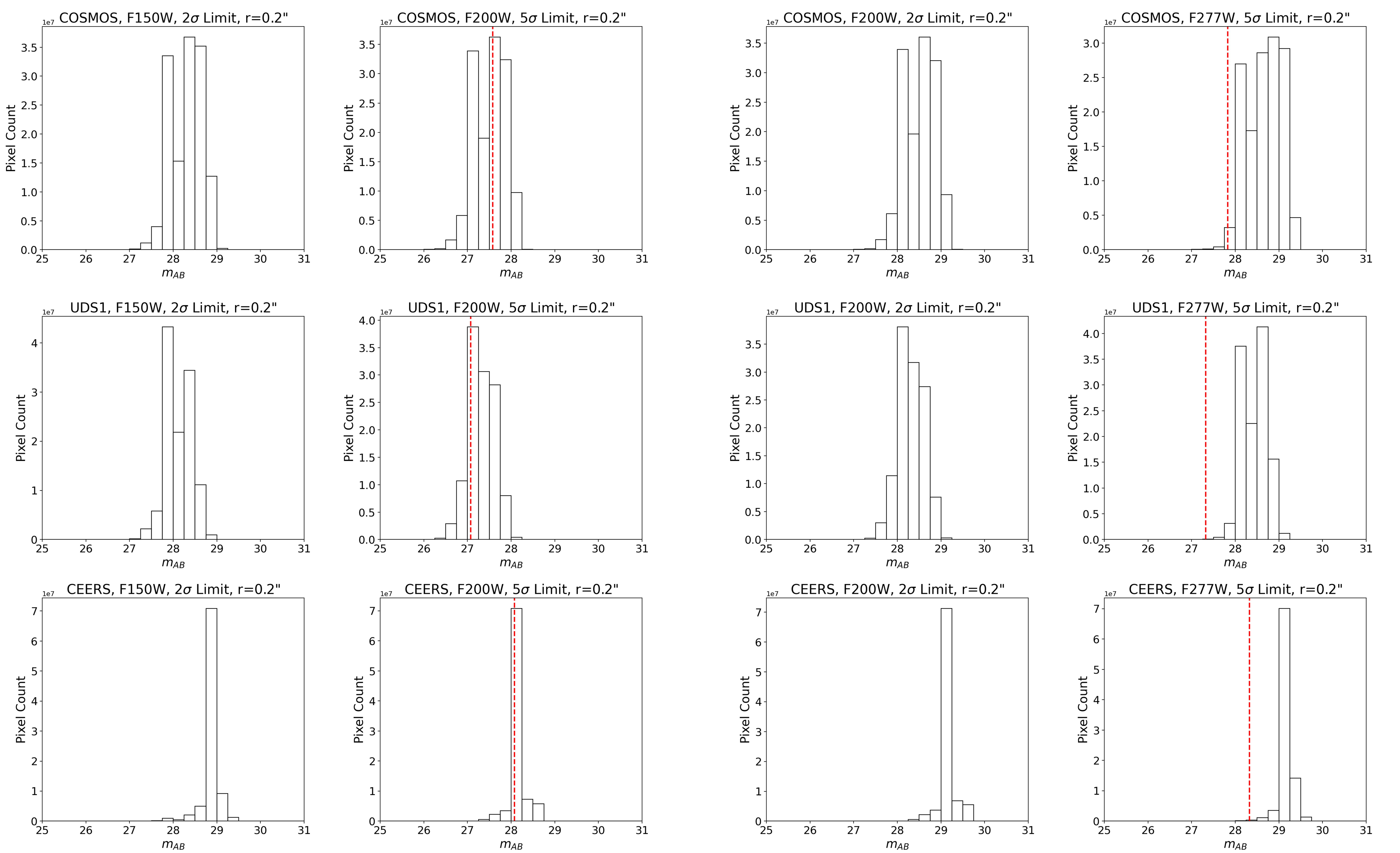}
    \caption{Similar to Figure~\ref{fig:slimitdeep} but for COSMOS (top row), 
    UDS1 (middle row) and CEERS (bottom row). The selection limits of these
    three shallow-and-wide fields explain the ``dip'' at the bright-end of
    the luminosity functions. See Section 6.4 for details.
}
    \label{fig:slimitwide}
\end{figure*}

    The most obvious one is the sharp ``drop-off'' at the faint-end of the 
$z\approx 12.7$ luminosity function based on the F150W dropouts. This happens
at $M_{UV}\gtrsim -18.1$~mag in both S1 and S2, which corresponds to the
$m_{356}$ bins of 29.5--30.0~mag and fainter. Figure~\ref{fig:slimitdeep} 
illustrates how the 
selection limits of F150W dropouts in JADES and NGDEEP could introduce the 
drop-off. The peak of 2~$\sigma$ limit (calculated within a circular aperture 
0.\arcsec2 in radius) of the F150W (the dropout band) mosaic in JADES 
(NGDEEP) is $m_{150}\approx 31.0$ (30.4)~mag, and the F150W dropout selection 
could reach $m_{200}\approx 30.2$ (29.6)~mag because the imposed amplitude 
for Lyman break is 0.8~mag. However, the peak of 5~$\sigma$ limit of the 
F200W (the shift-in band) mosaic in JADES (NGDEEP) is $m_{200}\approx 29.6$ 
(29.4)~mag, which means that the selection is largely limited at around this 
brighter level in most of the field. Assuming that the SEDs of a high-$z$
galaxy is mostly flat in $f_\nu$, this limit roughly correspond to the 
$m_{356}$ bin of 29.5--30.0~mag. Some F150W dropouts in our sample are fainter
than this limit, and this is because (1) the sizes of these objects are smaller
than the circular aperture used in Figure~\ref{fig:slimitdeep} and therefore
have S/N higher than the 
illustration and (2) they tend to locate in the regions deeper than the peak 
of the histogram in Figure~\ref{fig:slimitdeep}. 
Similarly, the same effect also impacts the 
$z\approx 17.3$ luminosity function based on the F200W dropouts. If we 
discard the data point at $M_{UV}=-16.1$~mag from a single, lensed object, 
this luminosity function truncates after $M_{UV}=-19.1$~mag, which 
also corresponds to $m_{356}$ bins of 29.5--30.0~mag and fainter. As shown in 
Figure~\ref{fig:slimitdeep}, the peak of 5-$\sigma$ limit of the F277W 
(the shift-in band) mosaic is $m_{277}\approx 31.1$ (30.6)~mag in JADES
(NGDEEP), however the peak of 2-$\sigma$ limit of the F200W (the dropout band)
mosaic is $m_{200}\approx 30.6$ (30.4)~mag, which means that the selection of 
F200W dropouts is largely confined to $m_{277}\lesssim 29.8$ (29.6) in most of
the JADES (NGDEEP) field.

   Another ``feature'' is the ``dip'' at $M_{UV} \approx -19.6$~mag in the
$z\approx 12.7$ luminosity function, which corresponds to the $m_{356}$ bin
of 28.0--28.5~mag. In S1, this dip is wider and includes the bin of
27.5--28.0~mag. Similar to the above, this is caused by the selection limits
in the three wide fields, which are illustrated in the left panels of 
Figure~\ref{fig:slimitwide}. The peaks of
2~$\sigma$ limits of the F150W mosaics in COSMOS, UDS1 and CEERS are
$\sim$28.5, 28.0 and 28.8~mag, respectively, and the peaks of 5~$\sigma$ 
limits of the F200W mosaics are $\sim$27.7, 27.5 and 28.1~mag, respectively. 
Therefore, the F150W dropout selection limits in these three fields are
$\sim$27.7, 27.2 and 28.0~mag, respectively, which largely explain the dip.
Such a dip also presents in the $z\approx 17.3$ luminosity function in S1 at
$M_{UV} \approx -20.1$~mag (corresponding to the same $m_{356}$ bin of
28.0--28.5~mag), which is caused by the similar reason (see the right panels of
Figure~\ref{fig:slimitwide}). 
The dip is less severe in S2, which can be explained if S2 accidentally 
rejects less contaminants in this bin.

   The magnitudes quoted above are the representative numbers around which
our dropout selections truncate in a given field. Due to the non-uniformity of
survey depth across a field, the actual limits are slightly different in 
different regions. Nonetheless, such selection limits impose systematic
effects that cannot be corrected by the incompleteness corrections as 
described in Section 6.2, because these are ``hard limits'' beyond which no 
objects would be selected.

\begin{figure*}[htbp]
    \centering
    \includegraphics[width=0.95\textwidth]{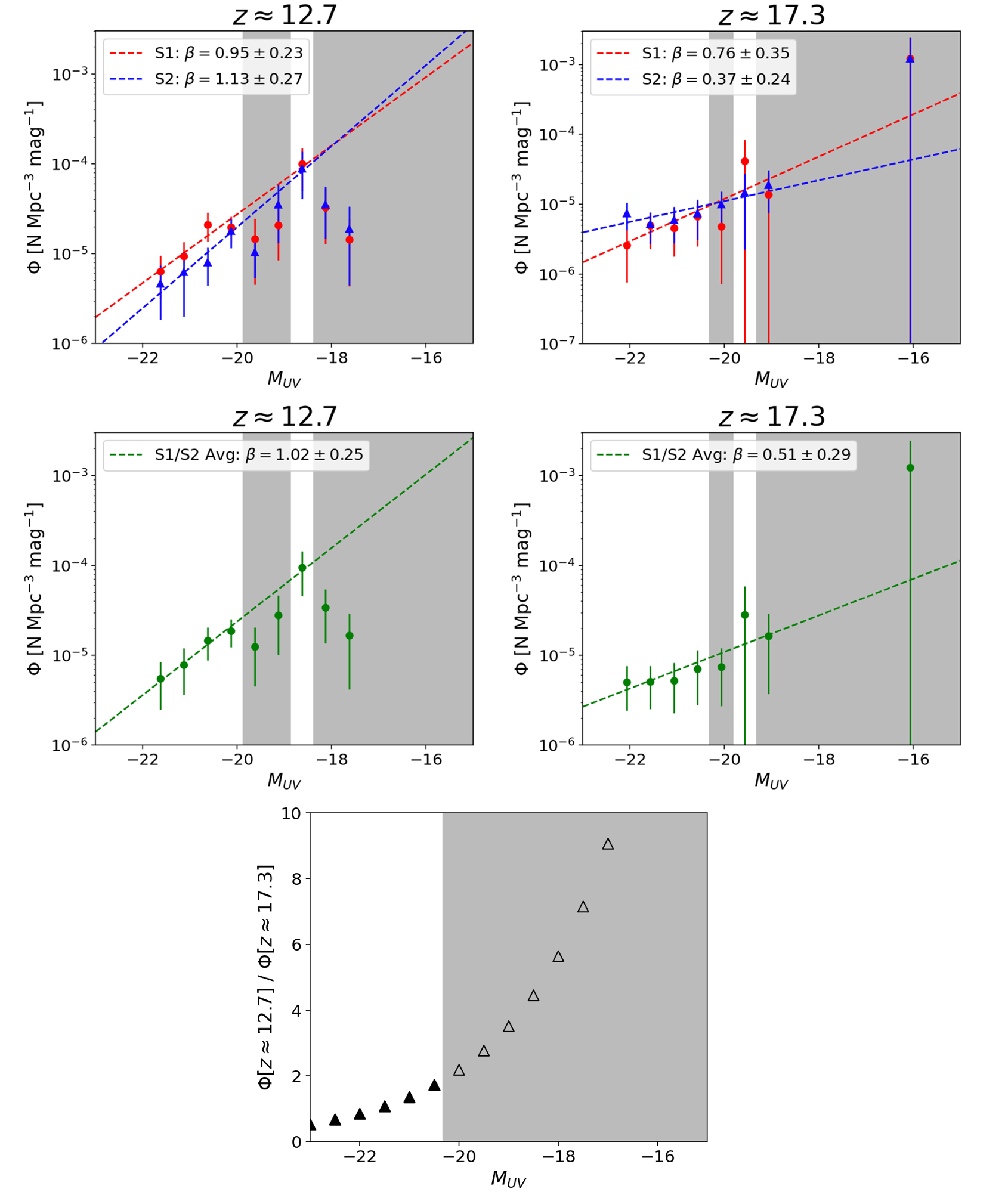}
    \caption{Power-law fits to the luminosity functions at $z\approx 12.7$ and
    17.3 as noted, following the formalism of Equation 2.
    The top two panels show the fits in S1 (red symbols)
    and S2 (blue symbols) separately, while the ones in the middle panel show
    the fits to the averaged results of S1 and S2 (green symbols). The grey 
    regions indicate the ranges where the luminosity functions are not reliable
    due to the hard limits in our dropout selections (see Section 6.4). The
    evolution between the two epochs is shown in the bottom panel as the ratio
    of the two luminosity functions, which is only a factor of $\lesssim 2$
    over the valid luminosity range (filled triangles).
}
    \label{fig:LFpw}
\end{figure*}


\section{Discussion}

   We first discuss the overall behavior of the luminosity functions obtained
above. Given the systematic effects described in Section 6.4, here we exclude
the faint-end data points at $M_{\rm UV}\gtrsim -18.1$~mag and the ``dip'' at 
$M_{\rm UV} \approx -19.6$~mag in the $z\approx 12.7$ luminosity function. 
For the $z\approx 17.3$ luminosity function, the ones at 
$M_{\rm UV}\gtrsim -19.1$~mag and $M_{\rm UV} \approx -20.1$~mag are 
excluded. 

   At both redshifts, the luminosity functions are better described by power
law than the conventional Schechter function. Therefore, we parameterize them
using the following form:
\begin{equation}
   \Phi = \Phi_0 L^{-\beta}.
\end{equation}
Considering that AB magnitudes are used and that 
${\rm AB}=-2.5\times\log(f_\nu)-48.6$, the above can be rewritten as

\begin{equation}
    \log(\Phi) = \log(\Phi_0) + 0.4 \beta (M + 40.85).
\end{equation}
The top panel of Figure \ref{fig:LFpw} shows the power-law fits to the 
luminosity function at both redshifts in S1 and S2 separately. As the 
behaviors in these two schemes are similar enough, we believe that the 
average of the two is more robust. This is shown in the middle panel of
Figure \ref{fig:LFpw}, and the best-fit parameters are

\begin{equation}\label{eq:coef}
    \beta, \log(\Phi_0) =
    \begin{cases}
        1.02\pm 0.25, -13.16\pm 2.10 \, & (z \approx 12.7) \\
        0.51\pm 0.29, -9.20\pm 2.56 \, & (z \approx 17.3)
    \end{cases}
\end{equation}

   Comparing the luminosity functions at these two redshifts, we find that
there is only a marginal evolution between them. If the difference is 
attributed to density evolution, the increase from $z\approx 17.3$ to 12.7 is 
only up to $\sim$2$\times$ over the luminosity range considered here. This is
demonstrated in the bottom panel of Figure \ref{fig:LFpw}, 
which shows the ratio of the best-fit power-law functions between the two
redshifts. Given that there is only 
$\sim$120 Myr in between the two epochs, such a small increase is probably 
not surprising.

\begin{figure*}[htbp]
    \centering
    \includegraphics[width=0.95\textwidth]{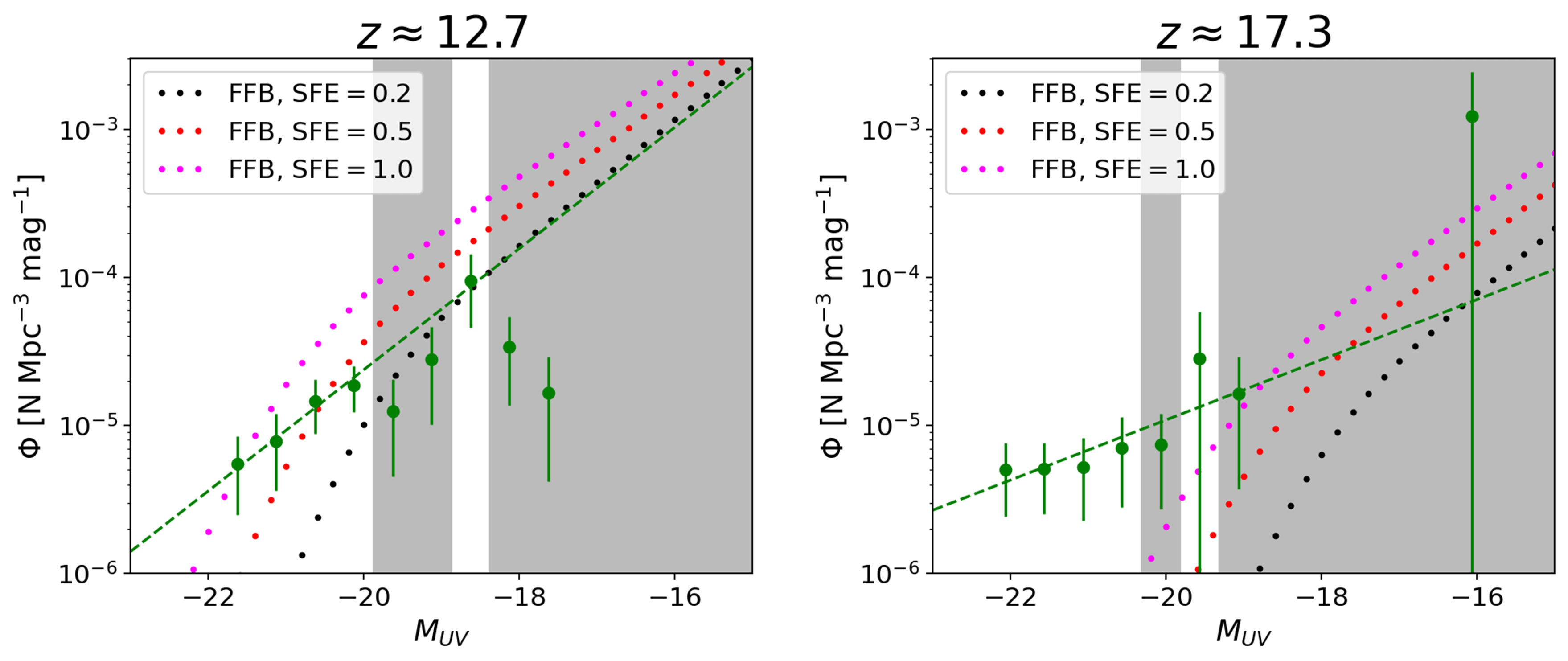}
    \caption{Comparison of our luminosity functions (green symbols) and the 
    predictions from the FFB models (dotted curves). At $z\approx 12.7$, the
    FFB predictions (different colors for different SFE as noted) broadly
    agree with the observations. The agreement in at $z\approx 17.3$ is poor,
    however it is possible to fine-tune the FFB parameters to reach better
    agreement (Li \& Dekel, private communications).
}
    \label{fig:ffb}
\end{figure*}

   No models in the pre-JWST era would predict such a large number of 
luminous galaxies in such early times in the universe. The most severe 
challenge, therefore, is whether their abundance can be explained within the 
$\rm \Lambda$CDM paradigm. A few new models have been proposed over the past 
year to address this problem, and here we compare our luminosity functions to 
the predictions from the ``feedback-free burst'' model 
\citep[FFB;][Li \& Dekel, in prep.]{Dekel2023}. 
In essence, this model suggests that FFBs occur in high-mass dark matter
halos ($M\gtrsim 10^{11}M_\odot$) in the early universe, where the free-fall 
time scale ($\lesssim$1~Myr) is shorter than that is needed for 
low-metallicity massive stars to develop winds and supernovae and therefore
the star formation process is not suppressed by feedback. The comparison is
shown in Figure \ref{fig:ffb}. Encouragingly, the FFB predictions are broadly
consistent with the $z\approx 12.7$ luminosity function when considering the
varying star formation efficiency (SFE). The current agreement at 
$z\approx 17.3$ is poor, however it is possible to fine-tune the FFB model 
(e.g., the amount of dust extinction) to match the observations
(Li \& Dekel, private communication). In other words, it is still possible to
produce a large number of luminous galaxies at $z>11$ in the 
$\rm \Lambda$CDM paradigm, and the basic framework of galaxy formation in the
early universe is not necessarily in crisis for the time being.

    On the other hand, the existence of galaxy population up to 
$z\approx 17.3$ does pose various challenges. One example is the epoch of 
hydrogen reionization ($z_{\rm re}$), for which the CMB anisotropy measurement 
of the Planck mission has determined $z_{\rm re}= 7.64\pm 0.74$
\citep[][]{Planck2018Par}. Our results, together with the many candidate
galaxies at $z>10$ from other teams (some of which have been confirmed), seem
to be in conflict with this value. On the other hand, \citet[][]{Bowman2018}
presented a tentative detection of global neutral hydrogen absorption at
78~MHz (FWHM of 19~MHz),
which suggests $z_{\rm re}=17.2$ (ranging from 13.6 to 23.1) and
coincides with the redshift range of our $z\approx 17.3$ galaxies selected 
as F200W dropouts. Spectroscopic identification of such objects will greatly
facilitate the resolution of this problem.

    Finally, we briefly note on the F277W dropouts. We refrain from carrying
out SED fitting for these objects because most of them have secure detections
in only two reddest bands (F356W and F444W) and the fitting would be highly
unconstrained. If any of these objects are indeed at $z\approx 24.7$, the
new picture of early galaxy formation that the community is starting to
rebuild will have to be changed once again. Spectroscopy is the only way to 
further investigate their nature and there does not seem to be any less 
expensive alternatives.

\section{Summary}

    In this work, we carry out a systematic search for galaxies at 
$z\gtrsim 11.3$ by applying the dropout method to the public NIRCam data in the 
JWST Cycle 1, which include six blank fields totalling 386~arcmin$^2$ and two 
lensing cluster fields totalling 48~arcmin$^2$. We select F150W, F200W and F277W
dropouts, which correspond to $z\approx 12.7$ ($11.3\lesssim z\lesssim 15.4$),
17.3 ($15.4\lesssim z\lesssim 21.8$) and 24.7 ($21.8\lesssim z\lesssim 28.3$),
respectively. In total, we have found 123 F150W dropouts, 52 F200W dropouts and
32 F277W dropouts. To our knowledge, this is the largest candidate galaxy sample
probing the highest redshift range to date.

   As most of the F227W dropouts are detected in only two bands, photometric
diagnostics of their nature would be highly uncertain. Therefore, our follow-up 
analysis of this sample focuses on the F150W dropouts and the F200W dropouts. 
To purify the sample, we fit their SEDs to derive their photometric redshifts 
and reject contaminants. Two very different SED fitting 
schemes are used: one utilizes Le Phare to fit to population synthesis models 
(S1) and the other utilizes EAZY to fit to PCA templates (S2). The purified 
sample consists of 54 F150W dropouts and 21 F200W dropouts in S1, and 62 F150W 
dropouts and 43 F200W dropouts in S2, respectively. Assuming that these 
purified dropouts are indeed at the expected redshifts, we derive galaxy 
luminosity functions at $z\approx 12.7$ and 17.3. As our selection
is largely limited to $m_{356}\leq 29.5$~mag for both the F150W and the F200W
dropouts, the derived luminosity functions are not secure at
$M_{\rm UV}> -18.1$~mag and $>-19.1$~mag for $z\approx 12.7$ and 17.3,
respectively. Within the range brighter than the limit, we find that the
luminosity functions at both redshifts follow power law instead of Schechter
function; in particular, the exponential cutoff at the bright-end is not
observed. The evolution from $z\approx 17.3$ to 12.7 is marginal and only
amounts to a factor of $\lesssim 2$ increase if attributed to density evolution.

   Most objects in our sample are bright enough for JWST spectroscopic 
confirmation, and it is imperative to identify at least a fraction of them so
that JWST's pursuit of ``first luminous objects'' can be put onto a more solid
footing. While no models in the pre-JWST era predicted the emergence of galaxy
population at a time as early as $z\approx 17.3$, it is not necessarily a crisis
for the $\Lambda$CDM paradigm if our candidates are proved to be at such high
redshifts. For example, the new ``feedback-free burst'' model within the same 
paradigm can now produce abundant luminous galaxies in very early time, and it 
is possible to fine-tune to match our derived luminosity functions. The 
existence of a large number of galaxies at $z\approx 17.3$ is in fact 
consistent with the suggestion that the hydrogen reionization began at 
$z\approx 17.2$ based on the detection of global neutral hydrogen absorption 
signal. It is, however, in conflict with the determination of $z_{\rm re}=7.64$
based on the CMB anisotrophy measurements of the Planck mission. To further 
address this problem, selections at fainter levels from deeper NIRCam surveys 
will be necessary.

All {\it JWST} non-proprietary data used in this paper can be found in MAST: 
\dataset[10.17909/qv8h-sz76]{http://dx.doi.org/10.17909/qv8h-sz76}. The JADES
DR1 data can also be found in MAST:
\dataset[10.17909/8tdj-8n28]{http://dx.doi.org/10.17909/8tdj-8n28}.

\begin{acknowledgements}
We are grateful to Avishai Dekel and Zhaozhou Li for providing their FFB model
predictions prior to the publication.
H.Y. and B.S. acknowledge the partial support from the University of Missouri 
Research Council Grant URC-23-029. 
This project is based on observations made with the NASA/ESA/CSA James
Webb Space Telescope and obtained from the Mikulski Archive for Space
Telescopes, which is a collaboration between the Space Telescope Science
Institute (STScI/NASA), the Space Telescope European Coordinating Facility
(ST-ECF/ESA), and the Canadian Astronomy Data Centre (CADC/NRC/CSA).

\end{acknowledgements}

\appendix 
\setcounter{figure}{0}
\renewcommand{\thefigure}{A\arabic{figure}}

\setcounter{table}{0}
\renewcommand{\thetable}{A\arabic{table}}

\section{Catalogs of dropouts, their images and SED fitting results}

   The source catalogs of the F150W, F200W and F277W dropouts in our sample are
given as tables below. Their NIRCam image cutouts are displayed in Figures A1
through A14, while their SED fitting results are show in Figures A15 through
A32, respectively.

\clearpage
\startlongtable
\begin{longrotatetable}
\movetabledown=7mm
\begin{deluxetable}{rccccccccccccc}
\tablecolumns{14}
\tabletypesize{\scriptsize}
\tablecaption{Catalog of F150W dropouts. }
\tablehead{
\colhead{SID} & 
\colhead{R.A.} & 
\colhead{Decl.} & 
\colhead{$m_{090}$} & 
\colhead{$m_{115}$} & 
\colhead{$m_{150}$} & 
\colhead{$m_{200}$} & 
\colhead{$m_{277}$} & 
\colhead{$m_{335}$} & 
\colhead{$m_{356}$} & 
\colhead{$m_{410}$} & 
\colhead{$m_{444}$} &
\colhead{S1} & 
\colhead{S2} \\
\colhead{} & 
\colhead{(degrees)} & 
\colhead{(degrees)} & 
\colhead{(mag)} & 
\colhead{(mag)} & 
\colhead{(mag)} & 
\colhead{(mag)} & 
\colhead{(mag)} & 
\colhead{(mag)} & 
\colhead{(mag)} & 
\colhead{(mag)} & 
\colhead{(mag)} & 
\colhead{} & 
\colhead{} 
}

\startdata
f150d\_cosmos\_010 & 150.0964901 & 2.1760648 & \textgreater27.59 & \textgreater27.64 & 27.51\textpm0.49 & 26.65\textpm0.18 & 26.43\textpm0.06 & ... & 26.58\textpm0.06 & 26.89\textpm0.14 & 26.91\textpm0.13 & 0 & 1 \\
f150d\_cosmos\_019 & 150.1204786 & 2.1876317 & \textgreater28.34 & \textgreater28.32 & 28.15\textpm0.54 & 27.35\textpm0.22 & 27.29\textpm0.08 & ... & 27.18\textpm0.06 & 27.26\textpm0.12 & 27.06\textpm0.09 & 0 & 0 \\
f150d\_cosmos\_021 & 150.0630699 & 2.1892588 & \textgreater27.70 & \textgreater27.68 & \textgreater27.84 & 26.75\textpm0.20 & 26.94\textpm0.09 & ... & 26.98\textpm0.08 & 27.05\textpm0.15 & 27.40\textpm0.17 & 1 & 1 \\
f150d\_cosmos\_030 & 150.0626968 & 2.2068158 & \textgreater27.74 & \textgreater27.73 & \textgreater27.93 & 26.63\textpm0.17 & 26.79\textpm0.08 & ... & 26.7\textpm0.07 & 27.11\textpm0.17 & 26.86\textpm0.12 & 1 & 0 \\
f150d\_cosmos\_032 & 150.1280942 & 2.2161717 & \textgreater28.45 & \textgreater28.47 & 28.18\textpm0.47 & 27.37\textpm0.19 & 27.44\textpm0.07 & ... & 27.4\textpm0.06 & 27.56\textpm0.14 & 27.77\textpm0.14 & 0 & 1 \\
f150d\_cosmos\_065 & 150.1532019 & 2.2483807 & \textgreater28.35 & \textgreater28.40 & \textgreater28.63 & 27.28\textpm0.16 & 27.57\textpm0.07 & ... & 27.35\textpm0.05 & 27.39\textpm0.11 & 27.38\textpm0.08 & 0 & 0 \\
f150d\_cosmos\_069 & 150.1506812 & 2.2492089 & \textgreater28.48 & \textgreater28.53 & \textgreater28.76 & 27.60\textpm0.20 & 27.65\textpm0.07 & ... & 27.66\textpm0.07 & 27.70\textpm0.13 & 27.91\textpm0.12 & 1 & 1 \\
f150d\_cosmos\_075 & 150.1121016 & 2.2578627 & \textgreater28.38 & \textgreater28.39 & \textgreater28.58 & 27.66\textpm0.21 & 27.79\textpm0.08 & ... & 27.94\textpm0.09 & 28.19\textpm0.22 & 27.95\textpm0.15 & 1 & 1 \\
f150d\_cosmos\_105 & 150.0780525 & 2.293364 & \textgreater28.12 & \textgreater28.18 & 28.21\textpm0.52 & 27.28\textpm0.18 & 27.39\textpm0.08 & ... & 27.70\textpm0.10 & 27.82\textpm0.21 & 28.00\textpm0.20 & 1 & 1 \\
f150d\_cosmos\_128 & 150.118757 & 2.3147096 & \textgreater28.54 & \textgreater28.56 & 28.41\textpm0.49 & 27.53\textpm0.18 & 27.49\textpm0.06 & ... & 27.59\textpm0.06 & 28.02\textpm0.17 & 27.66\textpm0.10 & 1 & 1 \\
f150d\_cosmos\_133 & 150.1411515 & 2.3181771 & \textgreater28.67 & \textgreater28.68 & \textgreater28.89 & 27.72\textpm0.21 & 28.00\textpm0.10 & ... & 27.88\textpm0.08 & 28.15\textpm0.19 & 28.01\textpm0.14 & 1 & 1 \\
f150d\_cosmos\_135 & 150.1729282 & 2.3184522 & \textgreater28.46 & \textgreater28.46 & \textgreater28.68 & 27.57\textpm0.20 & 27.83\textpm0.09 & ... & 27.97\textpm0.09 & 28.03\textpm0.17 & 28.26\textpm0.17 & 1 & 1 \\
f150d\_cosmos\_142 & 150.186187 & 2.3270214 & \textgreater28.34 & \textgreater28.38 & 27.92\textpm0.38 & 27.07\textpm0.14 & 26.73\textpm0.04 & ... & 26.75\textpm0.04 & 26.63\textpm0.06 & 26.72\textpm0.05 & 0 & 0 \\
f150d\_cosmos\_143 & 150.1150419 & 2.3277913 & \textgreater28.68 & \textgreater28.68 & \textgreater28.89 & 27.88\textpm0.16 & 28.00\textpm0.07 & ... & 28.30\textpm0.08 & 28.26\textpm0.14 & 28.20\textpm0.11 & 1 & 1 \\
f150d\_cosmos\_161 & 150.1981908 & 2.3418706 & \textgreater27.73 & \textgreater27.73 & \textgreater27.94 & 27.11\textpm0.19 & 26.95\textpm0.07 & ... & 27.31\textpm0.08 & 27.42\textpm0.16 & 27.33\textpm0.12 & 0 & 1 \\
f150d\_cosmos\_208 & 150.1794883 & 2.3745223 & \textgreater28.28 & \textgreater28.33 & \textgreater28.51 & 27.61\textpm0.21 & 28.23\textpm0.12 & ... & 28.25\textpm0.11 & 28.74\textpm0.33 & 28.45\textpm0.20 & 0 & 1 \\
f150d\_cosmos\_217 & 150.0637146 & 2.3796138 & \textgreater27.69 & \textgreater27.65 & \textgreater27.85 & 26.57\textpm0.19 & 26.55\textpm0.07 & ... & 26.51\textpm0.06 & 26.81\textpm0.14 & 26.54\textpm0.10 & 1 & 0 \\
f150d\_cosmos\_228 & 150.1798098 & 2.3888704 & \textgreater28.11 & \textgreater28.15 & 27.66\textpm0.49 & 26.60\textpm0.15 & 26.38\textpm0.05 & ... & 26.39\textpm0.04 & 26.67\textpm0.10 & 26.40\textpm0.06 & 1 & 1 \\
f150d\_cosmos\_238 & 150.1795179 & 2.3978379 & \textgreater28.32 & \textgreater28.34 & \textgreater28.58 & 27.56\textpm0.18 & 27.93\textpm0.09 & ... & 28.01\textpm0.09 & 28.09\textpm0.18 & 28.16\textpm0.15 & 1 & 1 \\
f150d\_cosmos\_247 & 150.1132498 & 2.4067704 & \textgreater28.41 & \textgreater28.46 & \textgreater28.49 & 27.23\textpm0.18 & 27.37\textpm0.06 & ... & 27.34\textpm0.05 & 27.21\textpm0.11 & 27.35\textpm0.09 & 1 & 0 \\
f150d\_cosmos\_265 & 150.1568953 & 2.419015 & \textgreater28.28 & \textgreater28.30 & \textgreater28.49 & 26.61\textpm0.12 & 26.26\textpm0.03 & ... & 26.05\textpm0.02 & 26.21\textpm0.05 & 26.03\textpm0.03 & 1 & 0 \\
f150d\_cosmos\_287 & 150.1323377 & 2.4418425 & \textgreater28.22 & \textgreater28.21 & 28.08\textpm0.51 & 27.22\textpm0.19 & 27.26\textpm0.08 & ... & 27.21\textpm0.07 & 27.5\textpm0.16 & 27.29\textpm0.11 & 1 & 1 \\
f150d\_cosmos\_309 & 150.1693524 & 2.4757285 & \textgreater27.76 & \textgreater27.75 & \textgreater27.93 & 26.87\textpm0.20 & 27.12\textpm0.10 & ... & 27.15\textpm0.09 & 27.50\textpm0.21 & 27.33\textpm0.15 & 1 & 1 \\
\hline
f150d\_uds1\_036 & 34.3652678 & $-$5.2949565 & \textgreater28.45 & \textgreater28.43 & \textgreater28.64 & 27.27\textpm0.19 & 27.22\textpm0.09 & ... & 26.77\textpm0.05 & 26.9\textpm0.11 & 27.01\textpm0.11 & 0 & 0 \\
f150d\_uds1\_037 & 34.3207228 & $-$5.2923562 & \textgreater28.05 & \textgreater28.08 & \textgreater28.25 & 27.38\textpm0.21 & 27.73\textpm0.11 & ... & 27.58\textpm0.09 & 27.96\textpm0.23 & 27.61\textpm0.15 & 0 & 1 \\
f150d\_uds1\_111 & 34.3568289 & $-$5.253295 & \textgreater28.27 & \textgreater28.30 & \textgreater28.51 & 27.09\textpm0.19 & 27.06\textpm0.07 & ... & 26.84\textpm0.06 & 27.22\textpm0.15 & 26.69\textpm0.08 & 1 & 0 \\
f150d\_uds1\_124 & 34.3381297 & $-$5.2463882 & \textgreater28.21 & \textgreater28.23 & 28.42\textpm0.51 & 27.57\textpm0.19 & 27.83\textpm0.09 & ... & 27.77\textpm0.08 & 27.64\textpm0.14 & 28.17\textpm0.19 & 0 & 0 \\
f150d\_uds1\_131 & 34.3426673 & $-$5.2391974 & \textgreater27.93 & \textgreater27.96 & \textgreater28.16 & 27.34\textpm0.21 & 27.57\textpm0.09 & ... & 27.54\textpm0.08 & 27.57\textpm0.17 & 27.64\textpm0.15 & 0 & 1 \\
f150d\_uds1\_140 & 34.319141 & $-$5.2301575 & \textgreater28.03 & \textgreater28.09 & \textgreater28.28 & 26.77\textpm0.2 & 26.82\textpm0.08 & ... & 26.64\textpm0.07 & 26.75\textpm0.14 & 26.74\textpm0.11 & 1 & 0 \\
f150d\_uds1\_155 & 34.2670458 & $-$5.2243054 & \textgreater27.54 & \textgreater27.61 & \textgreater27.82 & 26.41\textpm0.17 & 26.26\textpm0.05 & ... & 26.34\textpm0.05 & 26.35\textpm0.11 & 26.50\textpm0.10 & 1 & 1 \\
f150d\_uds1\_157 & 34.3275129 & $-$5.223183 & \textgreater28.02 & \textgreater28.05 & \textgreater28.25 & 27.44\textpm0.21 & 27.66\textpm0.09 & ... & 27.58\textpm0.08 & 27.94\textpm0.22 & 27.77\textpm0.16 & 0 & 1 \\
f150d\_uds1\_168 & 34.3827155 & $-$5.2106767 & \textgreater27.85 & \textgreater27.87 & 27.82\textpm0.46 & 26.99\textpm0.18 & 27.31\textpm0.07 & ... & 27.42\textpm0.08 & 27.61\textpm0.18 & 27.65\textpm0.15 & 0 & 1 \\
f150d\_uds1\_170 & 34.3516687 & $-$5.2068724 & \textgreater27.58 & \textgreater27.66 & 28.16\textpm0.51 & 27.28\textpm0.19 & 27.81\textpm0.11 & ... & 28.18\textpm0.14 & 27.89\textpm0.23 & 28.26\textpm0.26 & 0 & 1 \\
f150d\_uds1\_178 & 34.3365264 & $-$5.202591 & \textgreater27.48 & \textgreater27.54 & 27.23\textpm0.46 & 26.24\textpm0.15 & 26.42\textpm0.06 & ... & 26.46\textpm0.06 & 26.53\textpm0.13 & 26.25\textpm0.08 & 1 & 1 \\
f150d\_uds1\_179 & 34.3556961 & $-$5.2008919 & \textgreater27.54 & \textgreater27.57 & \textgreater27.80 & 26.86\textpm0.21 & 27.24\textpm0.09 & ... & 27.10\textpm0.08 & 27.61\textpm0.25 & 27.58\textpm0.19 & 0 & 1 \\
f150d\_uds1\_180 & 34.3419743 & $-$5.2002854 & \textgreater27.53 & \textgreater27.52 & 27.52\textpm0.35 & 26.71\textpm0.14 & 26.56\textpm0.04 & ... & 26.32\textpm0.03 & 26.33\textpm0.06 & 26.30\textpm0.05 & 0 & 0 \\
f150d\_uds1\_189 & 34.4532399 & $-$5.1642041 & \textgreater27.62 & \textgreater27.64 & 27.79\textpm0.49 & 26.95\textpm0.18 & 27.16\textpm0.08 & ... & 27.14\textpm0.07 & 27.16\textpm0.15 & 27.21\textpm0.12 & 0 & 1 \\
f150d\_uds1\_191 & 34.3380838 & $-$5.1555267 & \textgreater28.00 & \textgreater28.03 & \textgreater28.24 & 27.26\textpm0.17 & 27.88\textpm0.11 & ... & 27.97\textpm0.10 & 28.02\textpm0.20 & 27.91\textpm0.15 & 0 & 1 \\
\hline
f150d\_ceers\_003 & 215.0163153 & 52.8707935 & ... & \textgreater29.48 & \textgreater29.00 & 27.68\textpm0.19 & 27.68\textpm0.08 & ... & 27.54\textpm0.06 & 27.57\textpm0.14 & 27.45\textpm0.11 & 0 & 0 \\
f150d\_ceers\_014 & 214.8103744 & 52.7397808 & ... & \textgreater29.12 & 28.33\textpm0.49 & 27.39\textpm0.17 & 27.17\textpm0.06 & ... & 27.24\textpm0.06 & 27.07\textpm0.10 & 27.0\textpm0.08 & 1 & 0 \\
f150d\_ceers\_031 & 215.0666659 & 52.9333503 & ... & \textgreater29.30 & \textgreater28.87 & 27.24\textpm0.17 & 27.42\textpm0.07 & ... & 27.23\textpm0.05 & 27.42\textpm0.15 & 27.25\textpm0.10 & 1 & 0 \\
f150d\_ceers\_041 & 214.7737282 & 52.7404111 & ... & \textgreater29.05 & \textgreater28.85 & 28.01\textpm0.19 & 27.63\textpm0.05 & ... & 27.38\textpm0.04 & 27.25\textpm0.07 & 27.42\textpm0.07 & 0 & 0 \\
f150d\_ceers\_046 & 214.9739734 & 52.8847591 & ... & \textgreater29.29 & \textgreater28.85 & 27.95\textpm0.21 & 28.29\textpm0.11 & ... & 28.25\textpm0.08 & 28.76\textpm0.32 & 28.12\textpm0.15 & 0 & 1 \\
f150d\_ceers\_051 & 214.8876228 & 52.8296748 & ... & \textgreater29.05 & 28.67\textpm0.42 & 27.84\textpm0.15 & 27.53\textpm0.04 & ... & 27.64\textpm0.05 & 27.54\textpm0.08 & 27.47\textpm0.06 & 0 & 0 \\
f150d\_ceers\_060 & 214.941266 & 52.8805983 & ... & \textgreater29.44 & 27.82\textpm0.33 & 26.98\textpm0.13 & 27.27\textpm0.07 & ... & 27.18\textpm0.05 & 27.33\textpm0.14 & 27.23\textpm0.11 & 1 & 1 \\
f150d\_ceers\_079 & 214.9082292 & 52.8968555 & ... & \textgreater29.03 & 28.69\textpm0.47 & 27.74\textpm0.16 & 27.61\textpm0.06 & ... & 27.80\textpm0.06 & 28.07\textpm0.16 & 27.96\textpm0.13 & 1 & 1 \\
f150d\_ceers\_087 & 214.8740717 & 52.876717 & ... & \textgreater28.90 & \textgreater28.70 & 27.39\textpm0.14 & 26.92\textpm0.05 & ... & 27.29\textpm0.06 & 27.51\textpm0.15 & 27.62\textpm0.10 & 1 & 0 \\
f150d\_ceers\_088 & 214.8743088 & 52.8767738 & ... & \textgreater28.91 & \textgreater28.70 & 27.73\textpm0.17 & 27.68\textpm0.09 & ... & 27.87\textpm0.09 & 27.58\textpm0.14 & 28.86\textpm0.28 & 0 & 0 \\
f150d\_ceers\_089 & 214.9320161 & 52.9182491 & ... & \textgreater28.92 & \textgreater28.79 & 27.86\textpm0.20 & 27.82\textpm0.10 & ... & 27.64\textpm0.08 & 27.69\textpm0.17 & 27.71\textpm0.10 & 0 & 0 \\
f150d\_ceers\_125 & 214.9431552 & 52.9424513 & ... & \textgreater28.96 & \textgreater28.95 & 27.98\textpm0.15 & 28.34\textpm0.09 & ... & 28.41\textpm0.09 & 28.48\textpm0.19 & 28.28\textpm0.13 & 0 & 1 \\
f150d\_ceers\_150 & 214.9849543 & 52.990391 & ... & \textgreater29.09 & \textgreater28.92 & 27.89\textpm0.18 & 27.74\textpm0.07 & ... & 27.75\textpm0.07 & 28.01\textpm0.16 & 28.21\textpm0.16 & 1 & 1 \\
f150d\_ceers\_153 & 214.8567793 & 52.9030729 & ... & \textgreater28.95 & \textgreater28.80 & 27.77\textpm0.18 & 27.76\textpm0.08 & ... & 27.76\textpm0.08 & 27.99\textpm0.17 & 27.66\textpm0.11 & 1 & 1 \\
f150d\_ceers\_161 & 214.9066563 & 52.9455063 & ... & \textgreater29.00 & \textgreater28.83 & 27.48\textpm0.18 & 27.32\textpm0.06 & ... & 27.20\textpm0.05 & 27.21\textpm0.10 & 27.15\textpm0.08 & 1 & 0 \\
\hline
f150d\_glass\_003 & 3.5140762 & $-$30.3839153 & \textgreater29.66 & \textgreater29.62 & \textgreater29.42 & 28.29\textpm0.20 & 27.56\textpm0.05 & ... & 27.24\textpm0.03 & ... & 27.01\textpm0.02 & 0 & 0 \\
f150d\_glass\_024 & 3.4915046 & $-$30.362516 & \textgreater29.91 & \textgreater29.89 & \textgreater29.43 & 28.39\textpm0.20 & 27.87\textpm0.05 & ... & 27.81\textpm0.05 & ... & 27.83\textpm0.04 & 1 & 0 \\
f150d\_glass\_029 & 3.5321168 & $-$30.3614222 & \textgreater29.83 & \textgreater29.83 & \textgreater29.63 & 28.76\textpm0.21 & 28.93\textpm0.10 & ... & 28.25\textpm0.05 & ... & 28.27\textpm0.04 & 0 & 0 \\
f150d\_glass\_039 & 3.5135577 & $-$30.3567969 & \textgreater29.95 & \textgreater29.95 & \textgreater29.76 & 28.43\textpm0.15 & 28.91\textpm0.08 & ... & 29.24\textpm0.10 & ... & 30.30\textpm0.21 & 0 & 0 \\
f150d\_glass\_067 & 3.4506757 & $-$30.331406 & \textgreater29.57 & \textgreater29.59 & \textgreater29.39 & 27.62\textpm0.17 & 27.07\textpm0.04 & ... & 27.07\textpm0.04 & ... & 26.86\textpm0.03 & 1 & 0 \\
f150d\_glass\_075 & 3.4989394 & $-$30.3253288 & \textgreater29.87 & \textgreater29.72 & \textgreater29.69 & 28.38\textpm0.19 & 28.47\textpm0.09 & ... & 28.38\textpm0.07 & ... & 28.39\textpm0.06 & 1 & 0 \\
f150d\_glass\_076 & 3.4989889 & $-$30.3247604 & \textgreater29.88 & \textgreater29.75 & 28.69\textpm0.38 & 26.7\textpm0.05 & 27.04\textpm0.03 & ... & 27.09\textpm0.03 & ... & 27.10\textpm0.02 & 0 & 0 \\
\hline
f150d\_ngdeep\_014 & 53.2515487 & $-$27.8571346 & ... & \textgreater30.56 & \textgreater30.32 & 29.29\textpm0.14 & 30.74\textpm0.20 & ... & 30.28\textpm0.13 & ... & 30.55\textpm0.19 & 0 & 0 \\
f150d\_ngdeep\_018 & 53.2769065 & $-$27.8505161 & ... & \textgreater30.56 & 29.15\textpm0.21 & 28.15\textpm0.12 & 28.74\textpm0.05 & ... & 28.49\textpm0.05 & ... & 28.84\textpm0.06 & 0 & 0 \\
f150d\_ngdeep\_034 & 53.2266642 & $-$27.8210254 & ... & \textgreater30.26 & 29.96\textpm0.46 & 29.14\textpm0.19 & 28.7\textpm0.04 & ... & 28.46\textpm0.03 & ... & 28.87\textpm0.05 & 0 & 0 \\
f150d\_ngdeep\_038 & 53.2383719 & $-$27.8153782 & ... & \textgreater30.55 & 30.43\textpm0.45 & 29.46\textpm0.18 & 30.39\textpm0.14 & ... & 30.24\textpm0.13 & ... & 30.33\textpm0.16 & 0 & 1 \\
\hline
f150d\_jades\_A1\_005 & 53.1115765 & $-$27.8108276 & \textgreater30.21 & \textgreater30.61 & \textgreater30.41 & 29.45\textpm0.14 & 29.42\textpm0.05 & 29.60\textpm0.08 & 29.53\textpm0.05 & 30.00\textpm0.15 & 29.94\textpm0.12 & 1 & 1 \\
f150d\_jades\_A1\_007 & 53.1113635 & $-$27.8075635 & \textgreater30.21 & \textgreater30.61 & 30.91\textpm0.54 & 29.90\textpm0.17 & 30.62\textpm0.12 & 30.44\textpm0.13 & 30.19\textpm0.07 & 30.73\textpm0.23 & 30.83\textpm0.21 & 0 & 0 \\
f150d\_jades\_A1\_008 & 53.1686016 & $-$27.7939249 & \textgreater30.57 & \textgreater30.82 & 31.12\textpm0.43 & 30.21\textpm0.21 & 30.89\textpm0.11 & \textgreater31.38 & 30.48\textpm0.09 & 31.3\textpm0.27 & 30.89\textpm0.16 & 0 & 0 \\
f150d\_jades\_A1\_009 & 53.1686315 & $-$27.7927587 & \textgreater30.57 & \textgreater30.82 & 30.40\textpm0.28 & 29.23\textpm0.11 & 29.39\textpm0.03 & 29.59\textpm0.08 & 29.57\textpm0.05 & 29.83\textpm0.09 & 29.77\textpm0.07 & 1 & 1 \\
f150d\_jades\_A1\_012 & 53.1570142 & $-$27.7896965 & \textgreater30.65 & \textgreater30.95 & 31.30\textpm0.38 & 30.38\textpm0.19 & 30.77\textpm0.08 & \textgreater31.42 & 30.50\textpm0.08 & 31.82\textpm0.37 & 30.76\textpm0.12 & 0 & 0 \\
f150d\_jades\_A1\_018 & 53.1801703 & $-$27.7801054 & \textgreater30.37 & \textgreater30.72 & 29.31\textpm0.29 & 28.38\textpm0.10 & 27.96\textpm0.03 & 27.92\textpm0.04 & 27.78\textpm0.02 & 27.78\textpm0.04 & 28.05\textpm0.05 & 0 & 0 \\
f150d\_jades\_A1\_019 & 53.1803562 & $-$27.7800919 & \textgreater30.37 & \textgreater30.73 & 29.72\textpm0.46 & 27.97\textpm0.08 & 27.67\textpm0.03 & 27.61\textpm0.03 & 27.62\textpm0.02 & 27.69\textpm0.04 & 27.51\textpm0.03 & 1 & 0 \\
\hline
f150d\_jades\_A2\_001 & 53.1685698 & $-$27.8421018 & \textgreater30.23 & \textgreater30.65 & 30.34\textpm0.35 & 29.23\textpm0.10 & 30.35\textpm0.11 & 30.61\textpm0.18 & 29.82\textpm0.06 & 30.34\textpm0.19 & 29.58\textpm0.08 & 0 & 0 \\
f150d\_jades\_A2\_002 & 53.1504747 & $-$27.8350217 & \textgreater30.23 & \textgreater30.61 & \textgreater30.43 & 29.43\textpm0.15 & 29.14\textpm0.04 & 29.24\textpm0.07 & 28.99\textpm0.04 & 29.03\textpm0.07 & 29.14\textpm0.06 & 1 & 0 \\
f150d\_jades\_A2\_006 & 53.166346 & $-$27.8215581 & \textgreater30.87 & \textgreater31.18 & 31.03\textpm0.51 & 29.21\textpm0.09 & 29.07\textpm0.03 & 29.4\textpm0.06 & 29.16\textpm0.03 & 29.24\textpm0.05 & 29.37\textpm0.05 & 0 & 0 \\
f150d\_jades\_A2\_008 & 53.158609 & $-$27.8147378 & \textgreater30.79 & \textgreater31.10 & 30.91\textpm0.33 & 29.96\textpm0.13 & 30.44\textpm0.06 & 30.52\textpm0.11 & 30.33\textpm0.06 & 30.80\textpm0.15 & 30.57\textpm0.10 & 0 & 1 \\
f150d\_jades\_A2\_009 & 53.1662719 & $-$27.8137652 & \textgreater30.71 & \textgreater31.05 & 31.62\textpm0.50 & 30.35\textpm0.14 & 31.14\textpm0.09 & 31.00\textpm0.12 & 30.85\textpm0.07 & 30.64\textpm0.10 & 30.91\textpm0.10 & 0 & 0 \\
f150d\_jades\_A2\_015 & 53.1428281 & $-$27.8080515 & \textgreater30.86 & \textgreater31.18 & 31.62\textpm0.44 & 30.73\textpm0.19 & 31.19\textpm0.09 & 30.98\textpm0.13 & 31.15\textpm0.10 & 31.57\textpm0.23 & 31.27\textpm0.15 & 0 & 1 \\
f150d\_jades\_A2\_016 & 53.1612338 & $-$27.8074642 & \textgreater30.78 & \textgreater31.08 & 31.15\textpm0.44 & 30.24\textpm0.19 & 30.54\textpm0.07 & 30.33\textpm0.10 & 30.08\textpm0.06 & 30.39\textpm0.11 & 30.37\textpm0.09 & 0 & 0 \\
f150d\_jades\_A2\_021 & 53.195856 & $-$27.789175 & \textgreater30.36 & \textgreater30.70 & 30.53\textpm0.47 & 29.62\textpm0.16 & 29.86\textpm0.09 & 29.57\textpm0.09 & 29.37\textpm0.05 & 30.05\textpm0.17 & 30.49\textpm0.22 & 0 & 0 \\
f150d\_jades\_A2\_027 & 53.2148161 & $-$27.7743605 & \textgreater30.86 & \textgreater31.15 & 31.90\textpm0.50 & 31.03\textpm0.21 & 31.81\textpm0.16 & 32.50\textpm0.44 & 31.37\textpm0.10 & 32.17\textpm0.36 & 32.01\textpm0.28 & 0 & 0 \\
f150d\_jades\_A2\_030 & 53.1899349 & $-$27.7714981 & \textgreater30.90 & \textgreater31.17 & 30.42\textpm0.22 & 29.43\textpm0.09 & 30.00\textpm0.05 & 30.11\textpm0.09 & 30.14\textpm0.06 & 30.36\textpm0.12 & 30.71\textpm0.14 & 1 & 0 \\
f150d\_jades\_A2\_035 & 53.1811069 & $-$27.7567872 & \textgreater30.96 & \textgreater31.21 & \textgreater31.16 & 29.87\textpm0.16 & 29.67\textpm0.05 & 30.15\textpm0.11 & 29.76\textpm0.05 & 29.42\textpm0.06 & 29.63\textpm0.06 & 0 & 0 \\
f150d\_jades\_A2\_038 & 53.1805623 & $-$27.7454842 & \textgreater30.70 & \textgreater30.90 & 30.48\textpm0.28 & 29.51\textpm0.13 & 29.90\textpm0.06 & 29.91\textpm0.11 & 29.92\textpm0.07 & 30.14\textpm0.11 & 30.66\textpm0.16 & 0 & 0 \\
f150d\_jades\_A2\_042 & 53.1999861 & $-$27.7594299 & \textgreater30.70 & \textgreater30.77 & 31.06\textpm0.50 & 29.88\textpm0.17 & 30.44\textpm0.08 & 30.94\textpm0.24 & 30.16\textpm0.08 & 31.32\textpm0.29 & 31.41\textpm0.28 & 0 & 0 \\
\hline
f150d\_uncover\_006 & 3.5870168 & $-$30.4334048 & ... & \textgreater29.16 & \textgreater29.33 & 28.10\textpm0.21 & 28.07\textpm0.09 & ... & 28.55\textpm0.12 & 29.04\textpm0.31 & 29.07\textpm0.26 & 1 & 0 \\
f150d\_uncover\_010 & 3.6577825 & $-$30.4271833 & ... & \textgreater28.00 & \textgreater28.18 & 27.19\textpm0.19 & 27.31\textpm0.06 & ... & 27.63\textpm0.07 & ... & 27.64\textpm0.10 & 1 & 1 \\
f150d\_uncover\_012 & 3.6159364 & $-$30.4251697 & ... & \textgreater28.62 & \textgreater28.80 & 27.92\textpm0.18 & 27.97\textpm0.09 & ... & 28.22\textpm0.10 & ... & 28.33\textpm0.17 & 1 & 1 \\
f150d\_uncover\_014 & 3.5874615 & $-$30.4228282 & ... & \textgreater29.55 & \textgreater29.72 & 28.80\textpm0.18 & 28.79\textpm0.08 & ... & 28.67\textpm0.07 & 28.53\textpm0.10 & 28.57\textpm0.08 & 0 & 0 \\
f150d\_uncover\_037 & 3.563056 & $-$30.4080157 & ... & \textgreater29.48 & 28.34\textpm0.24 & 27.52\textpm0.11 & 27.58\textpm0.05 & ... & 28.03\textpm0.07 & 28.22\textpm0.14 & 28.01\textpm0.09 & 1 & 1 \\
f150d\_uncover\_042 & 3.6048043 & $-$30.402524 & ... & \textgreater29.70 & 30.28\textpm0.48 & 29.40\textpm0.21 & 29.23\textpm0.07 & ... & 29.79\textpm0.12 & 29.86\textpm0.21 & 29.67\textpm0.14 & 0 & 1 \\
f150d\_uncover\_046 & 3.5669842 & $-$30.3999489 & ... & \textgreater29.32 & \textgreater29.50 & 28.44\textpm0.20 & 28.34\textpm0.07 & ... & 28.43\textpm0.07 & 28.61\textpm0.14 & 28.26\textpm0.08 & 1 & 1 \\
f150d\_uncover\_050 & 3.5695468 & $-$30.3956585 & ... & \textgreater29.41 & \textgreater29.58 & 28.36\textpm0.21 & 28.20\textpm0.08 & ... & 28.26\textpm0.08 & 28.77\textpm0.21 & 28.44\textpm0.12 & 1 & 1 \\
f150d\_uncover\_066 & 3.5320761 & $-$30.3856738 & ... & \textgreater29.72 & \textgreater29.91 & 29.03\textpm0.21 & 29.51\textpm0.13 & ... & 29.43\textpm0.11 & 29.48\textpm0.22 & 29.47\textpm0.17 & 0 & 1 \\
f150d\_uncover\_073 & 3.5489514 & $-$30.3826595 & ... & \textgreater29.47 & \textgreater29.65 & 28.24\textpm0.19 & 28.32\textpm0.07 & ... & 28.60\textpm0.08 & 28.40\textpm0.13 & 28.18\textpm0.08 & 1 & 0 \\
f150d\_uncover\_075 & 3.5408473 & $-$30.3806381 & ... & \textgreater29.71 & \textgreater29.89 & 27.68\textpm0.13 & 27.33\textpm0.04 & ... & 27.14\textpm0.03 & 27.00\textpm0.05 & 27.05\textpm0.04 & 1 & 0 \\
f150d\_uncover\_076 & 3.5407993 & $-$30.3804695 & ... & \textgreater29.72 & 28.83\textpm0.38 & 27.20\textpm0.09 & 27.13\textpm0.03 & ... & 26.93\textpm0.03 & 27.07\textpm0.05 & 26.91\textpm0.04 & 1 & 0 \\
f150d\_uncover\_077 & 3.5882198 & $-$30.380086 & ... & \textgreater29.48 & 29.47\textpm0.39 & 28.47\textpm0.15 & 28.64\textpm0.07 & ... & 28.77\textpm0.09 & 28.92\textpm0.18 & 28.68\textpm0.10 & 0 & 1 \\
f150d\_uncover\_085 & 3.5854814 & $-$30.3769822 & ... & \textgreater29.42 & \textgreater29.60 & 28.69\textpm0.20 & 28.94\textpm0.10 & ... & 28.84\textpm0.10 & 29.16\textpm0.23 & 29.30\textpm0.19 & 1 & 1 \\
f150d\_uncover\_086 & 3.5127318 & $-$30.3760189 & ... & \textgreater29.24 & \textgreater29.35 & 27.58\textpm0.20 & 27.35\textpm0.06 & ... & 27.26\textpm0.05 & 27.38\textpm0.11 & 27.42\textpm0.09 & 1 & 0 \\
f150d\_uncover\_087 & 3.5998595 & $-$30.3752635 & ... & \textgreater29.56 & 29.09\textpm0.41 & 28.27\textpm0.19 & 27.77\textpm0.05 & ... & 27.85\textpm0.05 & 27.52\textpm0.07 & 27.68\textpm0.06 & 0 & 0 \\
f150d\_uncover\_101 & 3.5580068 & $-$30.3548082 & ... & \textgreater29.66 & 29.45\textpm0.45 & 28.33\textpm0.17 & 28.76\textpm0.09 & ... & 28.08\textpm0.05 & 28.69\textpm0.15 & 28.37\textpm0.09 & 0 & 0 \\
f150d\_uncover\_114 & 3.5734992 & $-$30.3382778 & ... & \textgreater29.52 & 29.75\textpm0.40 & 28.91\textpm0.19 & 28.78\textpm0.06 & ... & 28.80\textpm0.06 & 28.76\textpm0.10 & 28.99\textpm0.10 & 0 & 1 \\
\hline
f150d\_smacs0723\_A\_004 & 110.6870102 & $-$73.4982429 & \textgreater29.15 & ... & \textgreater29.34 & 28.46\textpm0.19 & 28.66\textpm0.08 & ... & 28.69\textpm0.08 & ... & 28.99\textpm0.16 & 1 & 1 \\
f150d\_smacs0723\_A\_006 & 110.6401082 & $-$73.4945819 & \textgreater28.60 & ... & 28.03\textpm0.32 & 27.1\textpm0.11 & 26.97\textpm0.04 & ... & 26.99\textpm0.04 & ... & 26.97\textpm0.06 & 0 & 1 \\
f150d\_smacs0723\_A\_013 & 110.7161536 & $-$73.4836327 & \textgreater29.29 & ... & \textgreater29.48 & 28.51\textpm0.19 & 28.72\textpm0.1 & ... & 28.7\textpm0.09 & ... & 29.28\textpm0.22 & 1 & 1 \\
f150d\_smacs0723\_A\_014 & 110.7348001 & $-$73.4834393 & \textgreater29.30 & ... & 29.4\textpm0.46 & 28.52\textpm0.18 & 28.8\textpm0.09 & ... & 28.96\textpm0.1 & ... & 28.88\textpm0.14 & 0 & 1 \\
f150d\_smacs0723\_A\_015 & 110.66939 & $-$73.4834712 & \textgreater29.26 & ... & \textgreater29.42 & 28.46\textpm0.21 & 28.6\textpm0.1 & ... & 28.57\textpm0.09 & ... & 28.75\textpm0.15 & 1 & 1 \\
f150d\_smacs0723\_A\_020 & 110.6401283 & $-$73.4800748 & \textgreater29.27 & ... & \textgreater29.42 & 27.17\textpm0.13 & 27.07\textpm0.05 & ... & 26.96\textpm0.04 & ... & 27.15\textpm0.07 & 1 & 0 \\
f150d\_smacs0723\_A\_022 & 110.6875291 & $-$73.4769086 & \textgreater29.31 & ... & \textgreater29.46 & 27.98\textpm0.17 & 28.06\textpm0.08 & ... & 27.98\textpm0.07 & ... & 28.38\textpm0.15 & 1 & 1 \\
f150d\_smacs0723\_A\_023 & 110.6025405 & $-$73.4749489 & \textgreater28.65 & ... & 28.68\textpm0.44 & 27.85\textpm0.18 & 28.35\textpm0.1 & ... & 28.71\textpm0.14 & ... & 28.57\textpm0.18 & 0 & 0 \\
f150d\_smacs0723\_A\_024 & 110.6716036 & $-$73.4744491 & \textgreater29.24 & ... & 29.5\textpm0.49 & 28.62\textpm0.18 & 29.14\textpm0.13 & ... & 28.93\textpm0.1 & ... & 28.94\textpm0.14 & 0 & 0 \\
f150d\_smacs0723\_A\_026 & 110.6598067 & $-$73.4658505 & \textgreater29.07 & ... & \textgreater29.23 & 28.39\textpm0.21 & 28.65\textpm0.11 & ... & 28.45\textpm0.09 & ... & 28.5\textpm0.13 & 0 & 1 \\
f150d\_smacs0723\_A\_028 & 110.6909711 & $-$73.4631649 & \textgreater29.29 & ... & 28.92\textpm0.4 & 27.95\textpm0.13 & 27.74\textpm0.05 & ... & 27.99\textpm0.06 & ... & 28.08\textpm0.09 & 0 & 1 \\
f150d\_smacs0723\_A\_029 & 110.682593 & $-$73.4587622 & \textgreater28.68 & ... & 28.73\textpm0.46 & 27.88\textpm0.17 & 27.88\textpm0.07 & ... & 28.07\textpm0.08 & ... & 27.87\textpm0.1 & 0 & 1 \\
f150d\_smacs0723\_B\_005 & 110.8540062 & $-$73.4647108 & \textgreater28.89 & ... & 28.55\textpm0.33 & 27.62\textpm0.12 & 28.2\textpm0.06 & ... & 28.71\textpm0.09 & ... & 29.99\textpm0.43 & 0 & 0 \\
f150d\_smacs0723\_B\_009 & 110.835143 & $-$73.4631564 & \textgreater29.15 & ... & 28.95\textpm0.42 & 27.93\textpm0.14 & 27.95\textpm0.05 & ... & 28.2\textpm0.06 & ... & 28.22\textpm0.09 & 0 & 1 \\
f150d\_smacs0723\_B\_019 & 110.7565239 & $-$73.4549995 & \textgreater29.04 & ... & 28.2\textpm0.49 & 26.96\textpm0.13 & 27.16\textpm0.06 & ... & 26.97\textpm0.06 & ... & 26.94\textpm0.08 & 0 & 0 \\
f150d\_smacs0723\_B\_028 & 110.8117102 & $-$73.4494943 & \textgreater29.20 & ... & 29.58\textpm0.47 & 28.71\textpm0.18 & 29.1\textpm0.1 & ... & 29.36\textpm0.12 & ... & \textgreater30.28 & 0 & 1 \\
f150d\_smacs0723\_B\_033 & 110.7961874 & $-$73.4446427 & \textgreater29.03 & ... & 29.1\textpm0.48 & 28.24\textpm0.18 & 28.79\textpm0.1 & ... & 29.07\textpm0.12 & ... & 29.21\textpm0.21 & 0 & 1 \\
f150d\_smacs0723\_B\_034 & 110.8377757 & $-$73.4410561 & \textgreater29.17 & ... & 27.94\textpm0.29 & 27.13\textpm0.11 & 26.83\textpm0.03 & ... & 26.82\textpm0.03 & ... & 27.08\textpm0.06 & 0 & 1 \\
f150d\_smacs0723\_B\_036 & 110.8711286 & $-$73.4408142 & \textgreater28.93 & ... & 28.92\textpm0.5 & 27.74\textpm0.14 & 28.06\textpm0.07 & ... & 27.93\textpm0.06 & ... & 28.09\textpm0.1 & 0 & 1 \\
f150d\_smacs0723\_B\_038 & 110.8084692 & $-$73.4381453 & \textgreater29.17 & ... & 29.63\textpm0.39 & 28.43\textpm0.11 & 28.38\textpm0.04 & ... & 28.41\textpm0.04 & ... & 28.43\textpm0.06 & 1 & 1 \\
f150d\_smacs0723\_B\_039 & 110.8130436 & $-$73.4362453 & \textgreater28.95 & ... & \textgreater29.19 & 27.34\textpm0.19 & 27.27\textpm0.07 & ... & 27.09\textpm0.06 & ... & 27.48\textpm0.12 & 1 & 0 \\
\enddata
\tablecomments{SID is the unique identifier of a given F150W dropout, which
has the field of discovery encoded. R.A. and Decl. are of J2000.0 equinox.
All magnitudes are SExtractor's \texttt{MAG\_ISO} whose aperture is defined in 
the F356W image. Columns S1 and S2 contain the flags of the purification in
S1 and S2, where ``1'' means that the given dropout is retained and ``0''
means that it is rejected.} 
\end{deluxetable}
\label{tab:catf150wd}
\end{longrotatetable}

\startlongtable
\begin{longrotatetable}
\begin{deluxetable}{cccccccccccccc}
\tablecolumns{14}
\tabletypesize{\scriptsize}
\tablecaption{Photometry table for F200W dropouts. }
\tablehead{
\colhead{SID} & 
\colhead{R.A.} & 
\colhead{Decl.} & 
\colhead{$m_{090}$} & 
\colhead{$m_{115}$} & 
\colhead{$m_{150}$} & 
\colhead{$m_{200}$} & 
\colhead{$m_{277}$} & 
\colhead{$m_{335}$} & 
\colhead{$m_{356}$} & 
\colhead{$m_{410}$} & 
\colhead{$m_{444}$} &
\colhead{S1} & 
\colhead{S2} \\
\colhead{} & 
\colhead{(degrees)} & 
\colhead{(degrees)} & 
\colhead{(mag)} & 
\colhead{(mag)} & 
\colhead{(mag)} & 
\colhead{(mag)} & 
\colhead{(mag)} & 
\colhead{(mag)} & 
\colhead{(mag)} & 
\colhead{(mag)} & 
\colhead{(mag)} & 
\colhead{} & 
\colhead{} 
}
\startdata
f200d\_cosmos\_032 & 150.1550473 & 2.2428093 & \textgreater27.79 & \textgreater27.82 & \textgreater28.01 & \textgreater28.28 & 27.29\textpm0.08 & ... & 27.16\textpm0.06 & 27.10\textpm0.11 & 27.30\textpm0.11 & 1 & 1 \\
f200d\_cosmos\_043 & 150.109837 & 2.2578077 & \textgreater28.41 & \textgreater28.43 & \textgreater28.62 & 28.06\textpm0.47 & 27.25\textpm0.08 & ... & 26.62\textpm0.05 & 26.23\textpm0.06 & 25.96\textpm0.04 & 0 & 0 \\
f200d\_cosmos\_065 & 150.0990525 & 2.287338 & \textgreater28.25 & \textgreater28.26 & \textgreater28.44 & 29.08\textpm0.53 & 28.27\textpm0.08 & ... & 28.68\textpm0.12 & 28.35\textpm0.17 & 28.95\textpm0.23 & 0 & 1 \\
f200d\_cosmos\_067 & 150.1653459 & 2.2885879 & \textgreater28.16 & \textgreater28.17 & \textgreater28.39 & \textgreater28.64 & 27.52\textpm0.11 & ... & 27.16\textpm0.07 & 27.98\textpm0.28 & 27.81\textpm0.19 & 0 & 1 \\
f200d\_cosmos\_079 & 150.1566124 & 2.3008789 & \textgreater28.17 & \textgreater28.19 & \textgreater28.43 & \textgreater28.62 & 27.72\textpm0.08 & ... & 27.58\textpm0.06 & 27.78\textpm0.14 & 27.69\textpm0.11 & 1 & 1 \\
f200d\_cosmos\_096 & 150.0921291 & 2.3270286 & \textgreater28.11 & \textgreater28.13 & \textgreater28.34 & \textgreater28.54 & 27.64\textpm0.10 & ... & 27.50\textpm0.08 & 27.88\textpm0.21 & 27.51\textpm0.12 & 1 & 1 \\
f200d\_cosmos\_099 & 150.1325098 & 2.328851 & \textgreater28.76 & \textgreater28.77 & \textgreater28.99 & \textgreater29.19 & 28.28\textpm0.10 & ... & 28.07\textpm0.08 & 28.39\textpm0.19 & 28.29\textpm0.14 & 1 & 1 \\
f200d\_cosmos\_131 & 150.1942323 & 2.3715004 & \textgreater27.73 & \textgreater27.71 & \textgreater27.89 & \textgreater28.12 & 27.11\textpm0.11 & ... & 26.99\textpm0.08 & 26.99\textpm0.15 & 27.25\textpm0.15 & 1 & 1 \\
f200d\_cosmos\_142 & 150.1752034 & 2.3844173 & \textgreater28.10 & \textgreater28.11 & \textgreater28.30 & \textgreater28.52 & 27.52\textpm0.14 & ... & 27.10\textpm0.08 & 27.36\textpm0.19 & 27.57\textpm0.18 & 0 & 1 \\
f200d\_cosmos\_166 & 150.1012306 & 2.4185747 & \textgreater27.92 & \textgreater27.97 & \textgreater28.19 & 27.43\textpm0.41 & 26.56\textpm0.06 & ... & 26.35\textpm0.05 & 26.40\textpm0.09 & 26.59\textpm0.09 & 0 & 1 \\
f200d\_cosmos\_167 & 150.1558489 & 2.4225843 & \textgreater28.40 & \textgreater28.43 & \textgreater28.64 & 28.41\textpm0.48 & 27.45\textpm0.08 & ... & 27.26\textpm0.06 & 27.31\textpm0.12 & 27.26\textpm0.09 & 0 & 1 \\
f200d\_cosmos\_168 & 150.1557862 & 2.4228785 & \textgreater28.34 & \textgreater28.34 & \textgreater28.57 & \textgreater28.76 & 26.80\textpm0.05 & ... & 26.59\textpm0.04 & 26.33\textpm0.06 & 26.69\textpm0.06 & 1 & 1 \\
f200d\_cosmos\_175 & 150.107368 & 2.4287796 & \textgreater27.63 & \textgreater27.68 & \textgreater27.90 & \textgreater28.09 & 26.69\textpm0.08 & ... & 26.49\textpm0.06 & 26.39\textpm0.11 & 26.49\textpm0.09 & 1 & 1 \\
f200d\_cosmos\_181 & 150.1044823 & 2.4354193 & \textgreater27.53 & \textgreater27.62 & \textgreater27.88 & \textgreater28.04 & 27.21\textpm0.14 & ... & 26.15\textpm0.05 & 25.99\textpm0.08 & 25.79\textpm0.05 & 0 & 1 \\
f200d\_cosmos\_191 & 150.1207149 & 2.4701671 & \textgreater27.64 & \textgreater27.64 & \textgreater27.84 & \textgreater28.06 & 26.72\textpm0.08 & ... & 26.32\textpm0.05 & 26.16\textpm0.07 & 26.13\textpm0.06 & 0 & 1 \\
\hline
f200d\_uds1\_003 & 34.2430186 & $-$5.319367 & \textgreater27.67 & \textgreater27.63 & \textgreater27.85 & \textgreater28.09 & 27.07\textpm0.09 & ... & 26.27\textpm0.04 & 26.27\textpm0.07 & 26.25\textpm0.06 & 0 & 1 \\
f200d\_uds1\_013 & 34.4008915 & $-$5.3110921 & \textgreater27.57 & \textgreater27.53 & \textgreater27.76 & \textgreater27.96 & 27.13\textpm0.07 & ... & 27.31\textpm0.08 & 26.38\textpm0.06 & 27.09\textpm0.10 & 0 & 0 \\
f200d\_uds1\_057 & 34.3013215 & $-$5.2880627 & \textgreater27.82 & \textgreater27.82 & \textgreater28.02 & 27.49\textpm0.41 & 26.51\textpm0.05 & ... & 26.35\textpm0.04 & 26.57\textpm0.09 & 26.53\textpm0.08 & 0 & 0 \\
f200d\_uds1\_058 & 34.3008324 & $-$5.2879215 & \textgreater27.99 & \textgreater27.99 & \textgreater28.18 & \textgreater28.42 & 26.69\textpm0.07 & ... & 26.19\textpm0.04 & 26.71\textpm0.12 & 26.61\textpm0.09 & 1 & 1 \\
f200d\_uds1\_071 & 34.3787906 & $-$5.2821969 & \textgreater28.10 & \textgreater28.10 & \textgreater28.29 & 28.39\textpm0.5 & 27.40\textpm0.08 & ... & 27.74\textpm0.10 & 27.79\textpm0.20 & 27.86\textpm0.18 & 0 & 1 \\
f200d\_uds1\_077 & 34.2324197 & $-$5.2806268 & \textgreater28.13 & \textgreater28.08 & \textgreater28.28 & \textgreater28.52 & 27.65\textpm0.10 & ... & 26.75\textpm0.04 & 26.81\textpm0.08 & 26.48\textpm0.05 & 0 & 1 \\
f200d\_uds1\_078 & 34.5031519 & $-$5.2796558 & \textgreater28.06 & \textgreater28.08 & \textgreater28.26 & 27.53\textpm0.27 & 26.70\textpm0.05 & ... & 26.80\textpm0.05 & 27.07\textpm0.12 & 27.06\textpm0.10 & 1 & 1 \\
f200d\_uds1\_089 & 34.3760515 & $-$5.2717343 & \textgreater28.02 & \textgreater28.01 & \textgreater28.24 & \textgreater28.47 & 27.58\textpm0.09 & ... & 27.51\textpm0.08 & 27.68\textpm0.18 & 27.56\textpm0.14 & 1 & 1 \\
f200d\_uds1\_105 & 34.3747928 & $-$5.2608881 & \textgreater28.07 & \textgreater28.06 & \textgreater28.26 & 28.36\textpm0.47 & 27.54\textpm0.09 & ... & 27.80\textpm0.10 & 28.48\textpm0.38 & 28.42\textpm0.29 & 0 & 1 \\
f200d\_uds1\_107 & 34.3890922 & $-$5.2591026 & \textgreater28.58 & \textgreater28.56 & \textgreater28.78 & \textgreater29.01 & 28.20\textpm0.09 & ... & 28.42\textpm0.11 & 29.06\textpm0.38 & 28.94\textpm0.28 & 1 & 1 \\
f200d\_uds1\_115 & 34.5019208 & $-$5.2521015 & \textgreater27.94 & \textgreater27.97 & \textgreater28.23 & \textgreater28.42 & 27.51\textpm0.07 & ... & 28.01\textpm0.10 & \textgreater29.15 & 28.30\textpm0.20 & 0 & 1 \\
f200d\_uds1\_116 & 34.3727848 & $-$5.2520965 & \textgreater28.22 & \textgreater28.27 & \textgreater28.47 & \textgreater28.67 & 27.25\textpm0.09 & ... & 27.30\textpm0.09 & 27.87\textpm0.32 & 28.64\textpm0.53 & 1 & 1 \\
f200d\_uds1\_119 & 34.3552164 & $-$5.2489235 & \textgreater28.27 & \textgreater28.30 & \textgreater28.50 & \textgreater28.72 & 27.73\textpm0.08 & ... & 27.86\textpm0.09 & 28.27\textpm0.26 & 28.13\textpm0.19 & 1 & 1 \\
f200d\_uds1\_158 & 34.2469537 & $-$5.2112719 & \textgreater27.60 & \textgreater27.58 & \textgreater27.85 & \textgreater28.08 & 27.04\textpm0.11 & ... & 26.64\textpm0.07 & 26.89\textpm0.19 & 26.72\textpm0.13 & 1 & 1 \\
f200d\_uds1\_175 & 34.3495507 & $-$5.1520157 & \textgreater28.08 & \textgreater28.09 & \textgreater28.32 & \textgreater28.50 & 27.34\textpm0.08 & ... & 27.55\textpm0.09 & 28.39\textpm0.35 & 27.68\textpm0.15 & 1 & 1 \\
\hline
f200d\_ceers\_008 & 215.0921882 & 52.9199048 & ... & \textgreater29.37 & \textgreater28.91 & 28.23\textpm0.42 & 27.03\textpm0.05 & ... & 26.69\textpm0.03 & 26.72\textpm0.08 & 26.73\textpm0.07 & 0 & 1 \\
f200d\_ceers\_054 & 214.8506944 & 52.7769011 & ... & \textgreater29.14 & \textgreater28.95 & \textgreater29.15 & 28.24\textpm0.13 & ... & 28.03\textpm0.09 & 27.59\textpm0.12 & 27.88\textpm0.14 & 0 & 1 \\
f200d\_ceers\_068 & 215.1373846 & 52.9885643 & ... & \textgreater29.32 & \textgreater28.84 & \textgreater29.06 & 28.20\textpm0.10 & ... & 28.10\textpm0.07 & 28.02\textpm0.16 & 28.28\textpm0.18 & 1 & 1 \\
\hline
f200d\_glass\_001 & 3.5142827 & $-$30.3837111 & \textgreater29.67 & \textgreater29.64 & \textgreater29.42 & \textgreater29.63 & 28.27\textpm0.09 & ... & 28.35\textpm0.08 & ... & 28.74\textpm0.10 & 1 & 1 \\
f200d\_glass\_008 & 3.5087307 & $-$30.3767261 & \textgreater29.90 & \textgreater29.88 & \textgreater29.69 & \textgreater29.89 & 28.71\textpm0.11 & ... & 28.83\textpm0.11 & ... & 29.23\textpm0.13 & 1 & 1 \\
f200d\_glass\_023 & 3.5274208 & $-$30.367695 & \textgreater29.83 & \textgreater29.86 & \textgreater29.66 & \textgreater29.83 & 27.94\textpm0.07 & ... & 27.84\textpm0.06 & ... & 27.80\textpm0.05 & 1 & 1 \\
f200d\_glass\_025 & 3.5234561 & $-$30.3667739 & \textgreater29.83 & \textgreater29.85 & \textgreater29.67 & 30.01\textpm0.50 & 28.90\textpm0.07 & ... & 29.65\textpm0.12 & ... & 29.21\textpm0.07 & 0 & 0 \\
f200d\_glass\_064 & 3.4970253 & $-$30.3498509 & \textgreater29.93 & \textgreater29.89 & \textgreater29.71 & 27.83\textpm0.20 & 26.96\textpm0.04 & ... & 26.97\textpm0.03 & ... & 27.11\textpm0.03 & 0 & 0 \\
f200d\_glass\_084 & 3.4767446 & $-$30.325534 & \textgreater29.87 & \textgreater29.87 & \textgreater29.39 & \textgreater29.89 & 29.00\textpm0.13 & ... & 28.59\textpm0.10 & ... & 29.20\textpm0.11 & 0 & 1 \\
\hline
f200d\_ngdeep\_004 & 53.2494577 & $-$27.8756523 & ... & \textgreater30.86 & \textgreater30.62 & 30.08\textpm0.36 & 29.27\textpm0.06 & ... & 29.30\textpm0.06 & ... & 29.55\textpm0.09 & 0 & 0 \\
\hline
f200d\_jades\_A1\_008 & 53.1677942 & $-$27.768485 & \textgreater30.69 & \textgreater30.86 & \textgreater30.84 & \textgreater30.71 & 29.54\textpm0.04 & 30.08\textpm0.11 & 29.57\textpm0.05 & 30.32\textpm0.14 & 30.27\textpm0.12 & 0 & 0 \\
\hline
f200d\_jades\_A2\_001 & 53.172829 & $-$27.8415601 & \textgreater30.25 & \textgreater30.65 & \textgreater30.46 & \textgreater30.70 & 29.88\textpm0.07 & 30.57\textpm0.17 & 29.89\textpm0.06 & 30.43\textpm0.20 & 30.70\textpm0.21 & 0 & 0 \\
f200d\_jades\_A2\_010 & 53.1269214 & $-$27.7910264 & \textgreater30.79 & \textgreater31.10 & \textgreater31.00 & 30.69\textpm0.32 & 29.76\textpm0.04 & 29.37\textpm0.05 & 29.35\textpm0.03 & 29.60\textpm0.06 & 29.58\textpm0.05 & 0 & 1 \\
f200d\_jades\_A2\_011 & 53.2005513 & $-$27.7849306 & \textgreater30.68 & \textgreater30.88 & \textgreater30.82 & \textgreater30.74 & 29.41\textpm0.04 & 29.31\textpm0.06 & 28.90\textpm0.03 & 29.70\textpm0.09 & 29.47\textpm0.06 & 0 & 0 \\
\hline
f200d\_uncover\_024 & 3.6338542 & $-$30.4374596 & ... & \textgreater28.29 & \textgreater28.45 & \textgreater28.65 & 27.60\textpm0.10 & ... & 27.68\textpm0.09 & ... & 28.02\textpm0.21 & 0 & 1 \\
f200d\_uncover\_147 & 3.6344337 & $-$30.3988995 & ... & \textgreater28.83 & \textgreater28.89 & 29.01\textpm0.51 & 28.10\textpm0.09 & ... & 28.47\textpm0.11 & 28.74\textpm0.31 & 28.67\textpm0.19 & 0 & 1 \\
f200d\_uncover\_189 & 3.5618012 & $-$30.3879883 & ... & \textgreater29.09 & \textgreater29.30 & \textgreater29.29 & 28.31\textpm0.10 & ... & 28.02\textpm0.07 & 28.23\textpm0.16 & 28.90\textpm0.23 & 0 & 1 \\
f200d\_uncover\_333 & 3.5782373 & $-$30.3309647 & ... & \textgreater29.16 & \textgreater29.42 & \textgreater29.42 & 28.38\textpm0.09 & ... & 28.72\textpm0.11 & 29.49\textpm0.41 & 29.15\textpm0.24 & 1 & 1 \\
\hline
f200d\_smacs0723\_A\_004 & 110.7335419 & $-$73.4894246 & \textgreater29.27 & ... & \textgreater29.46 & \textgreater29.66 & 28.47\textpm0.14 & ... & 27.73\textpm0.07 & ... & 27.93\textpm0.12 & 0 & 1 \\
f200d\_smacs0723\_A\_013 & 110.7067627 & $-$73.4743838 & \textgreater29.25 & ... & \textgreater29.41 & 28.68\textpm0.47 & 27.58\textpm0.07 & ... & 27.36\textpm0.05 & ... & 27.38\textpm0.08 & 0 & 1 \\
f200d\_smacs0723\_B\_010 & 110.8019147 & $-$73.4572911 & \textgreater29.00 & ... & \textgreater29.21 & \textgreater29.37 & 28.45\textpm0.09 & ... & 28.14\textpm0.06 & ... & 28.59\textpm0.14 & 1 & 1 \\
f200d\_smacs0723\_B\_023 & 110.8026074 & $-$73.442157 & \textgreater28.96 & ... & \textgreater29.21 & 29.01\textpm0.54 & 28.20\textpm0.08 & ... & 28.28\textpm0.08 & ... & 28.84\textpm0.20 & 0 & 1 \\
\hline
\enddata
\tablecomments{Similar to Table~\ref{tab:catf150wd} but for the F200W dropouts.}
\end{deluxetable}
\label{tab:catf200wd}
\end{longrotatetable}

\startlongtable
\begin{longrotatetable}
\begin{deluxetable}{cccccccccccccc}
\tablecolumns{14}
\tabletypesize{\scriptsize}
\tablecaption{Photometry table for F277W dropouts. }
\tablehead{
\colhead{SID} & 
\colhead{R.A.} & 
\colhead{Decl.} & 
\colhead{$m_{090}$} & 
\colhead{$m_{115}$} & 
\colhead{$m_{150}$} & 
\colhead{$m_{200}$} & 
\colhead{$m_{277}$} & 
\colhead{$m_{335}$} & 
\colhead{$m_{356}$} & 
\colhead{$m_{410}$} & 
\colhead{$m_{444}$} & \\
\colhead{} & 
\colhead{(degrees)} & 
\colhead{(degrees)} & 
\colhead{(mag)} & 
\colhead{(mag)} & 
\colhead{(mag)} & 
\colhead{(mag)} & 
\colhead{(mag)} & 
\colhead{(mag)} & 
\colhead{(mag)} & 
\colhead{(mag)} & 
\colhead{(mag)} 
}
\startdata
f277d\_cosmos\_026 & 150.1047951 & 2.264002 & \textgreater27.98 & \textgreater27.96 & \textgreater28.20 & \textgreater28.40 & 28.89\textpm0.17 & ... & 27.67\textpm0.06 & 27.45\textpm0.10 & 27.21\textpm0.07 \\
f277d\_cosmos\_038 & 150.0917946 & 2.2896575 & \textgreater28.06 & \textgreater28.05 & \textgreater28.26 & \textgreater28.47 & 28.74\textpm0.20 & ... & 27.91\textpm0.10 & 27.64\textpm0.15 & 27.39\textpm0.10 \\
f277d\_cosmos\_083 & 150.1319202 & 2.3999325 & \textgreater28.54 & \textgreater28.58 & \textgreater28.65 & \textgreater28.84 & \textgreater30.14 & ... & 29.04\textpm0.12 & 28.63\textpm0.18 & 28.83\textpm0.17 \\
f277d\_cosmos\_091 & 150.1044624 & 2.4353101 & \textgreater27.55 & \textgreater27.64 & \textgreater27.85 & \textgreater28.07 & 28.67\textpm0.32 & ... & 27.29\textpm0.09 & 27.15\textpm0.15 & 26.77\textpm0.08 \\
f277d\_cosmos\_092 & 150.1044823 & 2.4354193 & \textgreater27.53 & \textgreater27.62 & \textgreater27.88 & \textgreater28.04 & 27.21\textpm0.14 & ... & 26.15\textpm0.05 & 25.99\textpm0.08 & 25.79\textpm0.05 \\
f277d\_cosmos\_099 & 150.1510552 & 2.4737863 & \textgreater27.76 & \textgreater27.72 & \textgreater27.90 & \textgreater28.25 & 29.27\textpm0.48 & ... & 27.93\textpm0.12 & 27.5\textpm0.15 & 28.17\textpm0.22 \\
\hline
f277d\_uds1\_003 & 34.4769352 & $-$5.3142233 & \textgreater27.70 & \textgreater27.67 & \textgreater27.86 & \textgreater28.10 & 28.73\textpm0.31 & ... & 26.84\textpm0.05 & 26.12\textpm0.05 & 26.04\textpm0.04 \\
f277d\_uds1\_011 & 34.3655349 & $-$5.2856743 & \textgreater28.10 & \textgreater28.08 & \textgreater28.30 & \textgreater28.51 & \textgreater29.51 & ... & 28.12\textpm0.10 & 27.22\textpm0.09 & 27.05\textpm0.07 \\
f277d\_uds1\_040 & 34.4850689 & $-$5.2455844 & \textgreater28.28 & \textgreater28.34 & \textgreater28.53 & \textgreater28.72 & 29.34\textpm0.28 & ... & 28.43\textpm0.11 & 28.65\textpm0.28 & 28.22\textpm0.16 \\
f277d\_uds1\_056 & 34.4737407 & $-$5.2176787 & \textgreater28.19 & \textgreater28.21 & \textgreater28.44 & \textgreater28.64 & 29.76\textpm0.45 & ... & 27.82\textpm0.08 & 27.24\textpm0.09 & 27.44\textpm0.08 \\
f277d\_uds1\_064 & 34.3358389 & $-$5.2014531 & \textgreater27.51 & \textgreater27.57 & \textgreater27.77 & \textgreater27.98 & 29.43\textpm0.44 & ... & 28.23\textpm0.14 & 28.92\textpm0.58 & 27.66\textpm0.14 \\
f277d\_uds1\_069 & 34.3523512 & $-$5.1816831 & \textgreater27.93 & \textgreater27.97 & \textgreater28.19 & \textgreater28.42 & \textgreater29.63 & ... & 28.79\textpm0.15 & 26.88\textpm0.06 & 26.93\textpm0.04 \\
f277d\_uds1\_077 & 34.4294851 & $-$5.1288507 & \textgreater27.77 & \textgreater27.70 & \textgreater27.92 & \textgreater28.12 & 28.11\textpm0.22 & ... & 27.16\textpm0.08 & 26.42\textpm0.08 & 26.66\textpm0.08 \\
\hline
f277d\_ceers\_022 & 214.9298649 & 52.8072661 & ... & \textgreater29.62 & \textgreater28.96 & \textgreater29.22 & 29.47\textpm0.29 & ... & 28.01\textpm0.06 & 27.65\textpm0.11 & 27.68\textpm0.07 \\
f277d\_ceers\_095 & 214.8057858 & 52.7347332 & ... & \textgreater29.11 & \textgreater28.93 & \textgreater29.14 & 29.39\textpm0.25 & ... & 28.56\textpm0.11 & 29.17\textpm0.39 & 29.20\textpm0.33 \\
f277d\_ceers\_485 & 214.7148366 & 52.7434037 & ... & \textgreater29.06 & \textgreater28.87 & \textgreater29.07 & 29.67\textpm0.29 & ... & 28.51\textpm0.10 & 27.91\textpm0.12 & 27.94\textpm0.10 \\
\hline
f277d\_glass\_007 & 3.4891744 & $-$30.3645975 & \textgreater29.69 & ... & \textgreater29.41 & \textgreater29.67 & 30.33\textpm0.32 & ... & 29.44\textpm0.13 & ... & 30.20\textpm0.21 \\
f277d\_glass\_011 & 3.4949794 & $-$30.3614903 & \textgreater29.92 & ... & \textgreater29.71 & \textgreater29.91 & \textgreater30.90 & ... & 29.91\textpm0.17 & ... & 29.83\textpm0.13 \\
\hline
f277d\_ngdeep\_012 & 53.2519601 & $-$27.8732495 & ... & \textgreater30.86 & \textgreater30.63 & \textgreater30.68 & 31.72\textpm0.44 & ... & 30.49\textpm0.15 & ... & 30.46\textpm0.16 \\
f277d\_ngdeep\_053 & 53.2492831 & $-$27.824109 & ... & \textgreater30.79 & \textgreater30.57 & \textgreater30.63 & 31.27\textpm0.31 & ... & 30.15\textpm0.11 & ... & 30.48\textpm0.18 \\
\hline
f277d\_jades\_A1\_059 & 53.1622288 & $-$27.8025417 & \textgreater30.23 & \textgreater30.63 & \textgreater30.44 & \textgreater30.67 & 31.98\textpm0.21 & 31.03\textpm0.14 & 30.72\textpm0.07 & 29.94\textpm0.06 & 30.28\textpm0.07 \\
f277d\_jades\_A1\_122 & 53.1777365 & $-$27.774737 & \textgreater30.46 & \textgreater30.76 & \textgreater30.60 & \textgreater30.78 & 31.95\textpm0.30 & \textgreater31.68 & 30.57\textpm0.07 & 30.73\textpm0.15 & 31.88\textpm0.40 \\
f277d\_jades\_A1\_125 & 53.1591677 & $-$27.7731353 & \textgreater30.37 & \textgreater30.69 & \textgreater30.49 & \textgreater30.73 & 32.03\textpm0.34 & 32.02\textpm0.47 & 30.76\textpm0.09 & \textgreater31.14 & 31.01\textpm0.18 \\
f277d\_jades\_A1\_993 & 53.1727658 & $-$27.7918711 & \textgreater30.81 & \textgreater31.11 & \textgreater31.02 & \textgreater31.06 & \textgreater32.22 & 31.32\textpm0.18 & 31.3\textpm0.12 & \textgreater31.64 & \textgreater31.81 \\
\hline
f277d\_jades\_A2\_056 & 53.1976644 & $-$27.7700823 & \textgreater30.60 & \textgreater30.76 & \textgreater30.73 & \textgreater30.51 & 32.32\textpm0.26 & 31.52\textpm0.19 & 31.24\textpm0.10 & \textgreater31.72 & \textgreater31.83 \\
f277d\_jades\_A2\_064 & 53.1318805 & $-$27.766531 & \textgreater30.73 & \textgreater31.03 & \textgreater30.92 & \textgreater30.94 & 31.38\textpm0.24 & 31.38\textpm0.32 & 30.38\textpm0.08 & \textgreater31.19 & 30.52\textpm0.16 \\
f277d\_jades\_A2\_074 & 53.1817889 & $-$27.7584684 & \textgreater31.24 & \textgreater31.51 & \textgreater31.38 & \textgreater31.44 & \textgreater32.48 & \textgreater32.00 & 31.26\textpm0.09 & 30.69\textpm0.08 & 30.60\textpm0.07 \\
f277d\_jades\_A2\_104 & 53.1717078 & $-$27.7442152 & \textgreater30.70 & \textgreater30.90 & \textgreater30.83 & \textgreater30.70 & 32.06\textpm0.28 & \textgreater31.26 & 30.73\textpm0.10 & 31.06\textpm0.17 & \textgreater31.58 \\
f277d\_jades\_A2\_112 & 53.172755 & $-$27.739298 & \textgreater30.69 & \textgreater30.90 & \textgreater30.82 & \textgreater30.70 & 33.10\textpm0.51 & 31.22\textpm0.17 & 31.51\textpm0.14 & \textgreater31.50 & \textgreater31.64 \\
f277d\_jades\_A2\_118 & 53.1878657 & $-$27.7420226 & \textgreater30.78 & \textgreater30.94 & \textgreater30.90 & \textgreater30.72 & 33.38\textpm0.51 & 32.23\textpm0.33 & 32.07\textpm0.18 & 32.43\textpm0.34 & 31.87\textpm0.18 \\
\hline
f277d\_uncover\_011 & 3.6369782 & $-$30.4211829 & ... & \textgreater28.63 & \textgreater28.83 & \textgreater29.05 & 27.09\textpm0.08 & ... & 25.94\textpm0.02 & ... & 25.48\textpm0.03 \\
\hline
f277d\_smacs0723\_B\_001 & 110.78867 & $-$73.4559726 & \textgreater29.26 & ... & \textgreater29.46 & \textgreater29.62 & \textgreater30.64 & ... & 29.75\textpm0.15 & ... & 28.78\textpm0.09 \\
\hline
\enddata
\tablecomments{Similar to Table~\ref{tab:catf150wd} but for the F277W dropouts.
Note that these dropouts are not purified due to the limitation in their SEDs.}
\end{deluxetable}
\label{tab:catf277wd}
\end{longrotatetable}


\begin{figure*}[htbp]
    \centering
    \includegraphics[width=0.85\textwidth]{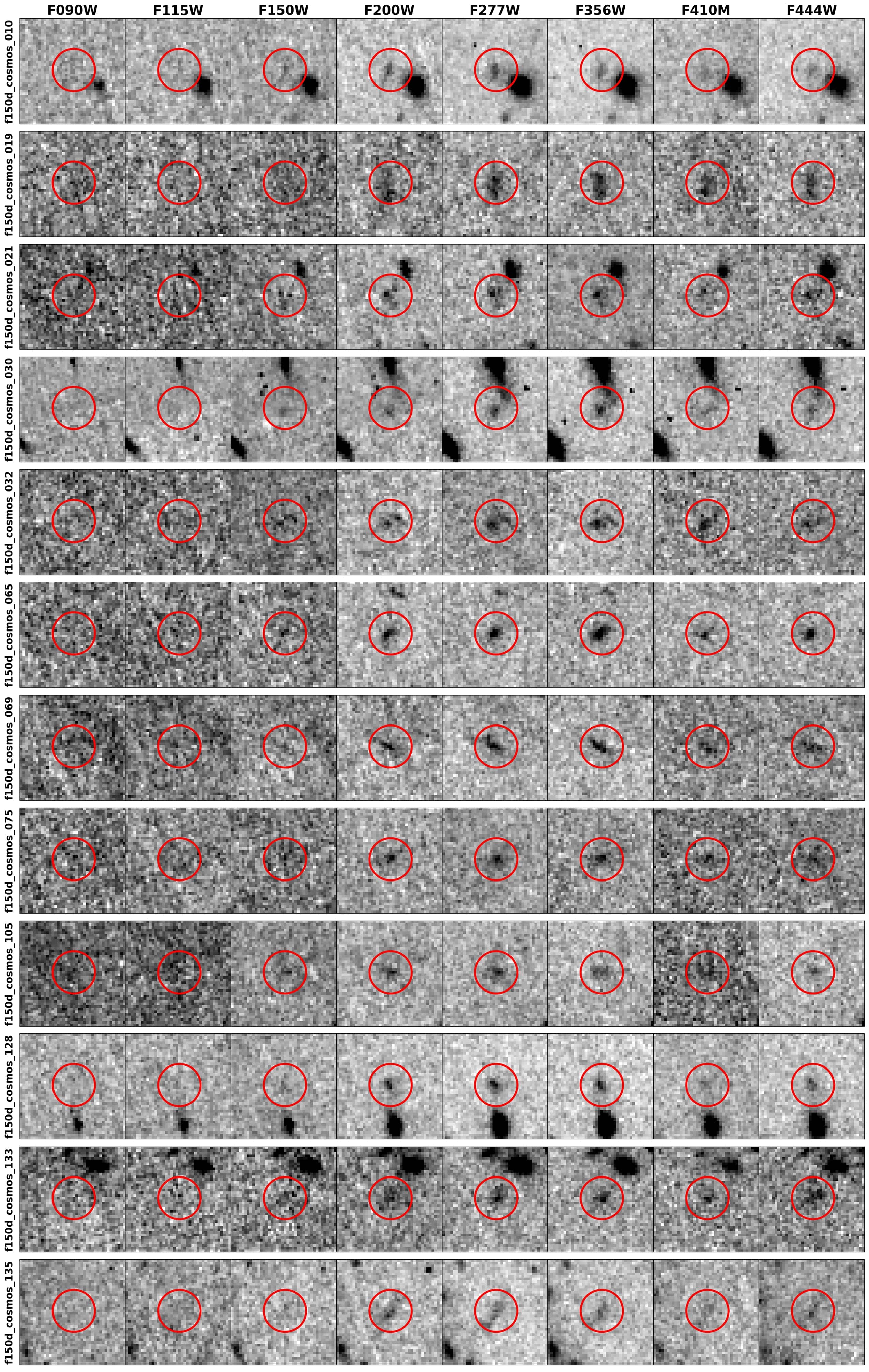}
    \caption{Similar to Figure~\ref{fig:dropoutdemo}; for the F150W dropouts
     in COSMOS.
   }
    \label{fig:f150d_cosmos_1}
\end{figure*}

\renewcommand{\thefigure}{A\arabic{figure} (Cont.)}
\addtocounter{figure}{-1}

\begin{figure*}[htbp]
    \centering
    \includegraphics[width=0.85\textwidth]{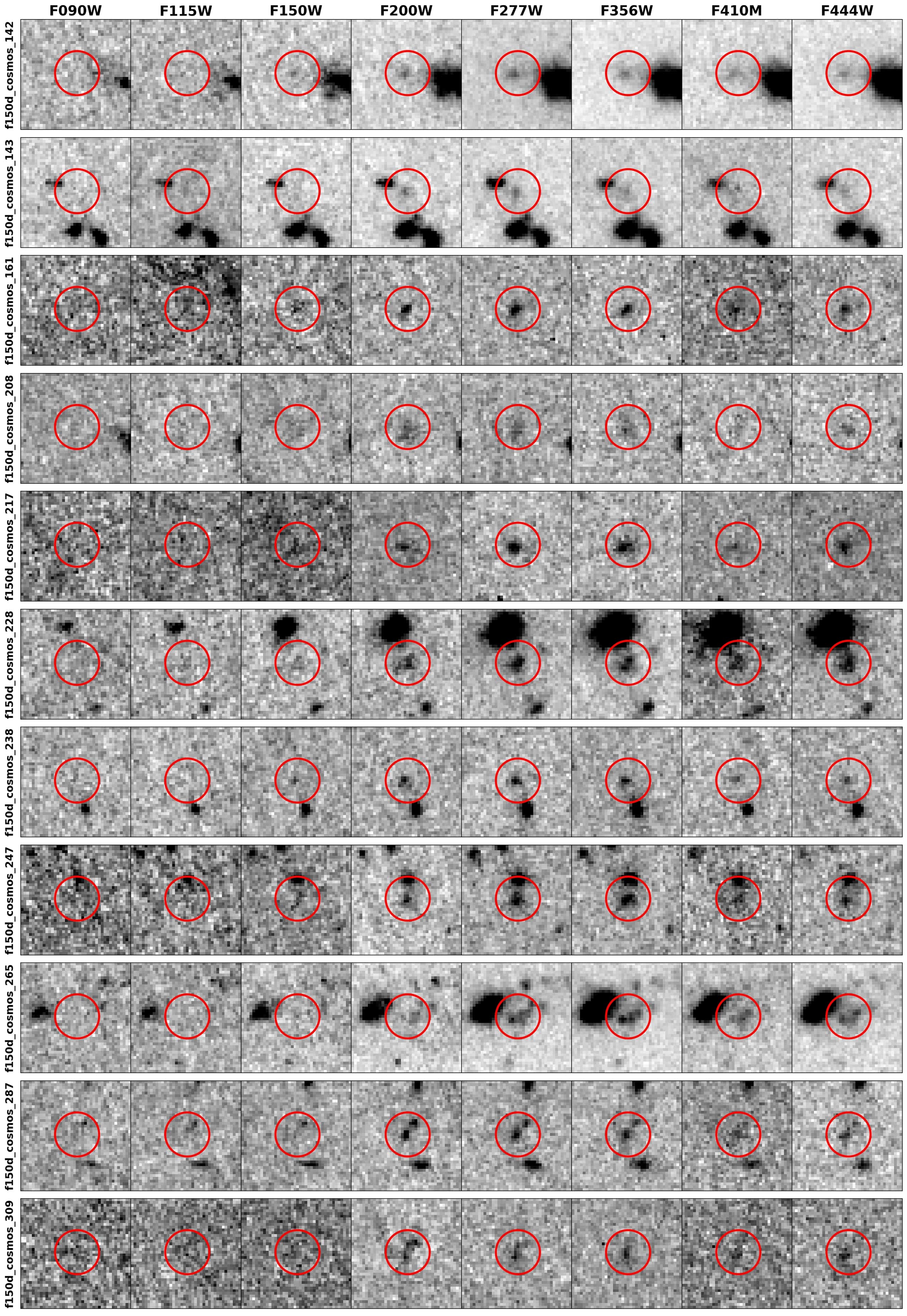}
    \caption{Similar to Figure~\ref{fig:dropoutdemo}; for the F150W dropouts 
     in COSMOS.
   }
    \label{fig:f150d_cosmos_2}
\end{figure*}

\renewcommand{\thefigure}{A\arabic{figure}}

\begin{figure*}[htbp]
    \centering
    \includegraphics[width=0.65\textwidth]{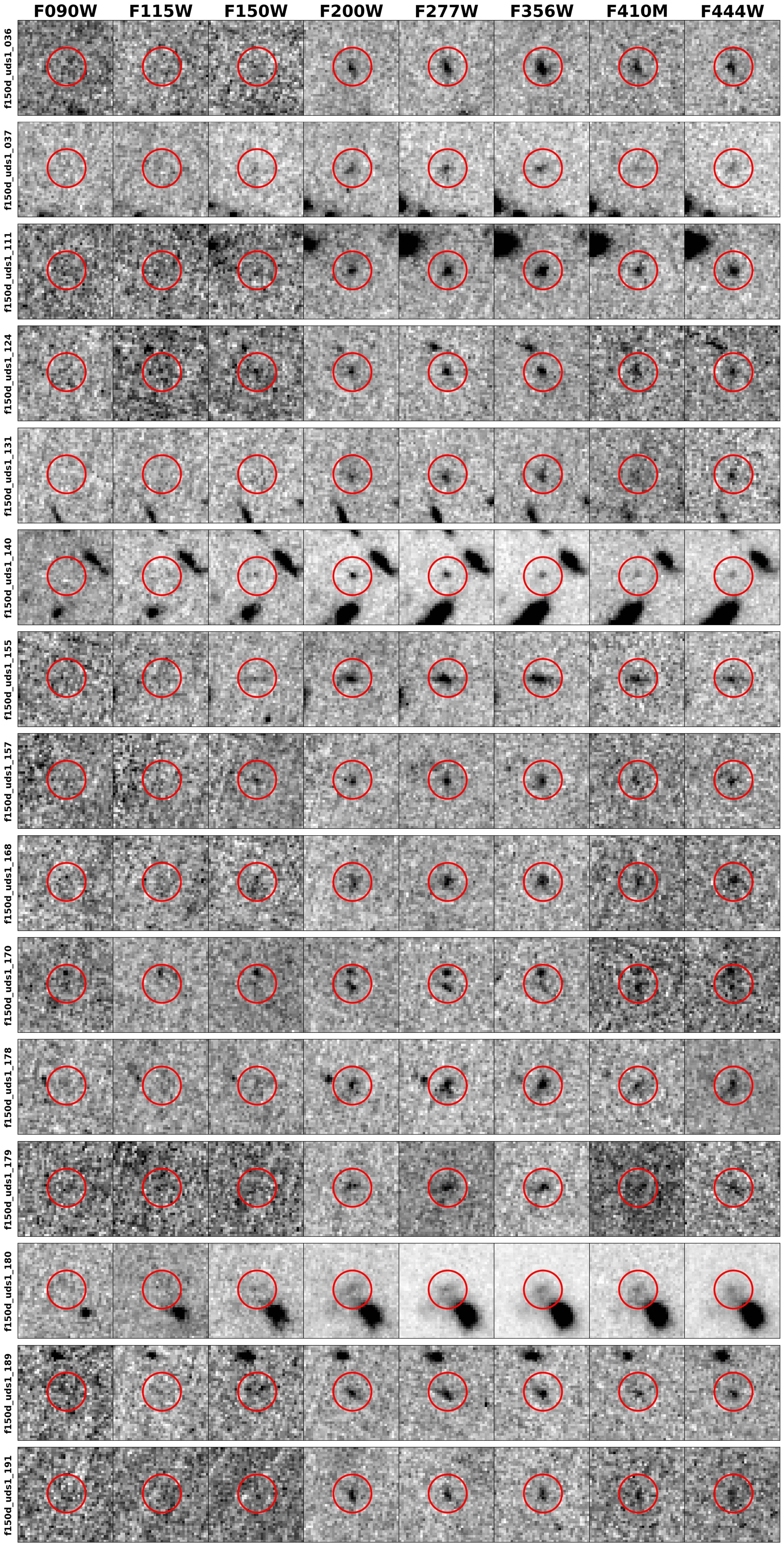}
    \caption{Similar to Figure~\ref{fig:dropoutdemo}; for the F150W dropouts 
     in UDS1.
   }
    \label{fig:f150d_uds1}
\end{figure*}

\begin{figure*}[htbp]
    \centering
    \includegraphics[width=\textwidth]{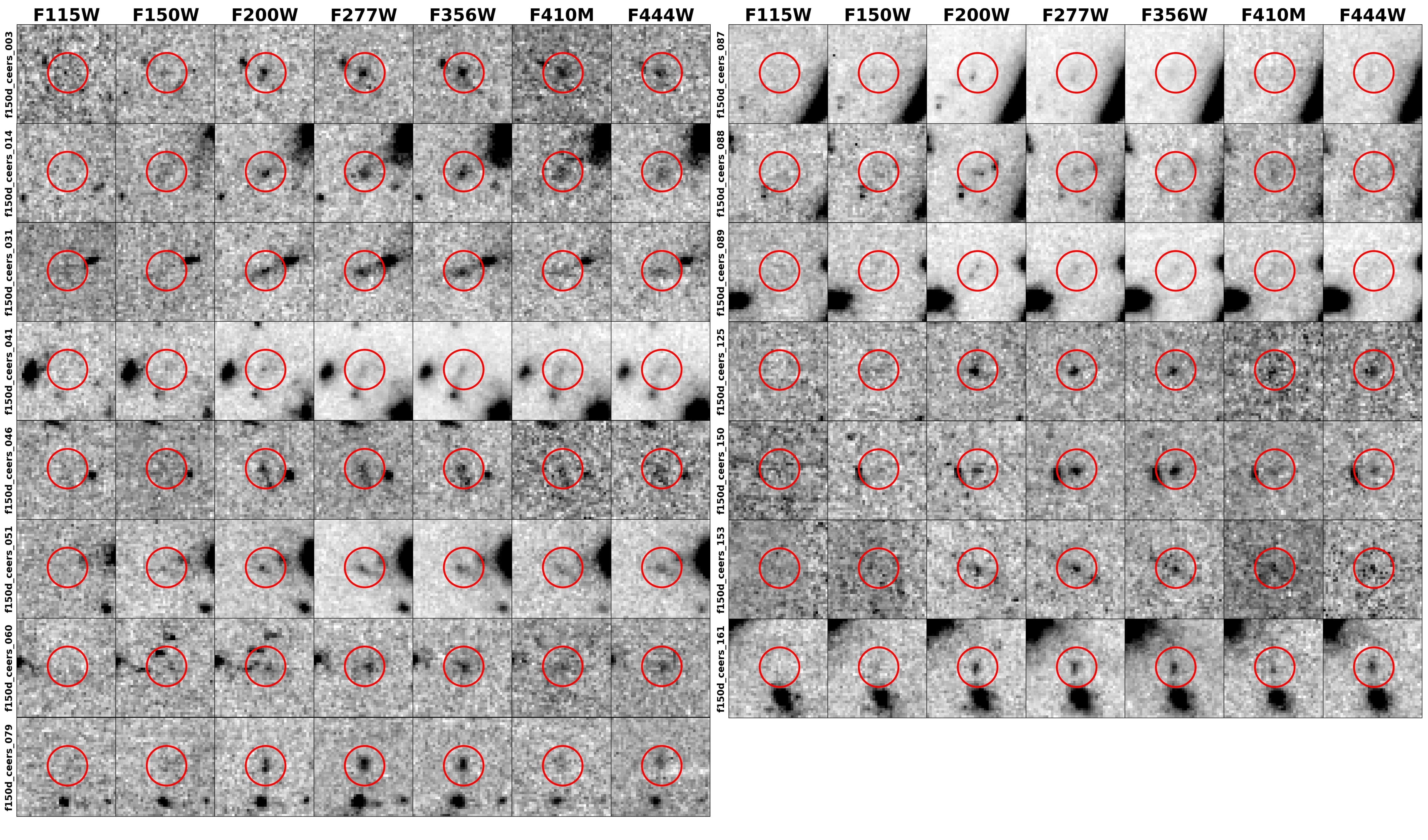}
    \caption{Similar to Figure~\ref{fig:dropoutdemo}; for the F150W dropouts
     in CEERS.
   }
    \label{fig:f150d_ceers}
\end{figure*}

\begin{figure*}[htbp]
    \centering
    \includegraphics[width=\textwidth]{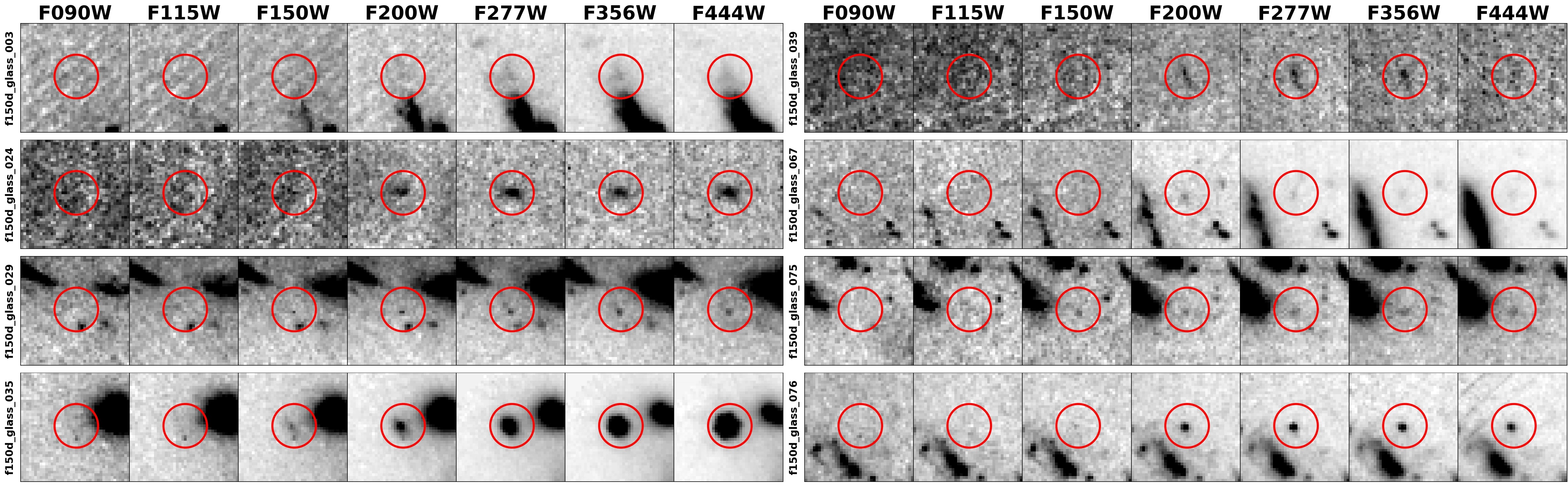}
    \includegraphics[width=\textwidth]{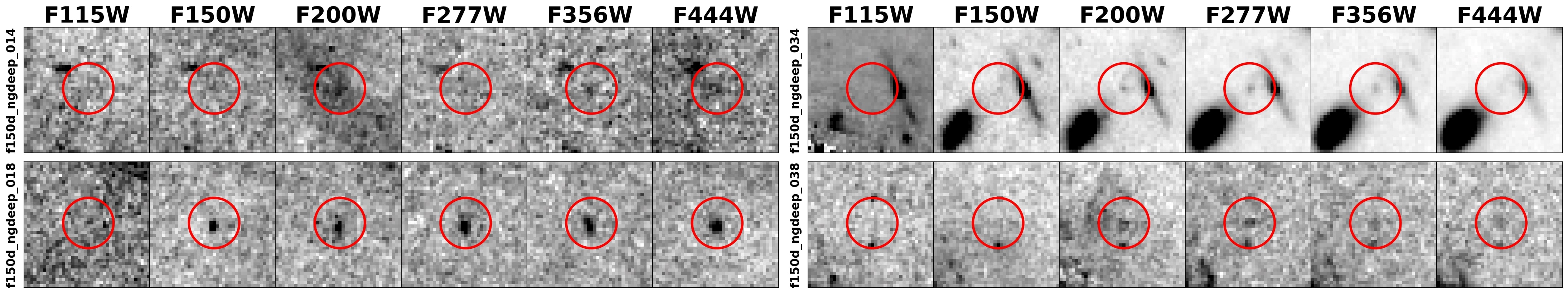}
    \caption{Similar to Figure~\ref{fig:dropoutdemo}; for the F150W dropouts
     in GLASS and NGDEEP
   }
    \label{fig:f150d_glass_ngdeep}
\end{figure*}

\begin{figure*}[htbp]
    \centering
    \includegraphics[width=\textwidth]{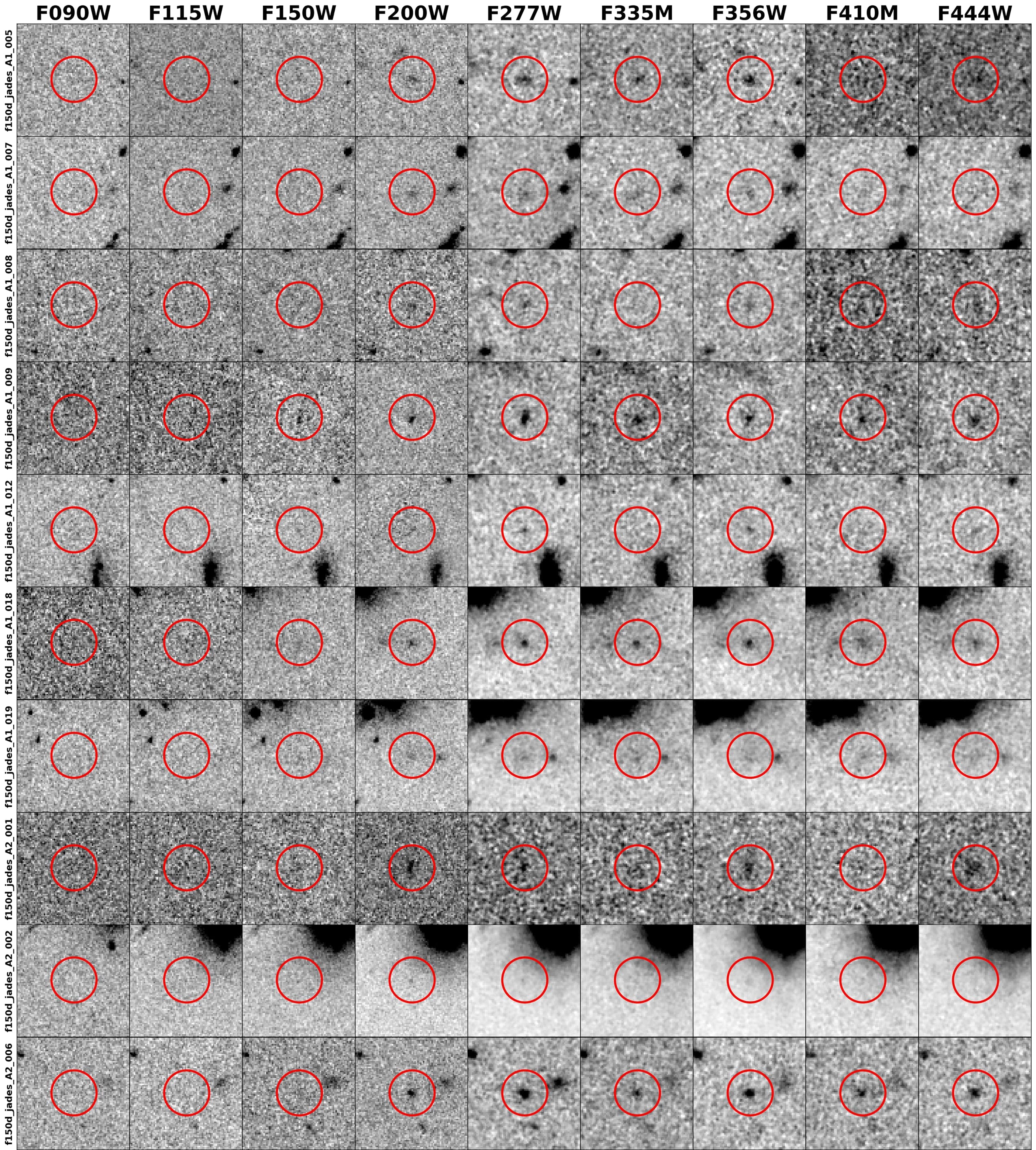}
    \caption{Similar to Figure~\ref{fig:dropoutdemo}; for the F150W dropouts
     in JADES GOODS-S.
   }
    \label{fig:f150d_jades_1}
\end{figure*}

\renewcommand{\thefigure}{A\arabic{figure} (Cont.)}
\addtocounter{figure}{-1}

\begin{figure*}[htbp]
    \centering
    \includegraphics[width=\textwidth]{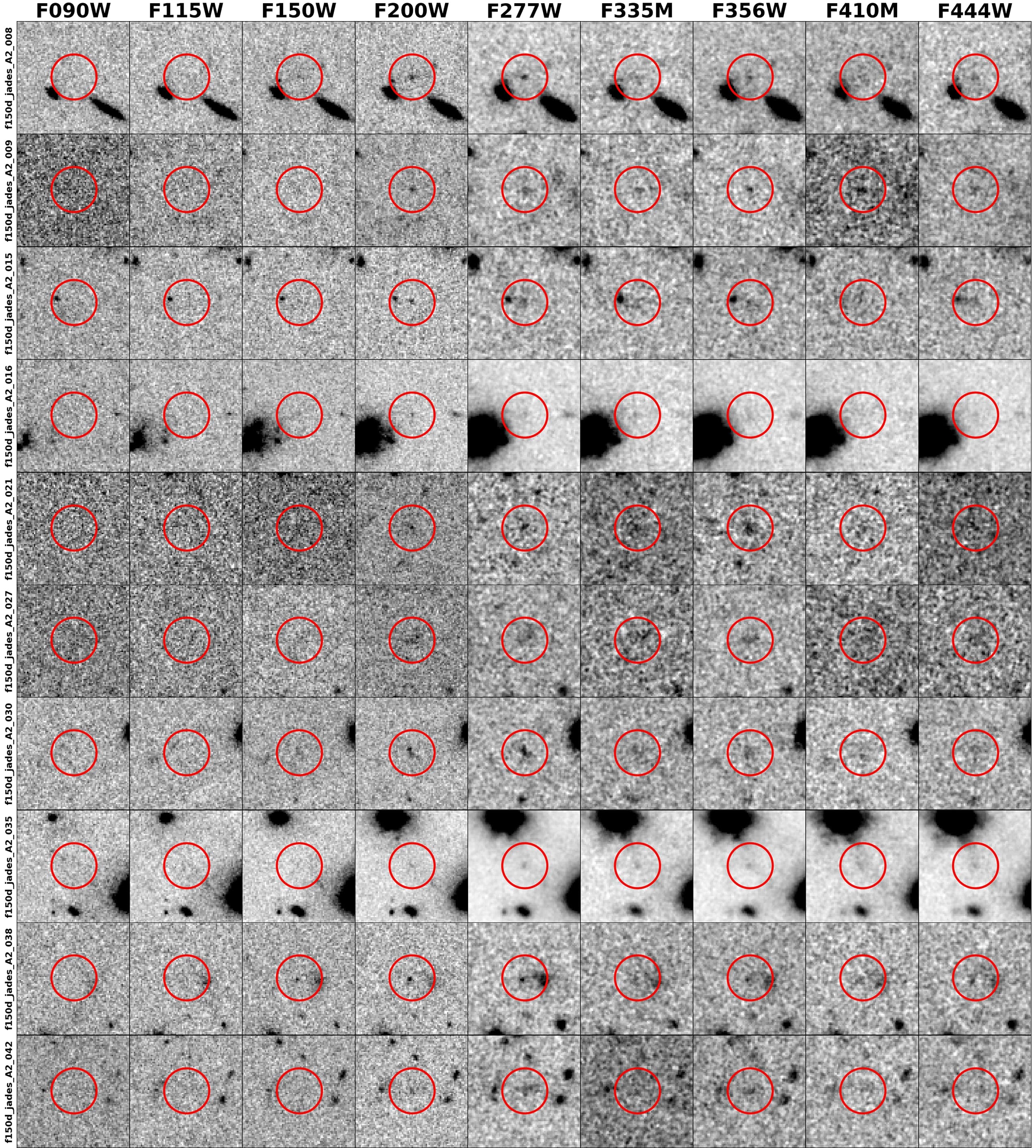}
    \caption{Similar to Figure~\ref{fig:dropoutdemo}; for the F150W dropouts
     in JADES GOODS-S.
   }
    \label{fig:f150d_jades_1}
\end{figure*}

\renewcommand{\thefigure}{A\arabic{figure}}

\begin{figure*}[htbp]
    \centering
    \includegraphics[width=\textwidth]{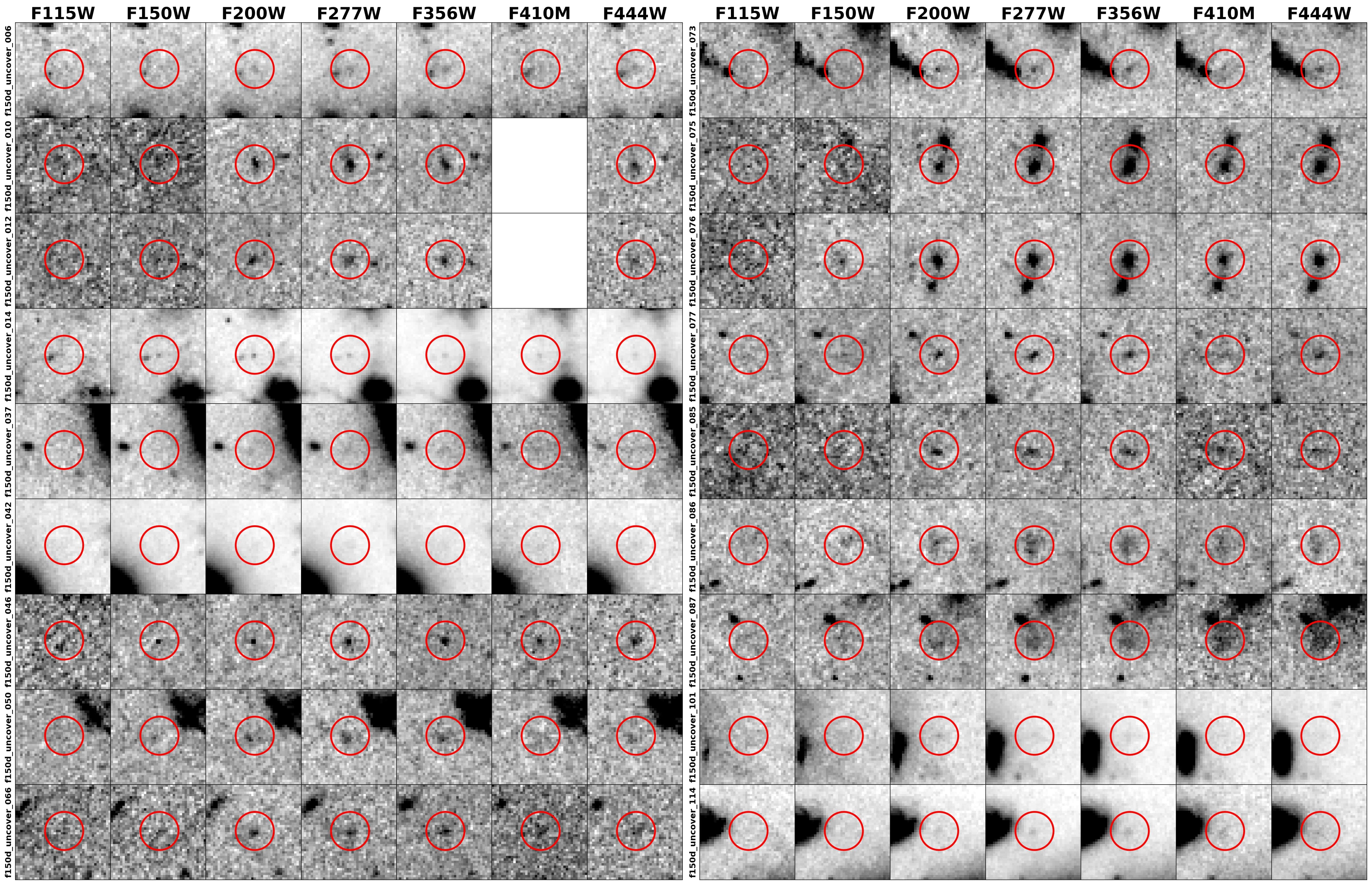}
    \caption{Similar to Figure~\ref{fig:dropoutdemo}; for the F150W dropouts
     in UNCOVER.
   }
    \label{fig:f150d_uncover}
\end{figure*}

\begin{figure*}[htbp]
    \centering
    \includegraphics[width=\textwidth]{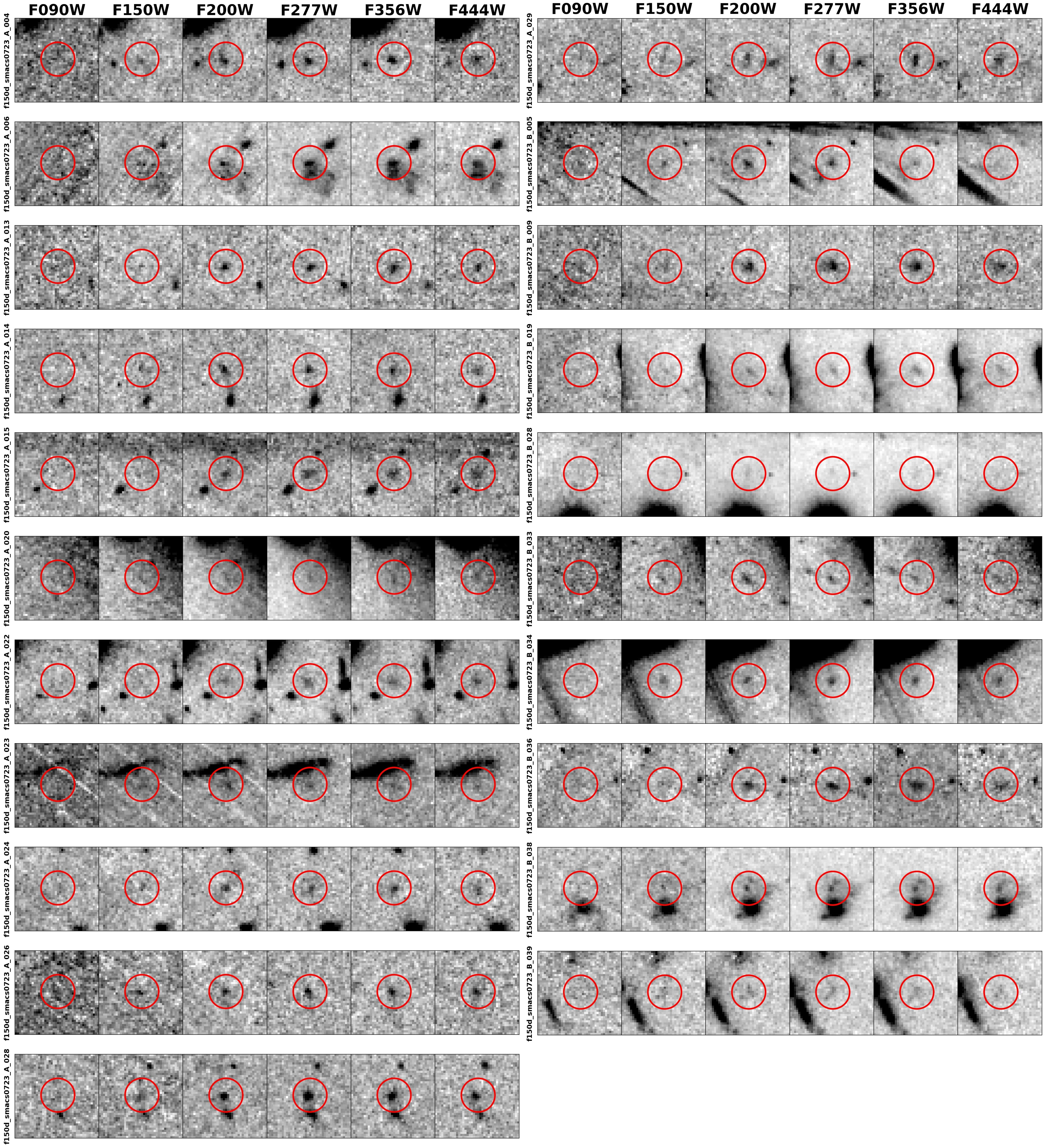}
    \caption{Similar to Figure~\ref{fig:dropoutdemo}; for the F150W dropouts
     in SMACS0723.
   }
    \label{fig:f150d_smacs}
\end{figure*}


\begin{figure*}[htbp]
    \centering
    \includegraphics[width=0.65\textwidth]{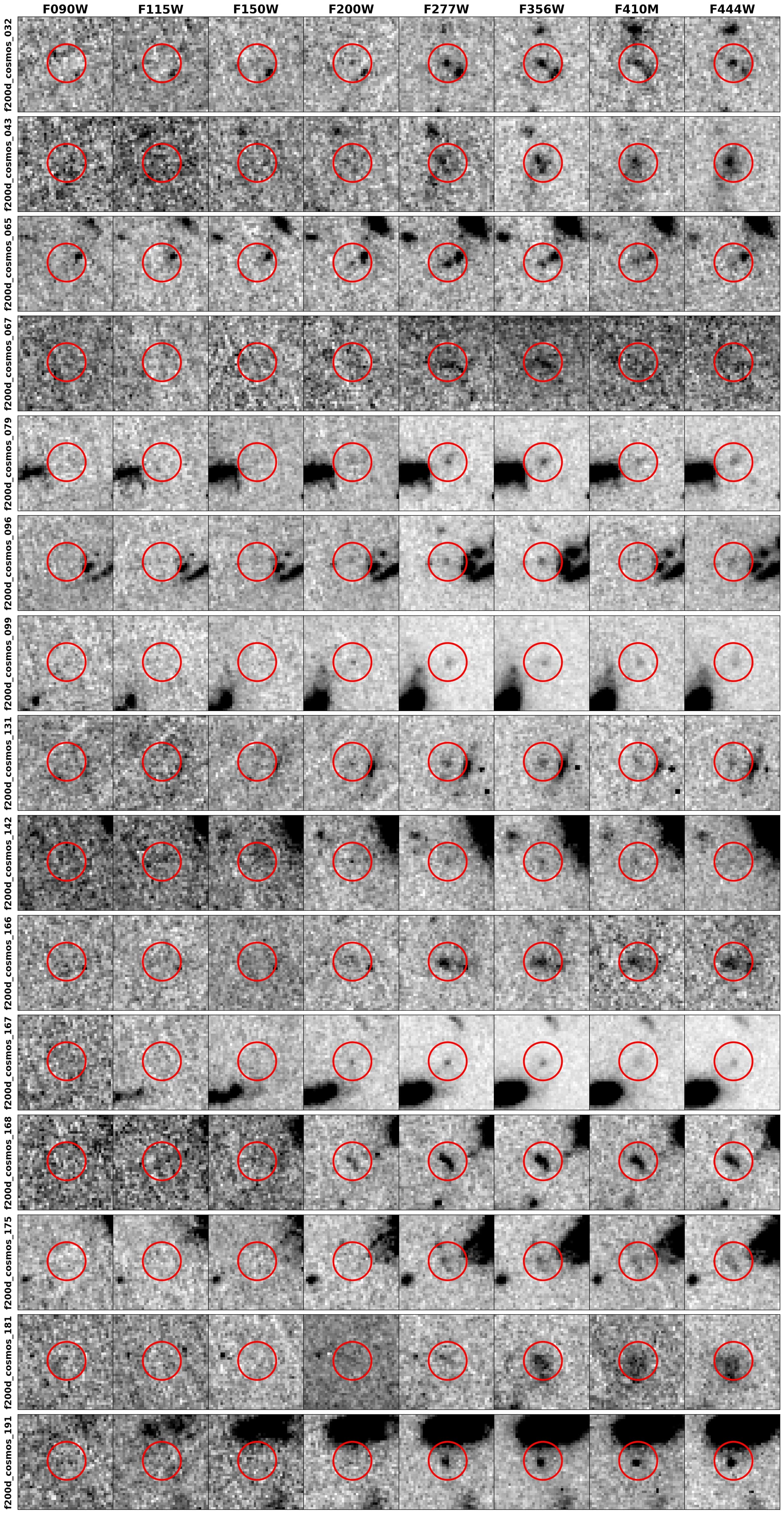}
    \caption{Similar to Figure~\ref{fig:dropoutdemo}; for the F200W dropouts 
     in COSMOS.
   }
    \label{fig:f200d_cosmos}
\end{figure*}

\begin{figure*}[htbp]
    \centering
    \includegraphics[width=0.65\textwidth]{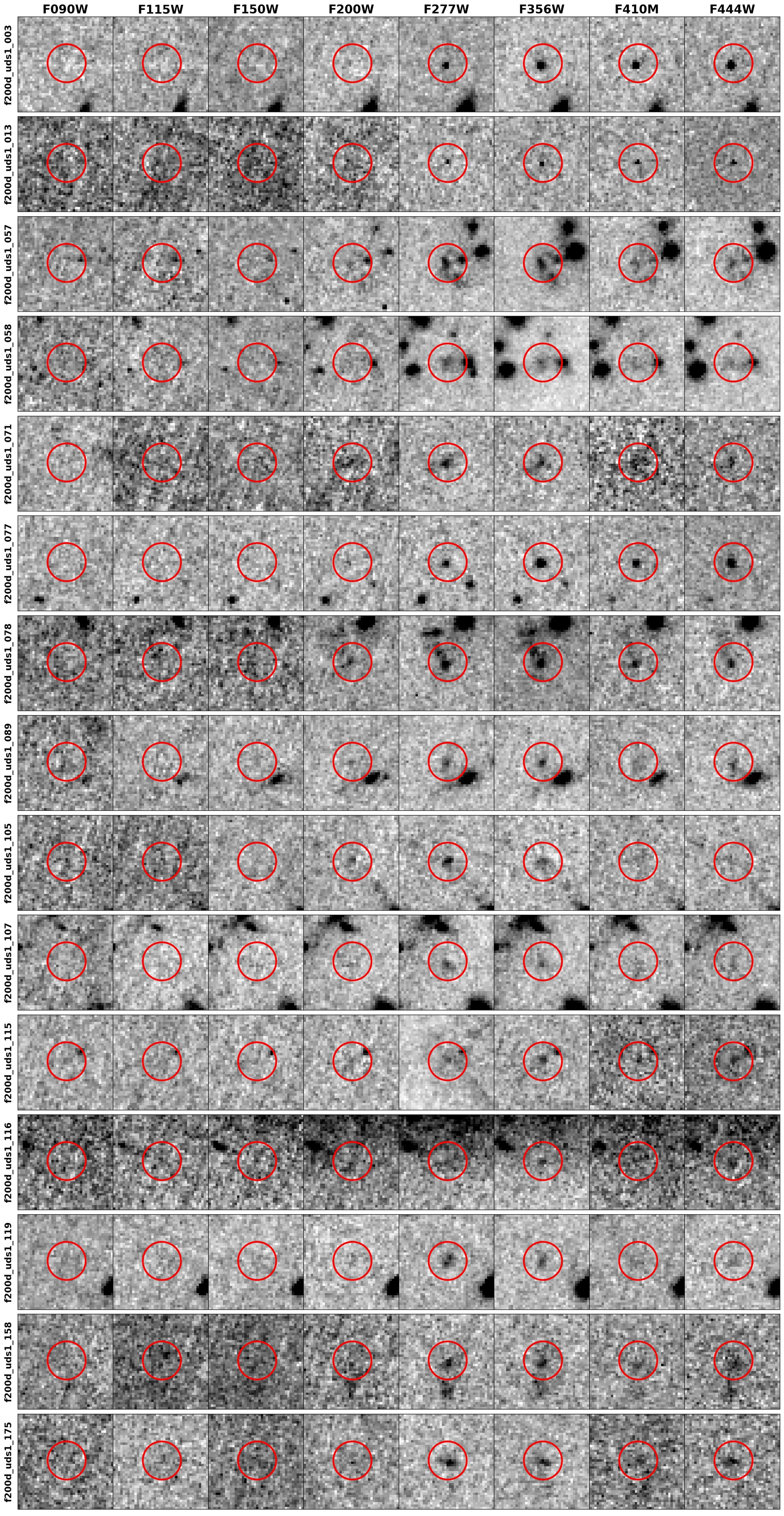}
    \caption{Similar to Figure~\ref{fig:dropoutdemo}; for the F200W dropouts 
     in UDS1.
   }
    \label{fig:f200d_uds}
\end{figure*}

\begin{figure*}[htbp]
    \centering
    \includegraphics[width=0.65\textwidth]{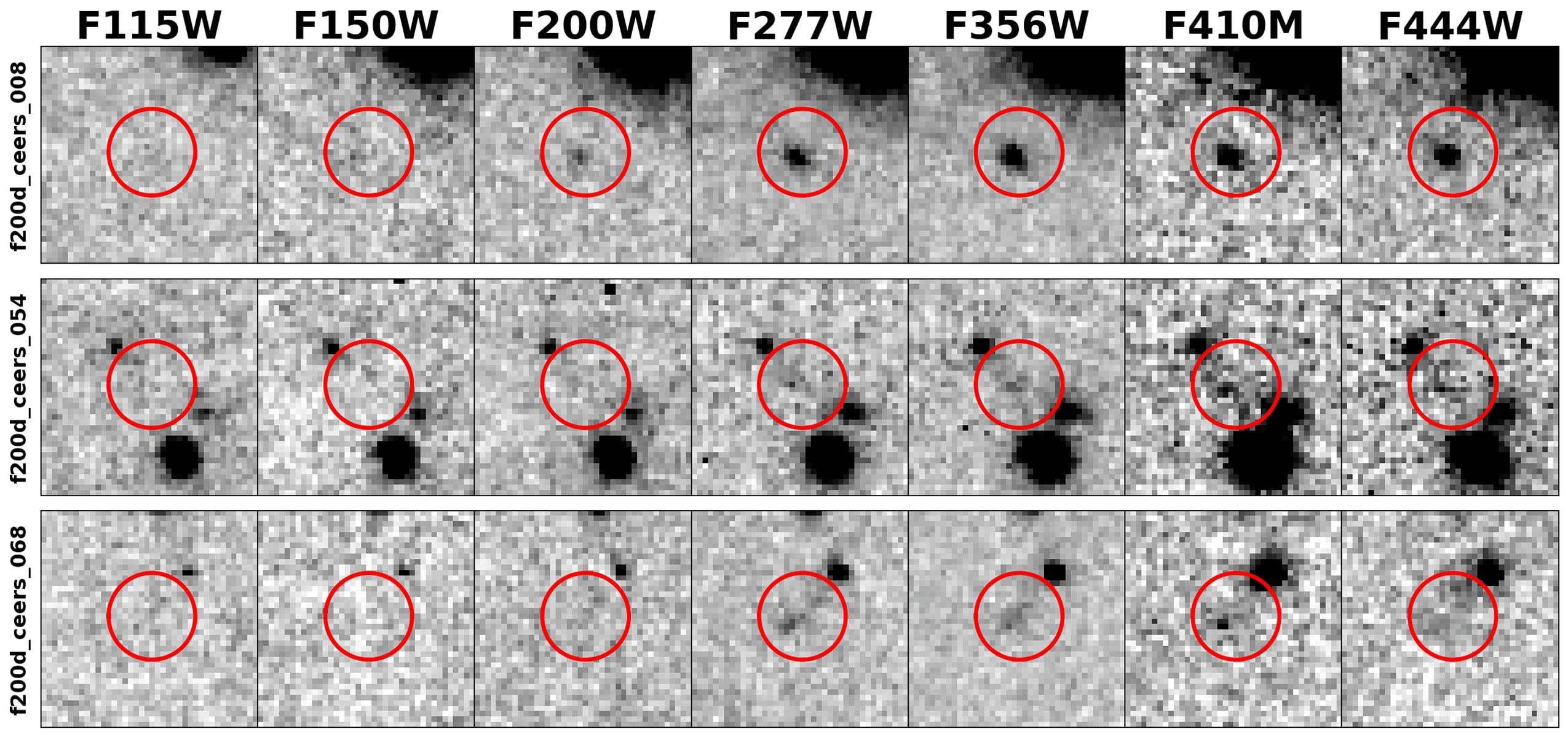}
    \includegraphics[width=0.65\textwidth]{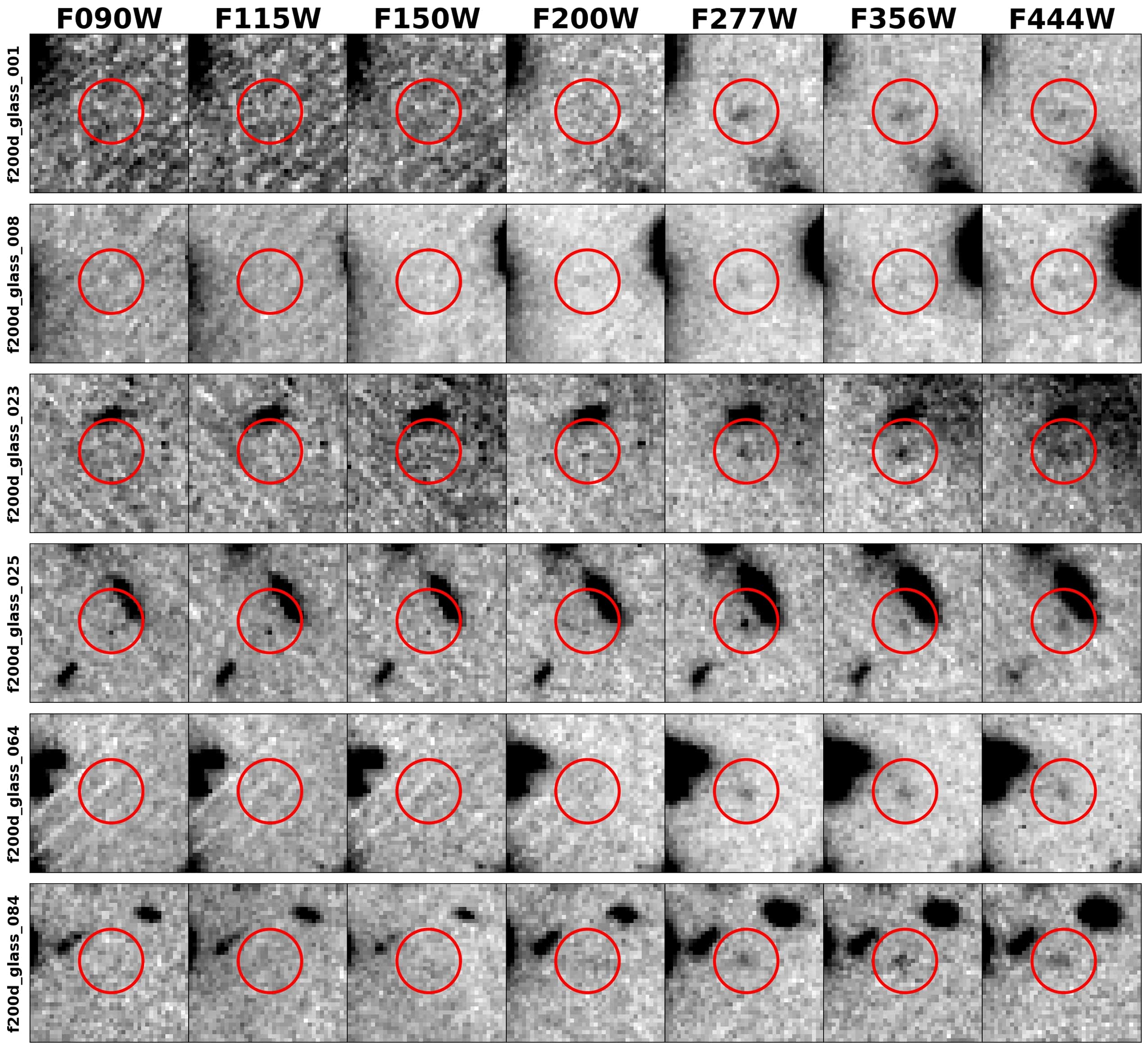}
    \includegraphics[width=0.65\textwidth]{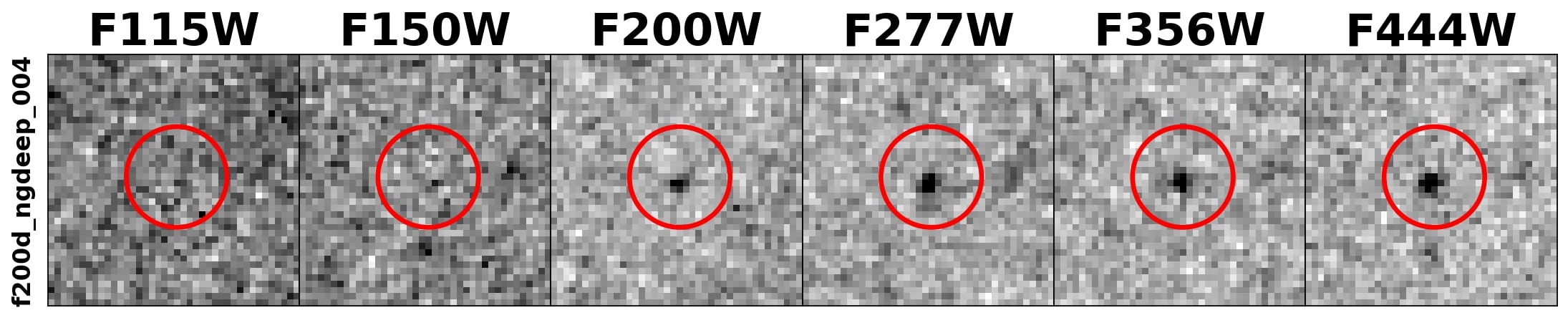}
    \caption{Similar to Figure~\ref{fig:dropoutdemo}; for the F200W dropouts 
     in CEERS, GLASS and NGDEEP.
   }
    \label{fig:f200d_ceers_glass_ngdeep}
\end{figure*}

\begin{figure*}[htbp]
    \centering
    \includegraphics[width=0.65\textwidth]{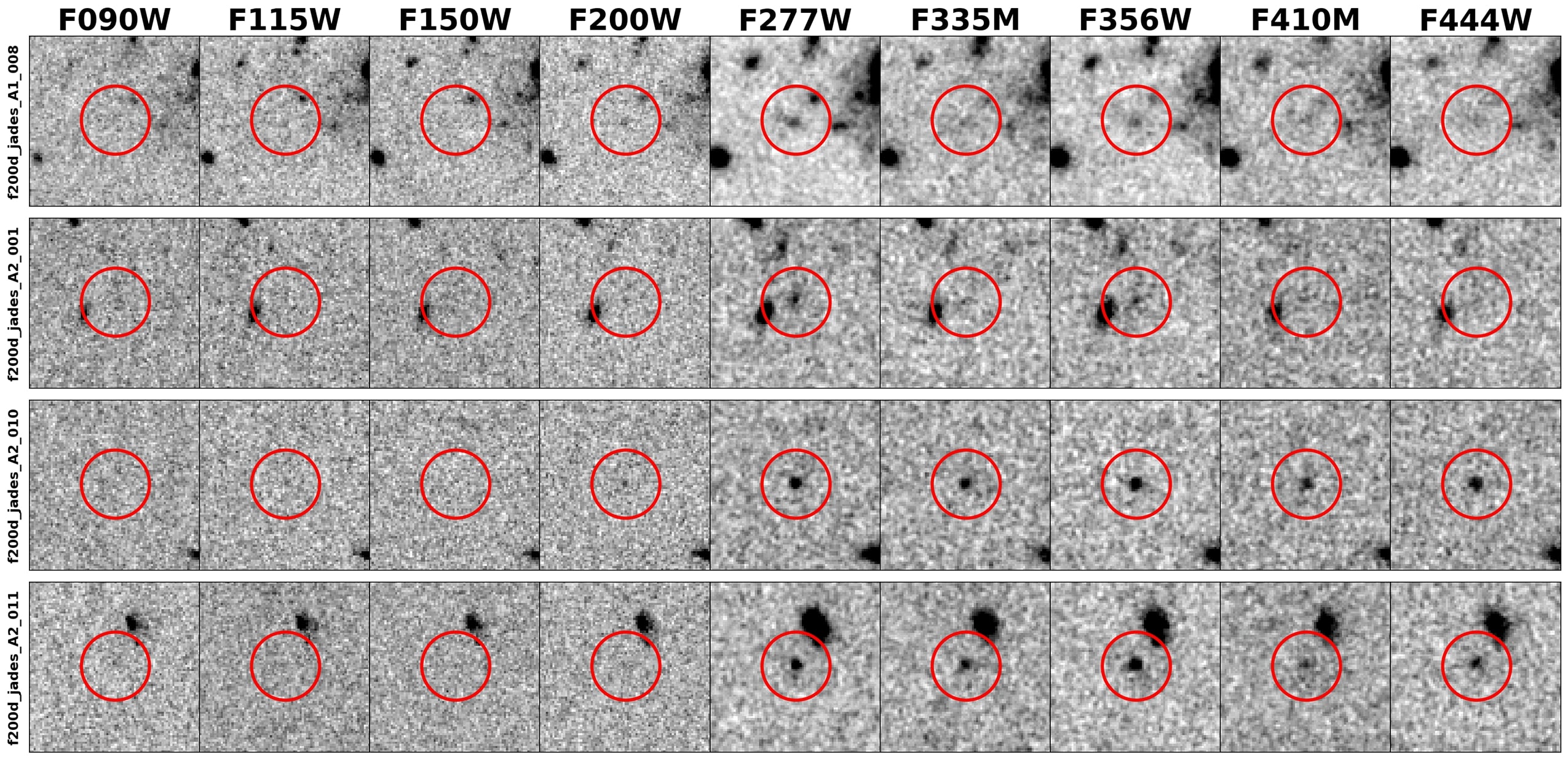}
    \includegraphics[width=0.65\textwidth]{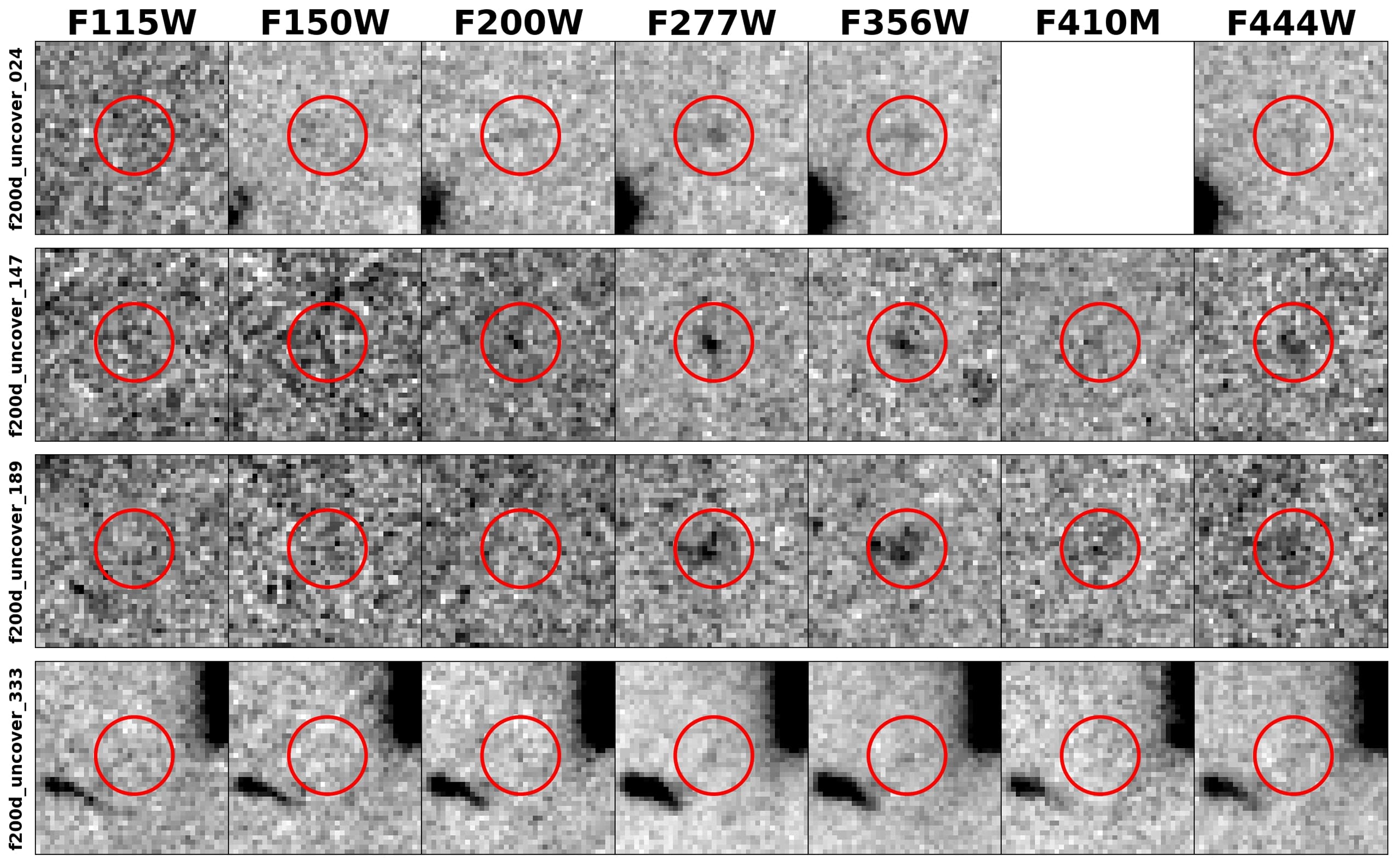}
    \includegraphics[width=0.65\textwidth]{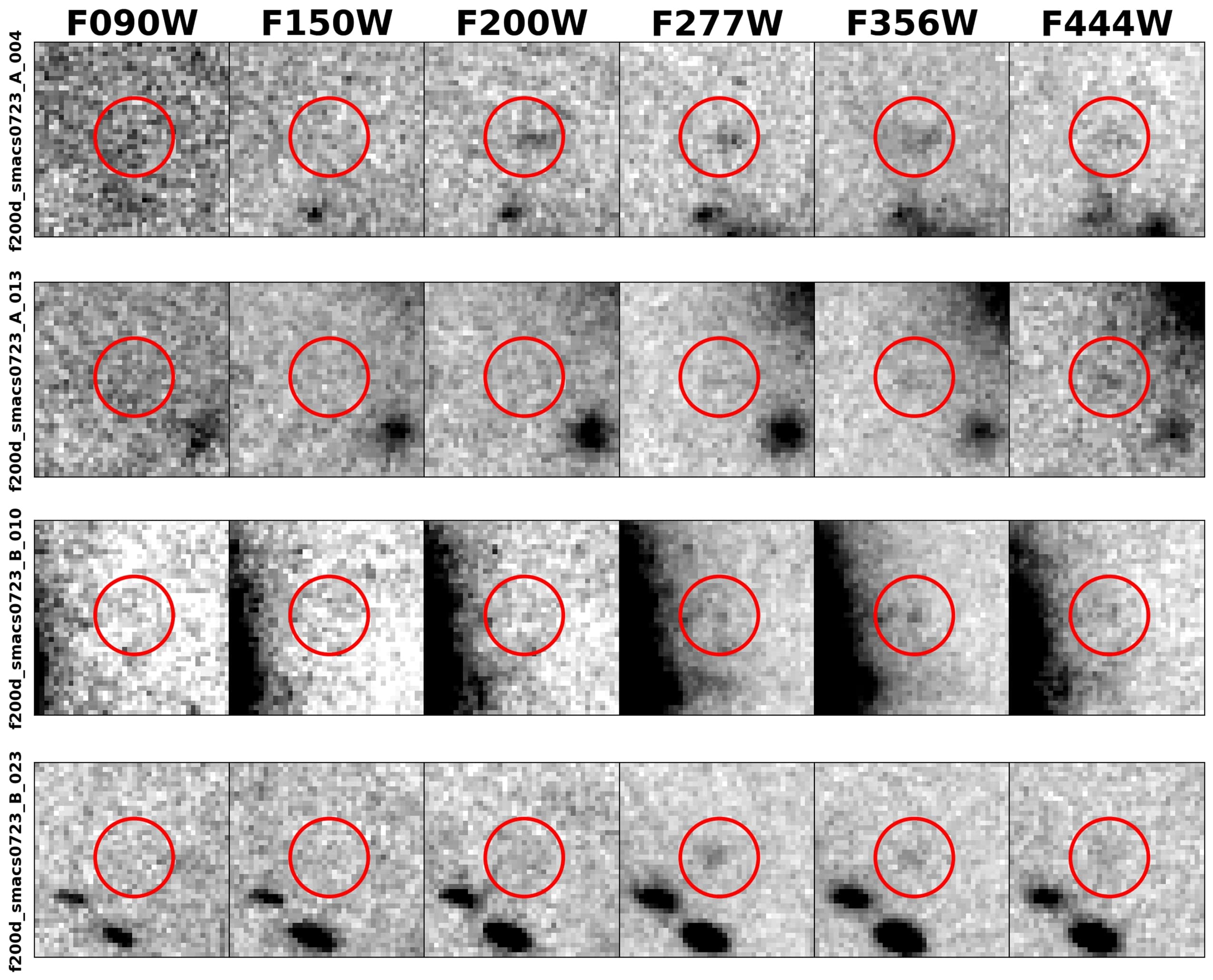}
    \caption{Similar to Figure~\ref{fig:dropoutdemo}; for the F200W dropouts 
     in JADES GOODS-S, UNCOVER and SMACS0723.
   }
    \label{fig:f200d_jades_uncover_smacs}
\end{figure*}

\begin{figure*}[htbp]
    \centering
    \includegraphics[width=0.65\textwidth]{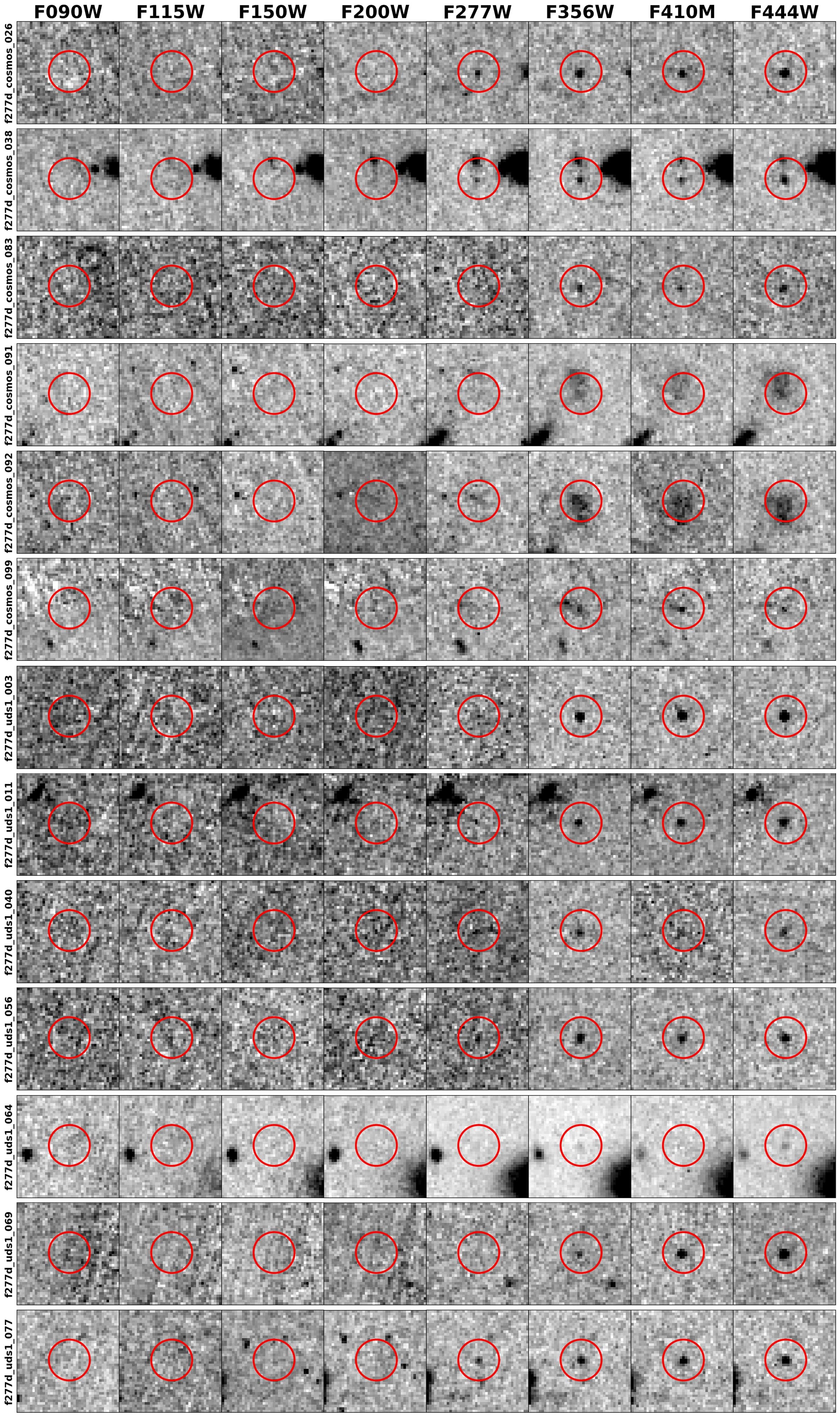}
    \caption{Similar to Figure~\ref{fig:dropoutdemo}; for the F277W dropouts
     in COSMOS and UDS1.
   }
    \label{fig:f277d_cosmos_uds1}
\end{figure*}

\begin{figure*}[htbp]
    \centering
    \includegraphics[width=\textwidth]{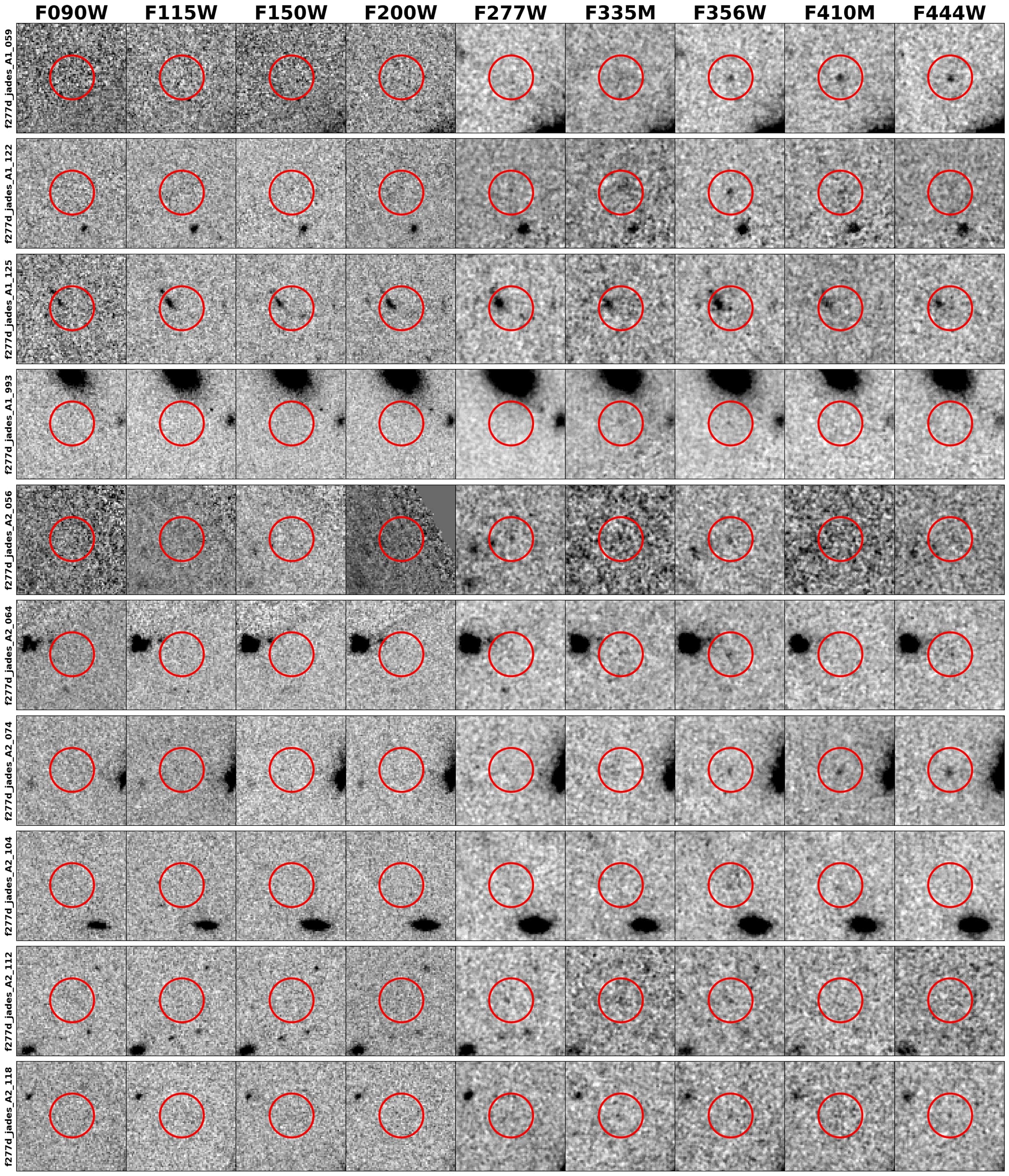}
    \caption{Similar to Figure~\ref{fig:dropoutdemo}; for the F277W dropouts
     in JADES GOODS-S.
   }
    \label{fig:f277d_jades}
\end{figure*}

\begin{figure*}[htbp]
    \centering
    \includegraphics[width=0.65\textwidth]{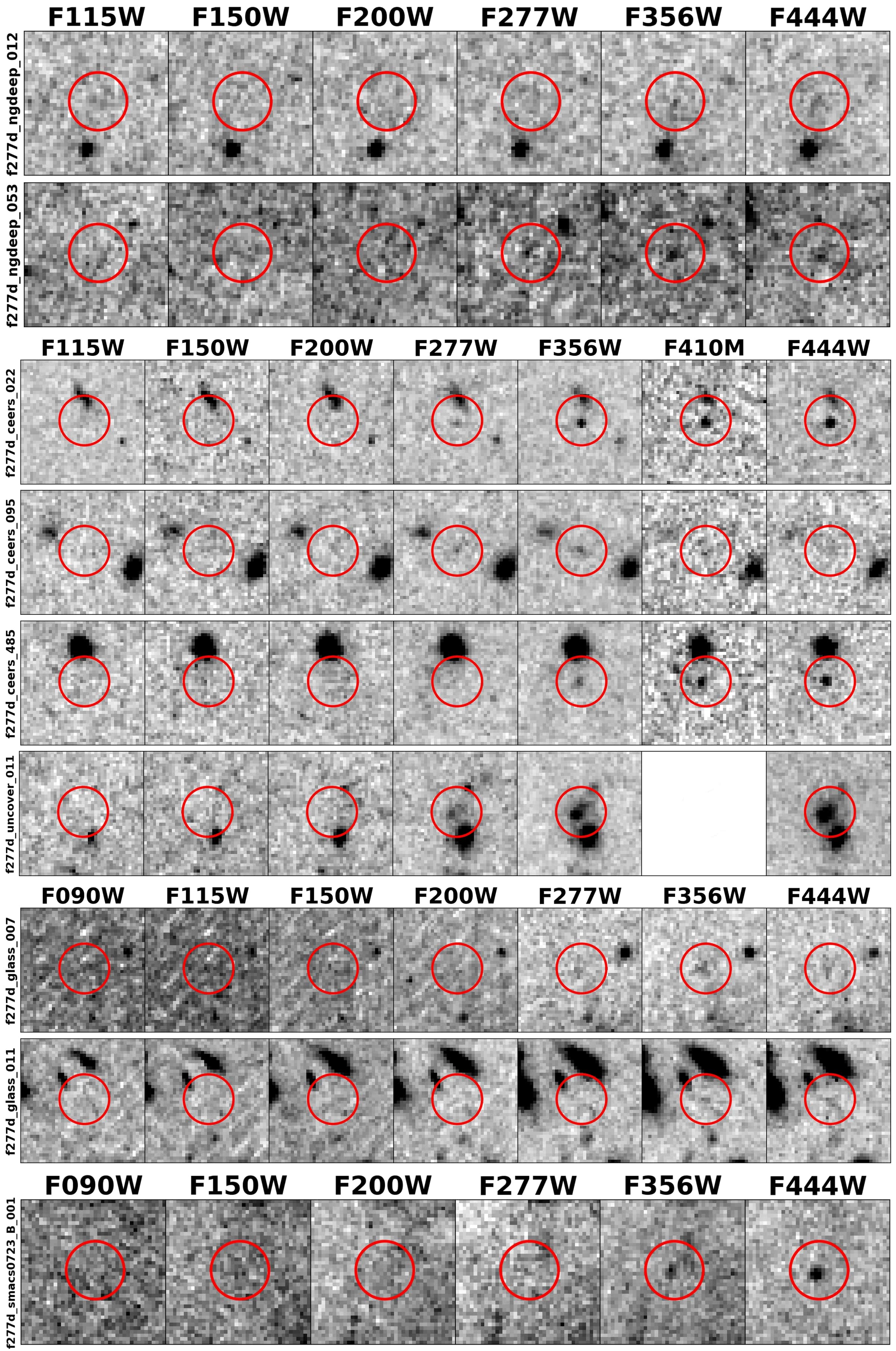}
    \caption{Similar to Figure~\ref{fig:dropoutdemo}; for the F277W dropouts
     in NGDEEP, CEERS, UNCOVER, GLASS, and SMACS0723.
   }
    \label{fig:f277d_others}
\end{figure*}


\begin{figure*}[htbp]
    \centering
    \includegraphics[width=0.85\textwidth]{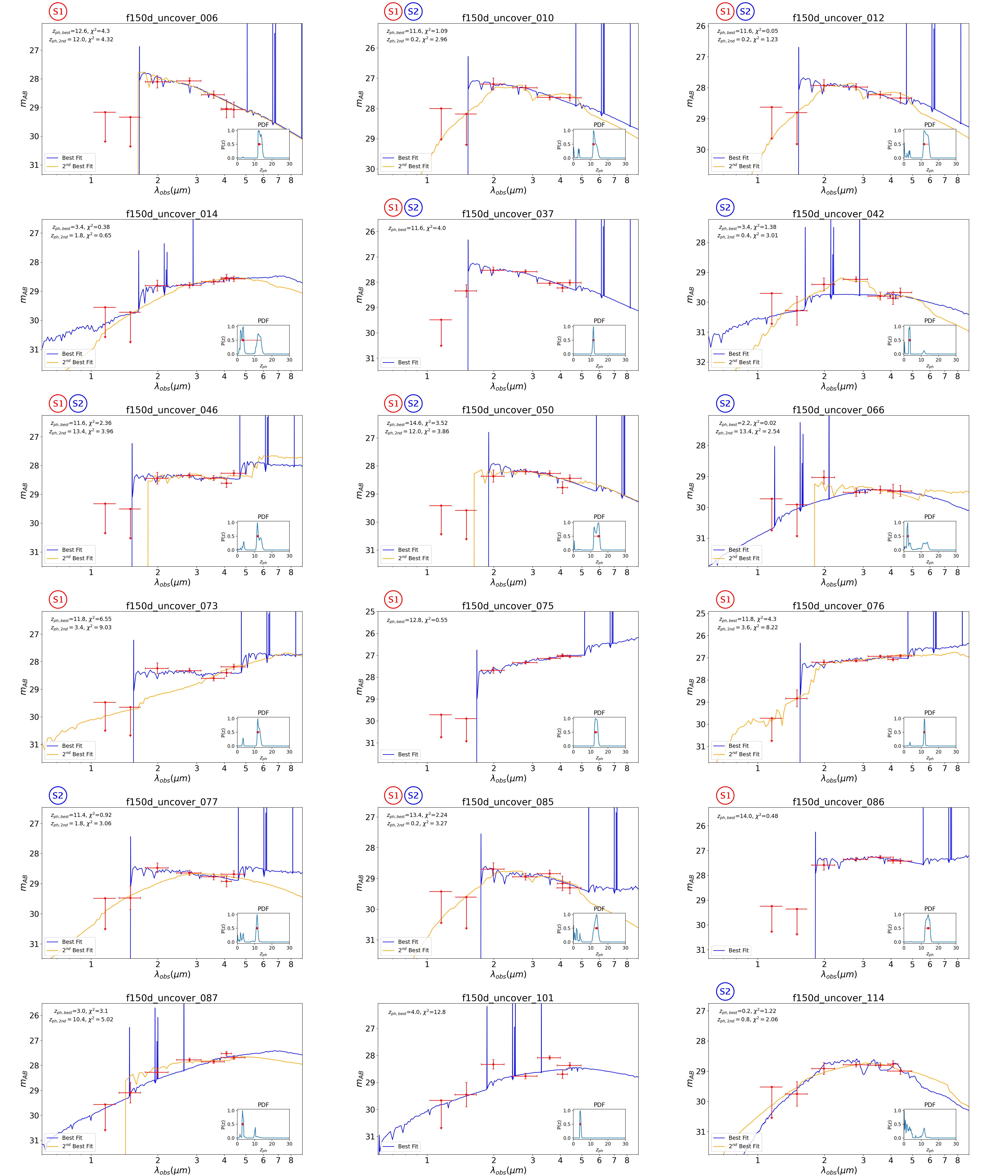}
    \caption{Scheme 1 (S1; using Le Phare and BC03 models) SED fitting results
    for F150W dropouts in UNCOVER. Each panel shows one F150W dropout, whose
    SID is labeled to left. The PDF is shown in the inset. The blue curve is
    the spectrum of the best-fit model corresponding to the first peak of the
    PDF, which also gives $z_{\rm ph}$ as labeled. If the secondary peak 
    exists, the corresponding model spectrum is also shown as the orange curve.
    If an object is retained in S1 and/or S2, the plot is marked with
    \textcircled{{\tiny S1}} and/or \textcircled{{\tiny S2}}.
   } 
    \label{fig:f150d_lp_uncover}
\end{figure*}

\begin{figure*}[htbp]
    \centering
    \includegraphics[width=0.85\textwidth]{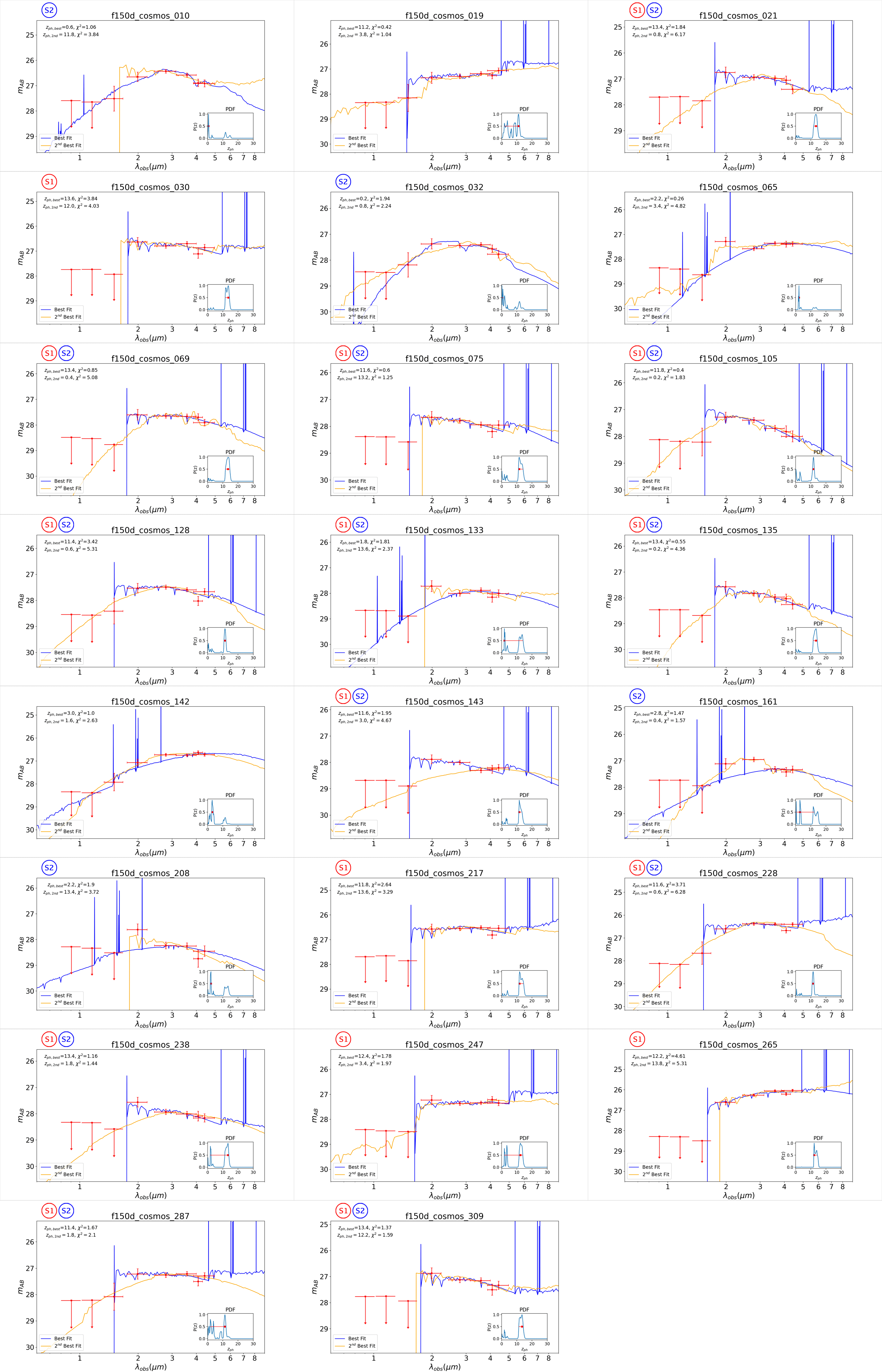}
    \caption{Same as Figure~\ref{fig:f150d_lp_uncover}; for the F150W
    dropouts in COSMOS.
   } 
    \label{fig:f150d_lp_cosmos}
\end{figure*}

\begin{figure*}[htbp]
    \centering
    \includegraphics[width=0.85\textwidth]{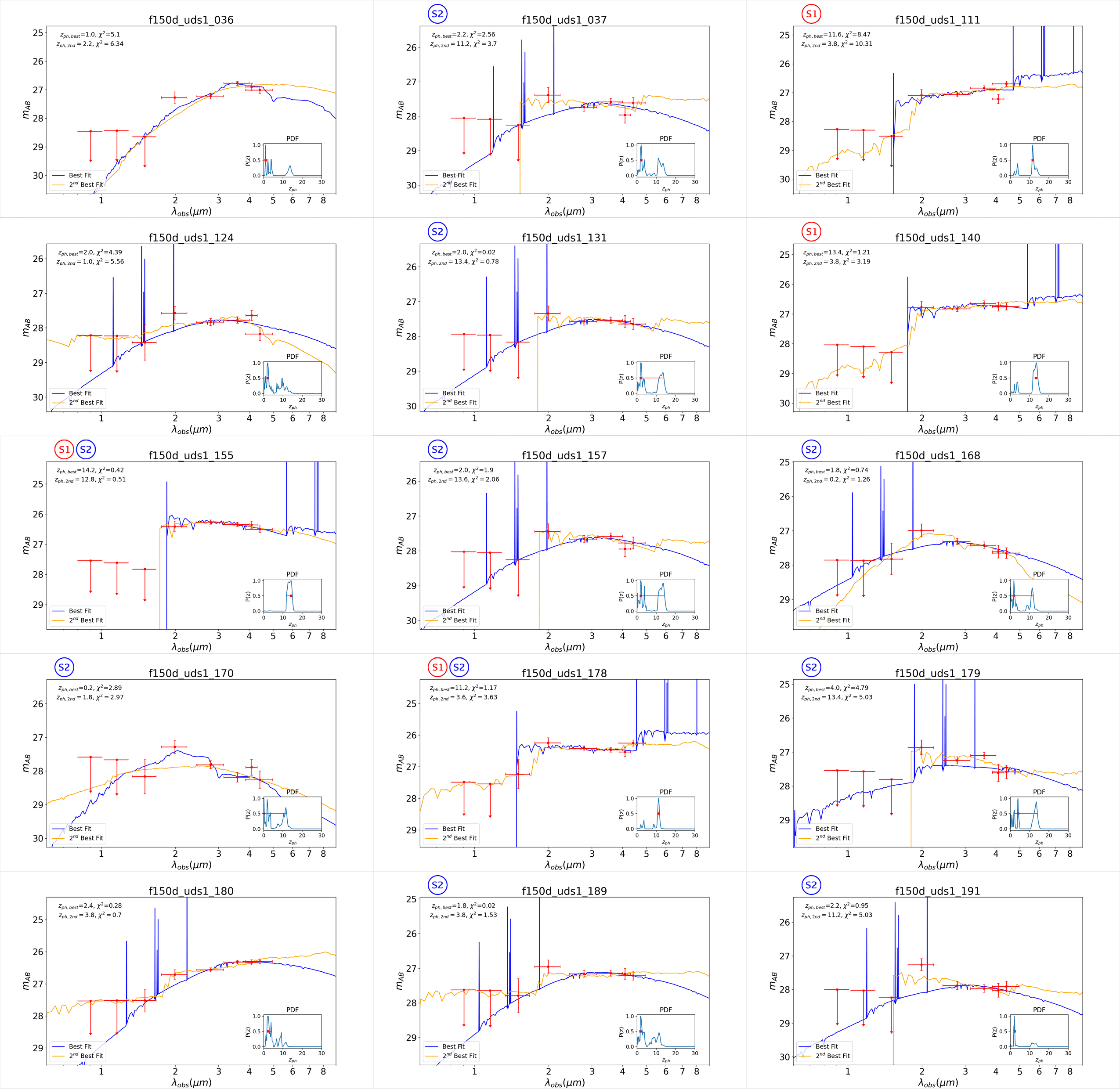}
    \includegraphics[width=0.85\textwidth]{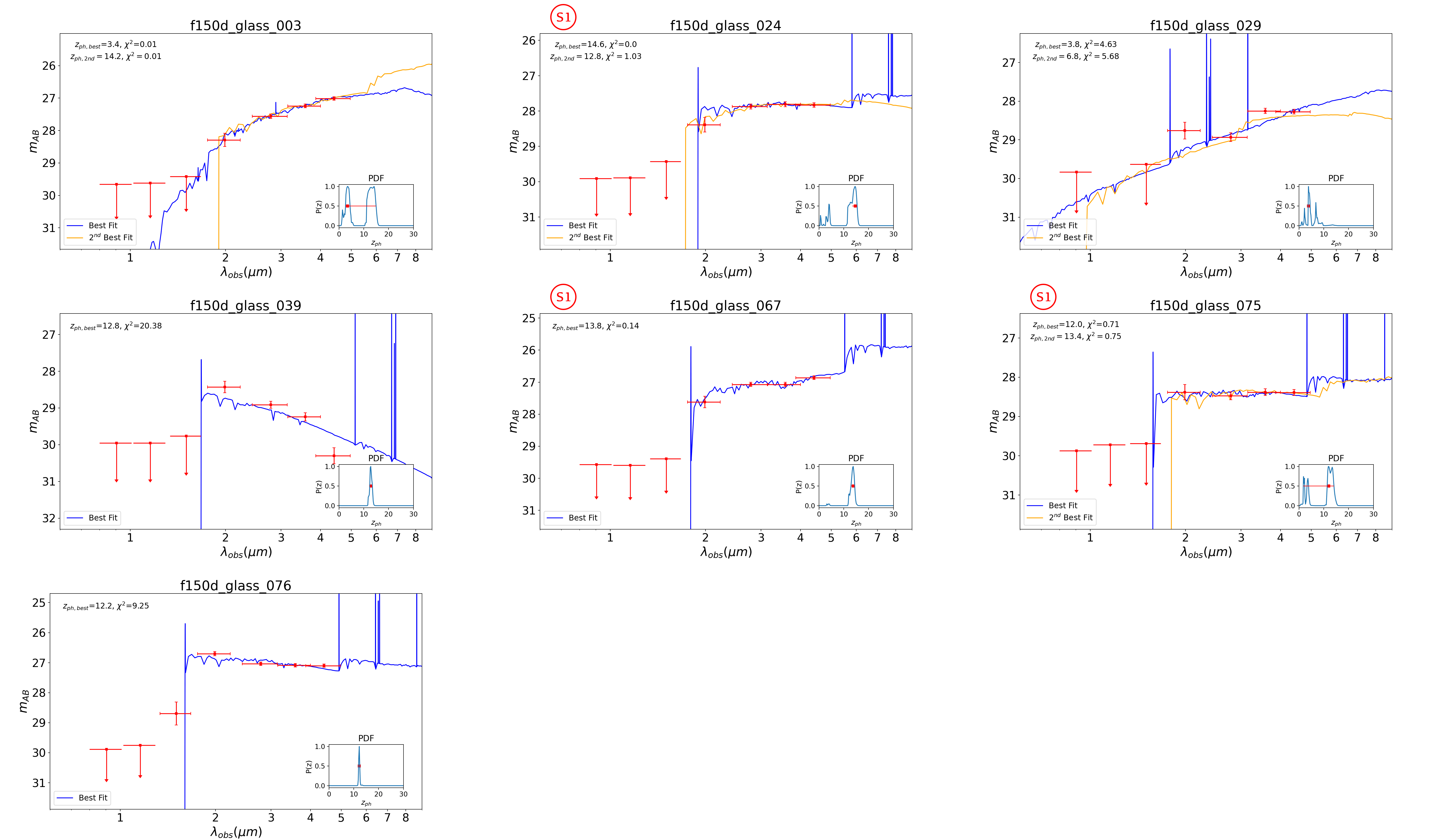}
    \caption{Same as Figure~\ref{fig:f150d_lp_uncover}; for the F150W
    dropouts in UDS1 and GLASS.
   } 
    \label{fig:f150d_lp_uds1_glass}
\end{figure*}

\begin{figure*}[htbp]
    \centering
    \includegraphics[width=0.85\textwidth]{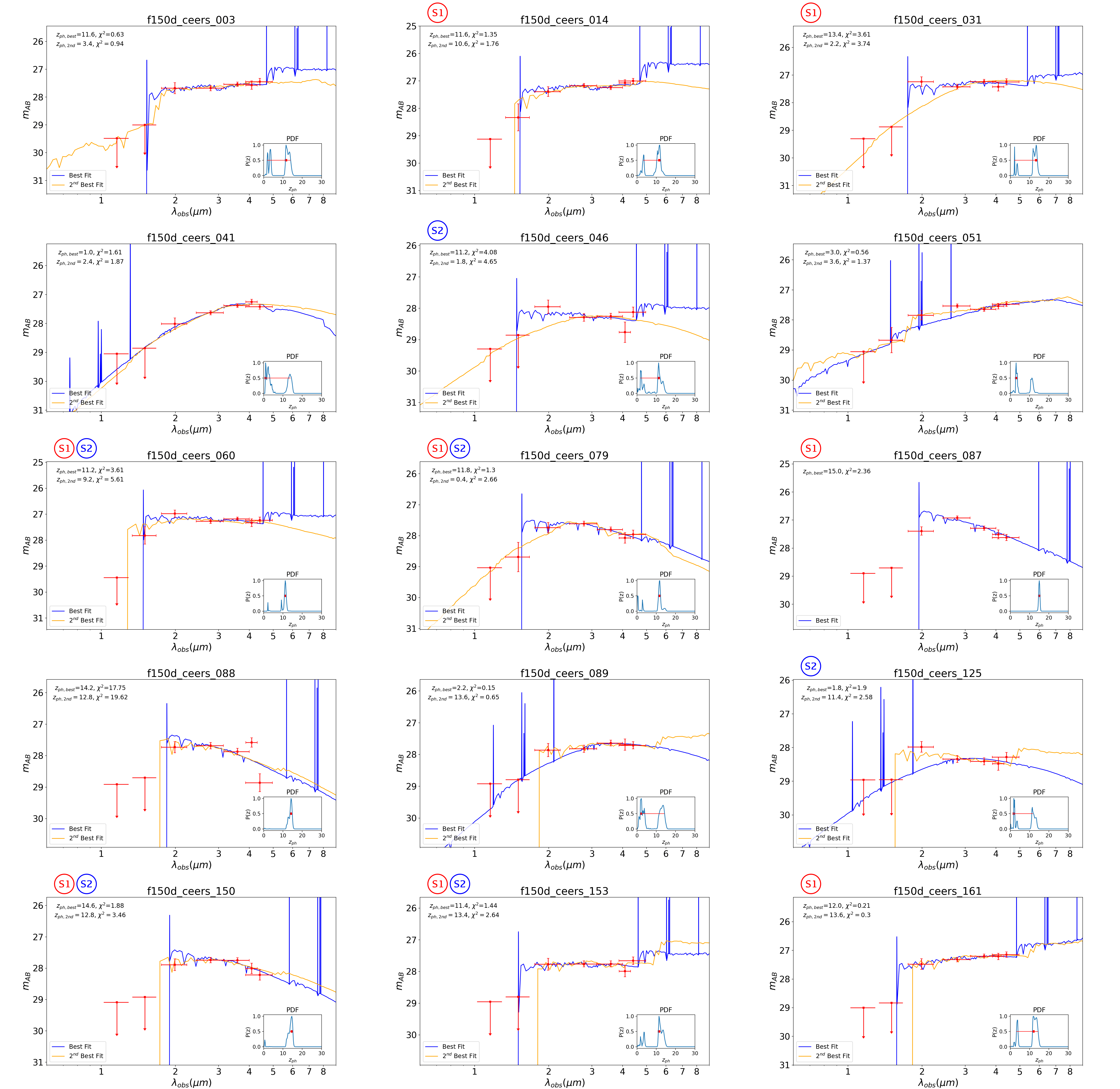}
    \includegraphics[width=0.85\textwidth]{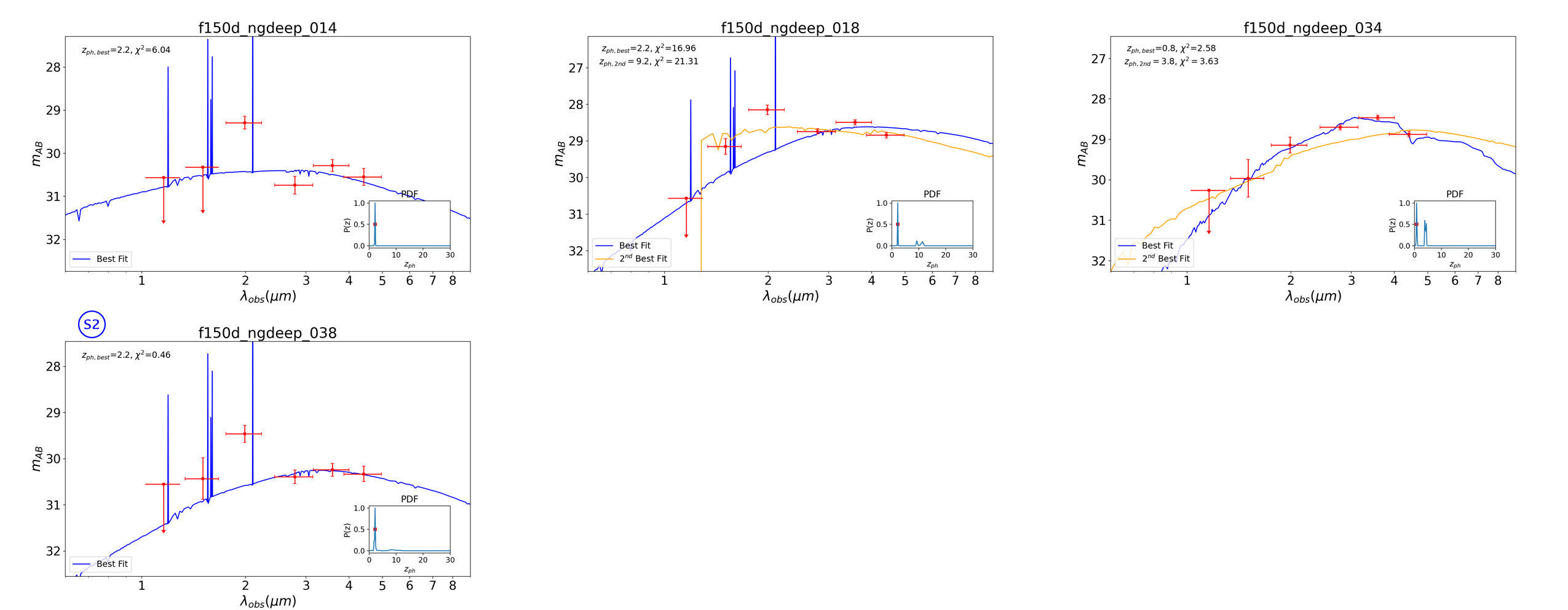}
    \caption{Same as Figure~\ref{fig:f150d_lp_uncover}; for the F150W
    dropouts in CEERS and NGDEEP.
   } 
    \label{fig:f150d_lp_ceers_ngdeep}
\end{figure*}

\begin{figure*}[htbp]
    \centering
    \includegraphics[width=0.85\textwidth]{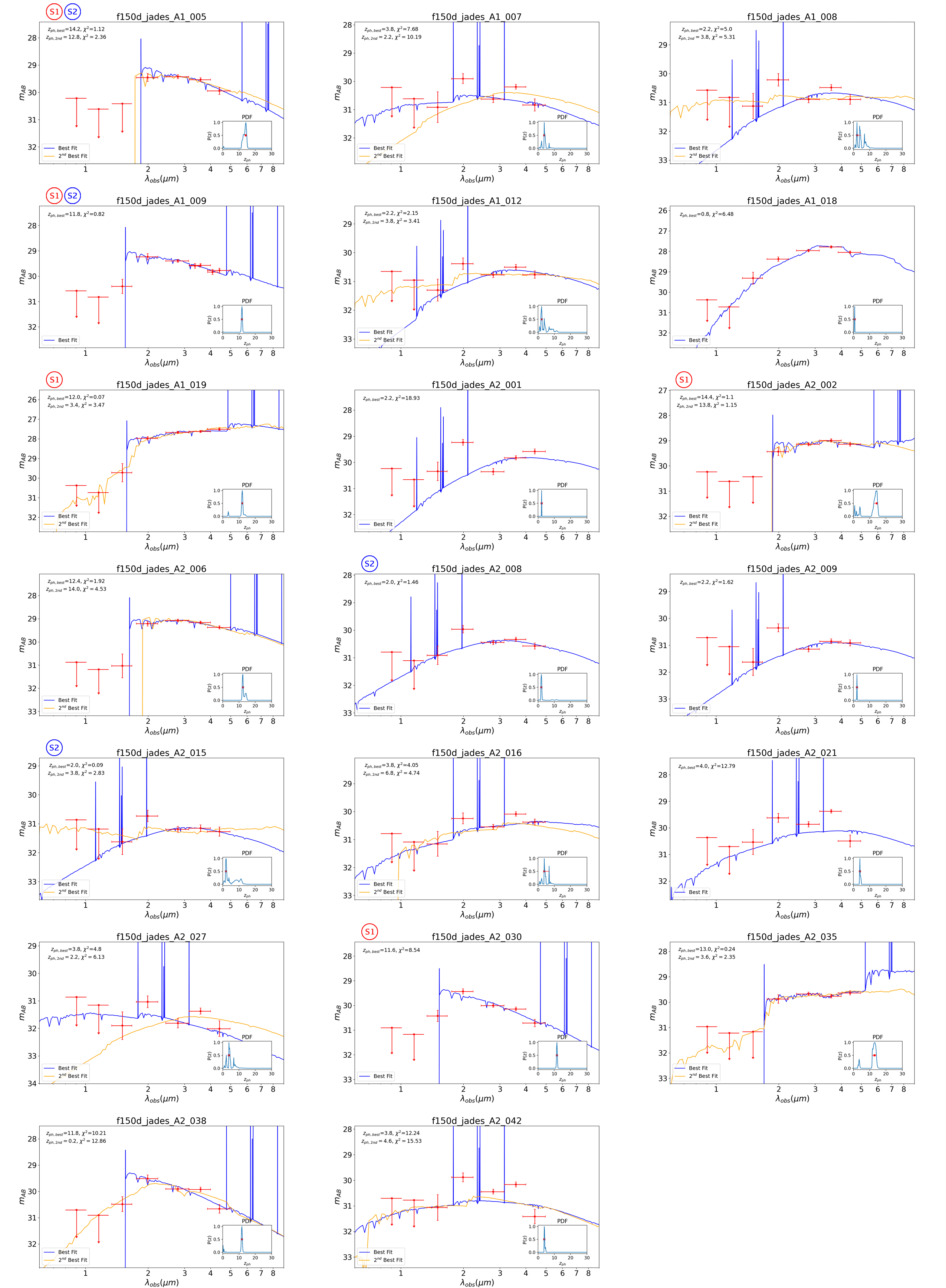}
    \caption{Same as Figure~\ref{fig:f150d_lp_uncover}; for the F150W
    dropouts in JADES GOODS-S.
   } 
    \label{fig:f150d_lp_jades}
\end{figure*}

\begin{figure*}[htbp]
    \centering
    \includegraphics[width=0.85\textwidth]{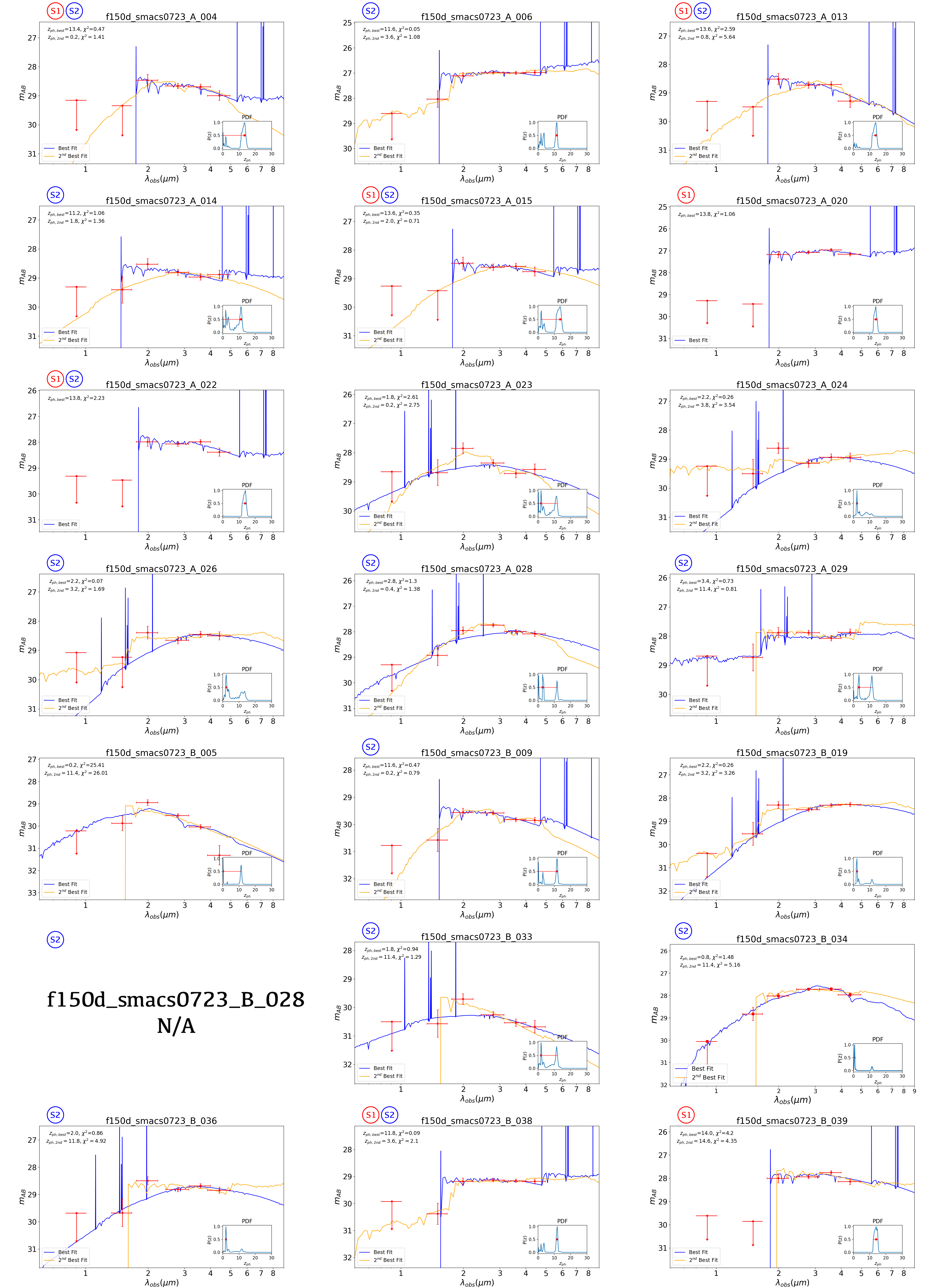}
    \caption{Same as Figure~\ref{fig:f150d_lp_uncover}; for the F150W
    dropouts in SMACS0723. The fitting fails for 
    \texttt{f150d\_smacs0723\_B\_028}, which is not shown.
   } 
    \label{fig:f150d_lp_smacs}
\end{figure*}


\begin{figure*}[htbp]
    \centering
    \includegraphics[width=0.85\textwidth]{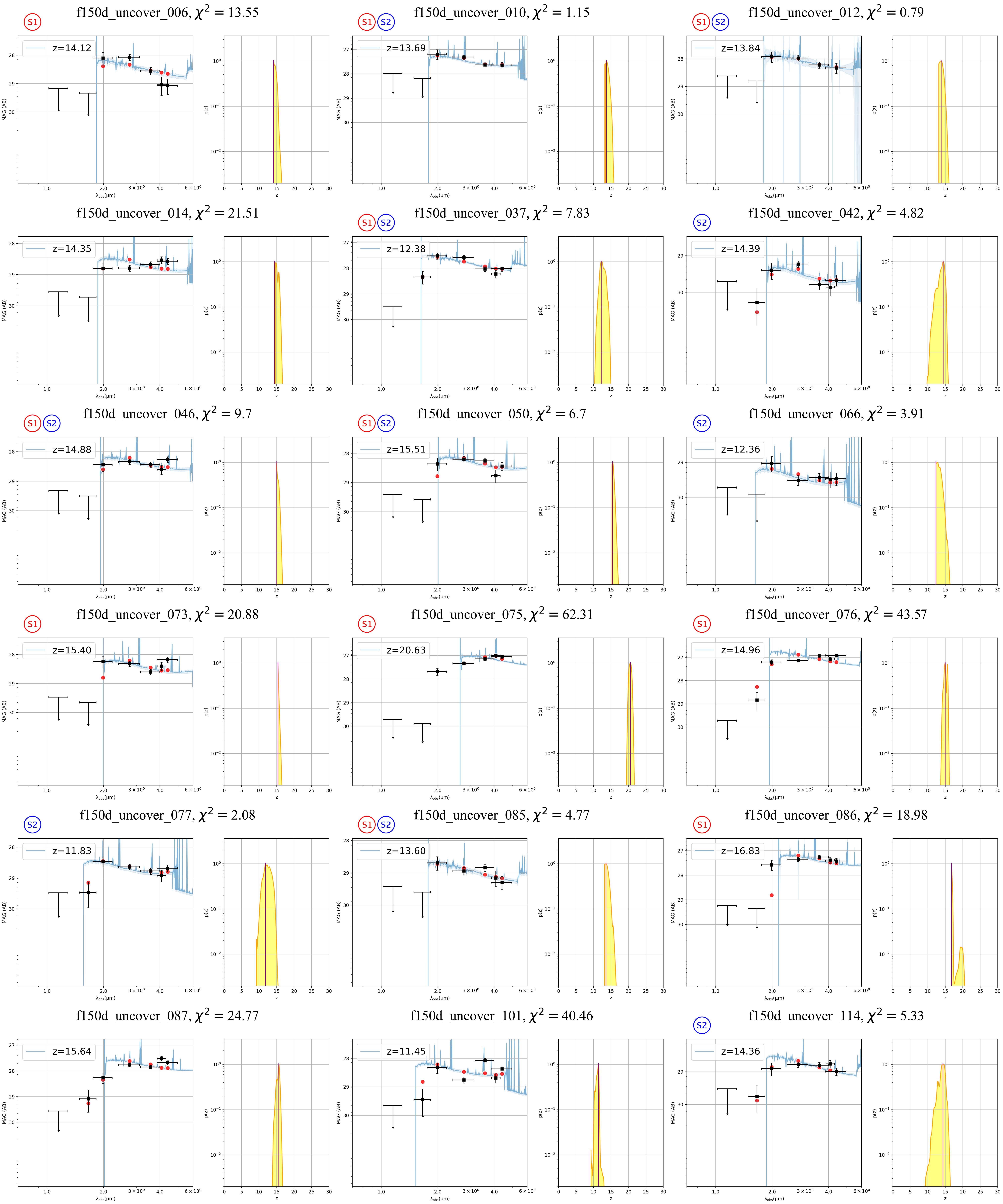}
    \caption{Scheme 2 (S2; using EAZY and the ``set 3+4'' templates ) SED 
    fitting results
    for F150W dropouts in UNCOVER. Each panel shows one F150W dropout, whose
    SID is labeled to left. The PDF is shown in the smaller panel to right. 
    The dark blue curve is
    the spectrum of the best-fit model corresponding to the first peak of the
    PDF, which also gives $z_{\rm ph}$ as labeled. 
    If an object is retained in S1 and/or S2, the plot is marked with
    \textcircled{{\tiny S1}} and/or \textcircled{{\tiny S2}}.
   } 
    \label{fig:f150d_ez_uncover}
\end{figure*}

\begin{figure*}[htbp]
    \centering
    \includegraphics[width=0.85\textwidth]{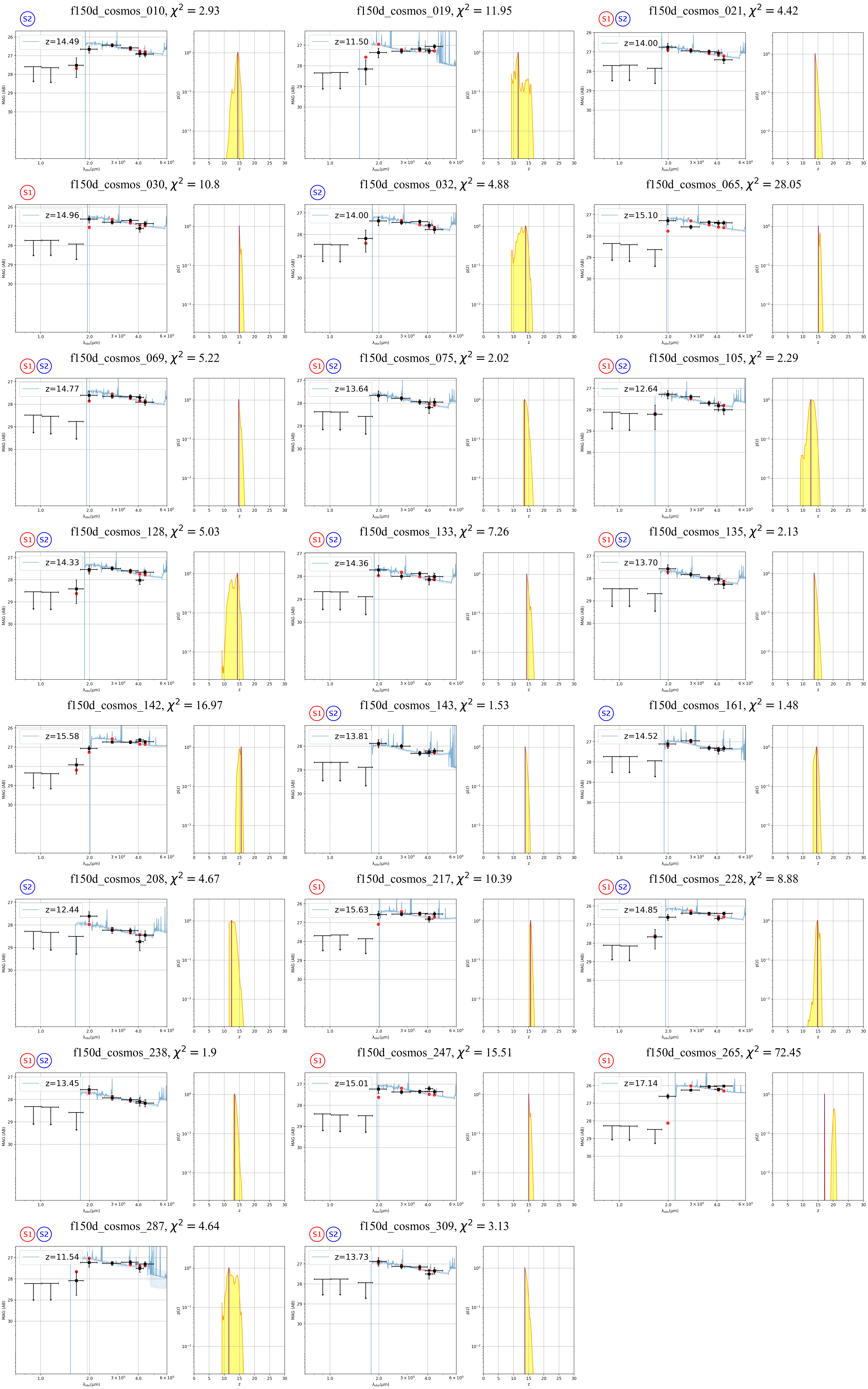}
    \caption{Same as Figure~\ref{fig:f150d_ez_uncover}; for the F150W
    dropouts in COSMOS.
   } 
    \label{fig:f150d_ez_cosmos}
\end{figure*}

\begin{figure*}[htbp]
    \centering
    \includegraphics[width=0.85\textwidth]{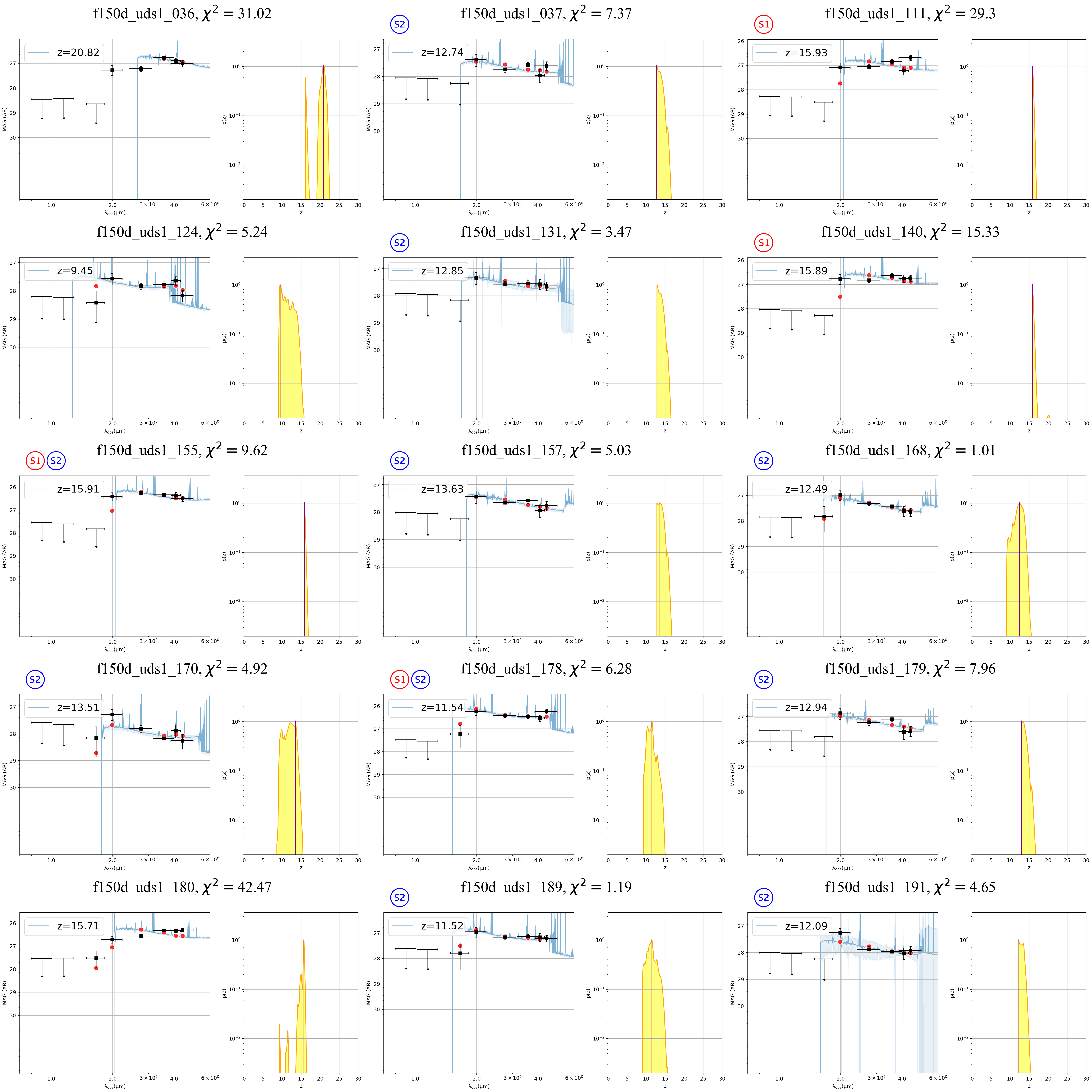}
    \includegraphics[width=0.85\textwidth]{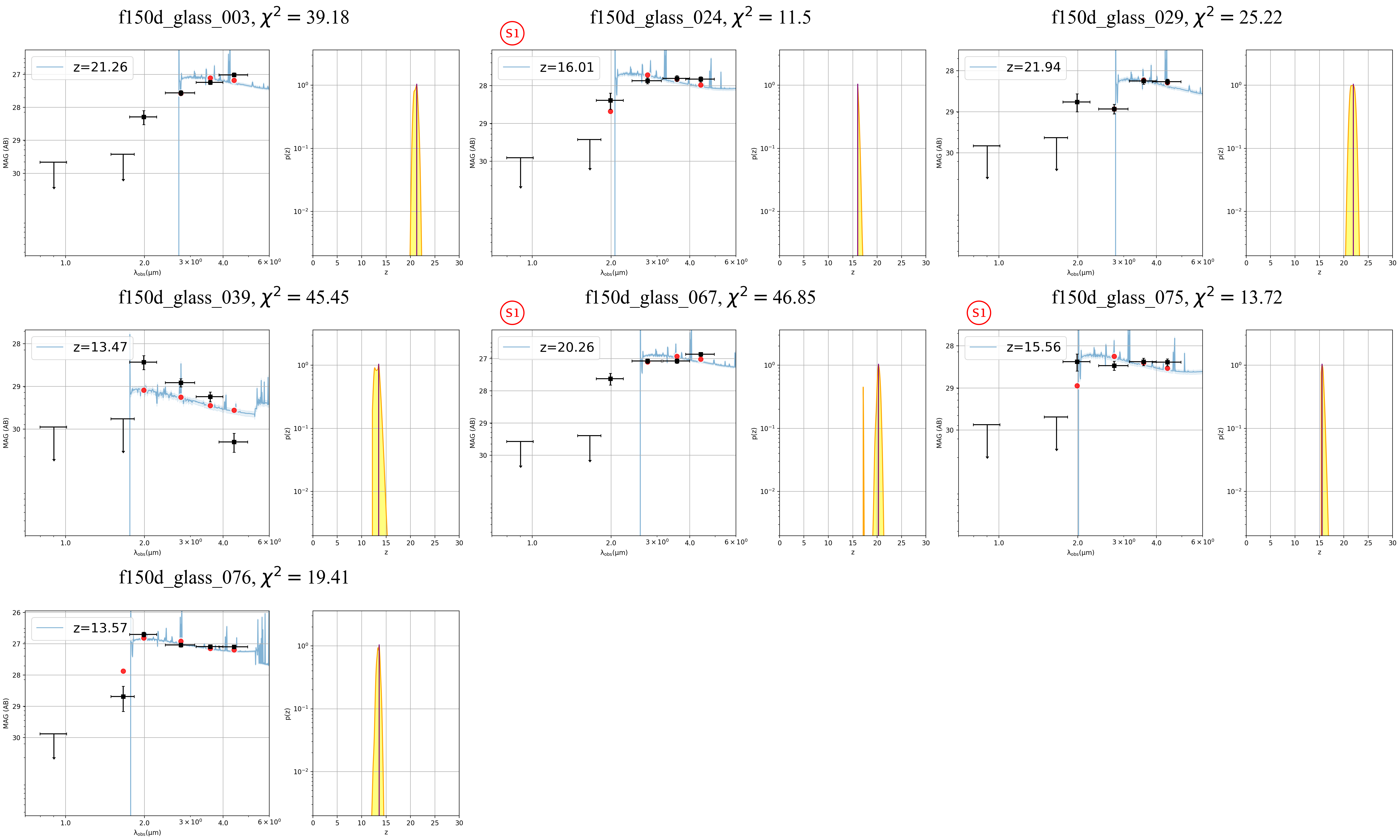}
    \caption{Same as Figure~\ref{fig:f150d_ez_uncover}; for the F150W
    dropouts in UDS1 and GLASS.
   } 
    \label{fig:f150d_ez_uds1_glass}
\end{figure*}

\begin{figure*}[htbp]
    \centering
    \includegraphics[width=0.85\textwidth]{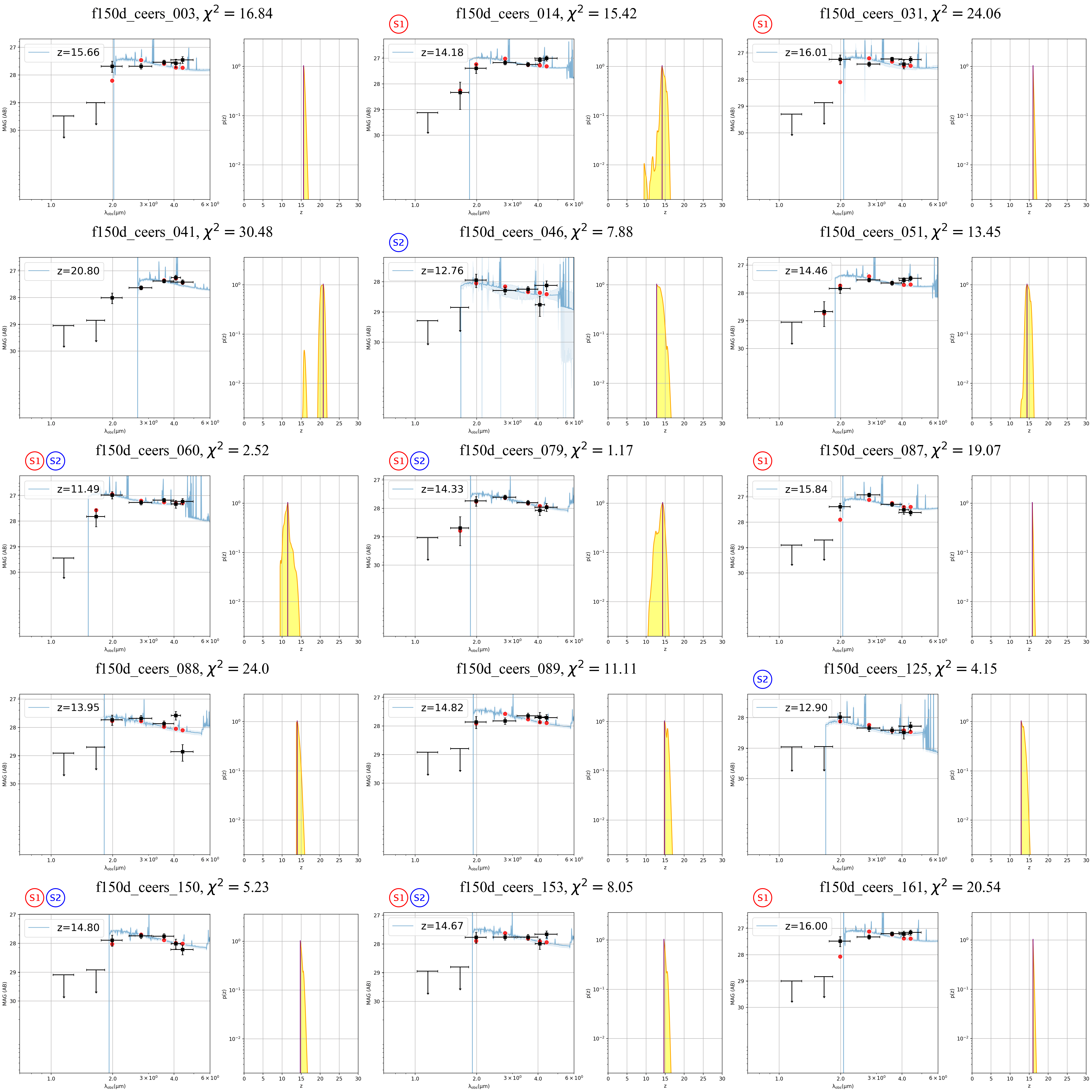}
    \includegraphics[width=0.85\textwidth]{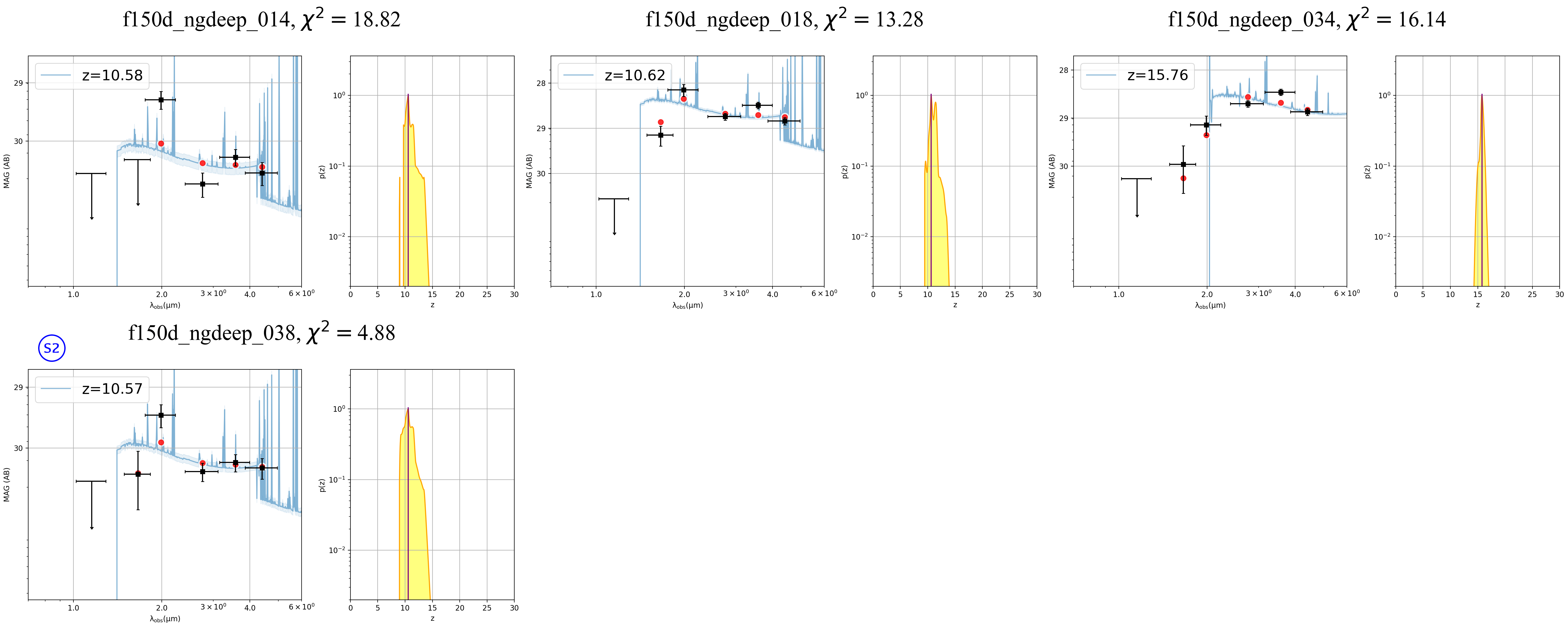}
    \caption{Same as Figure~\ref{fig:f150d_ez_uncover}; for the F150W
    dropouts in CEERS and NGDEEP.
   } 
    \label{fig:f150d_ez_ceers_ngdeep}
\end{figure*}

\begin{figure*}[htbp]
    \centering
    \includegraphics[width=0.85\textwidth]{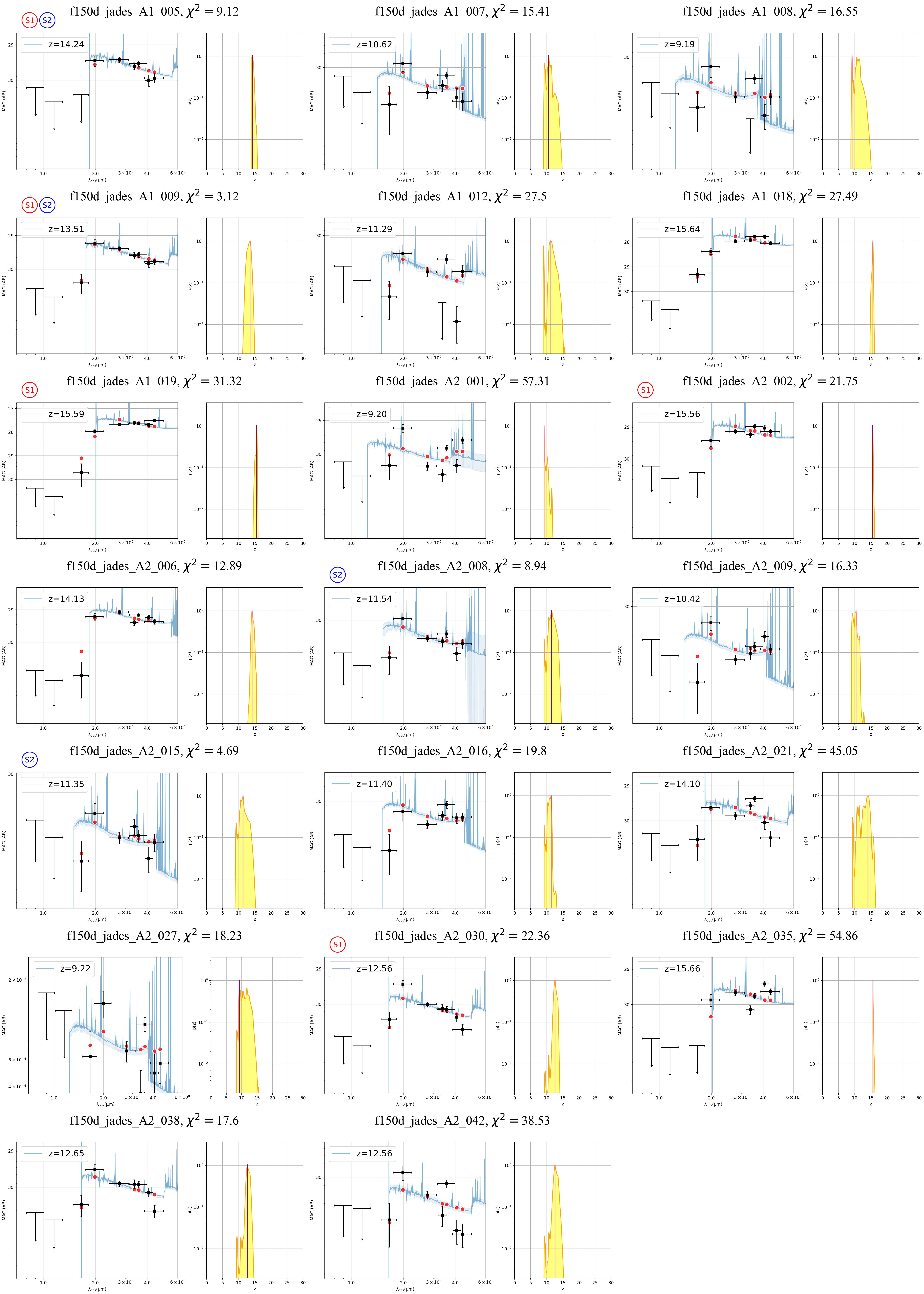}
    \caption{Same as Figure~\ref{fig:f150d_ez_uncover}; for the F150W
    dropouts in JADES GOODS-S.
   } 
    \label{fig:f150d_ez_jades}
\end{figure*}

\begin{figure*}[htbp]
    \centering
    \includegraphics[width=0.85\textwidth]{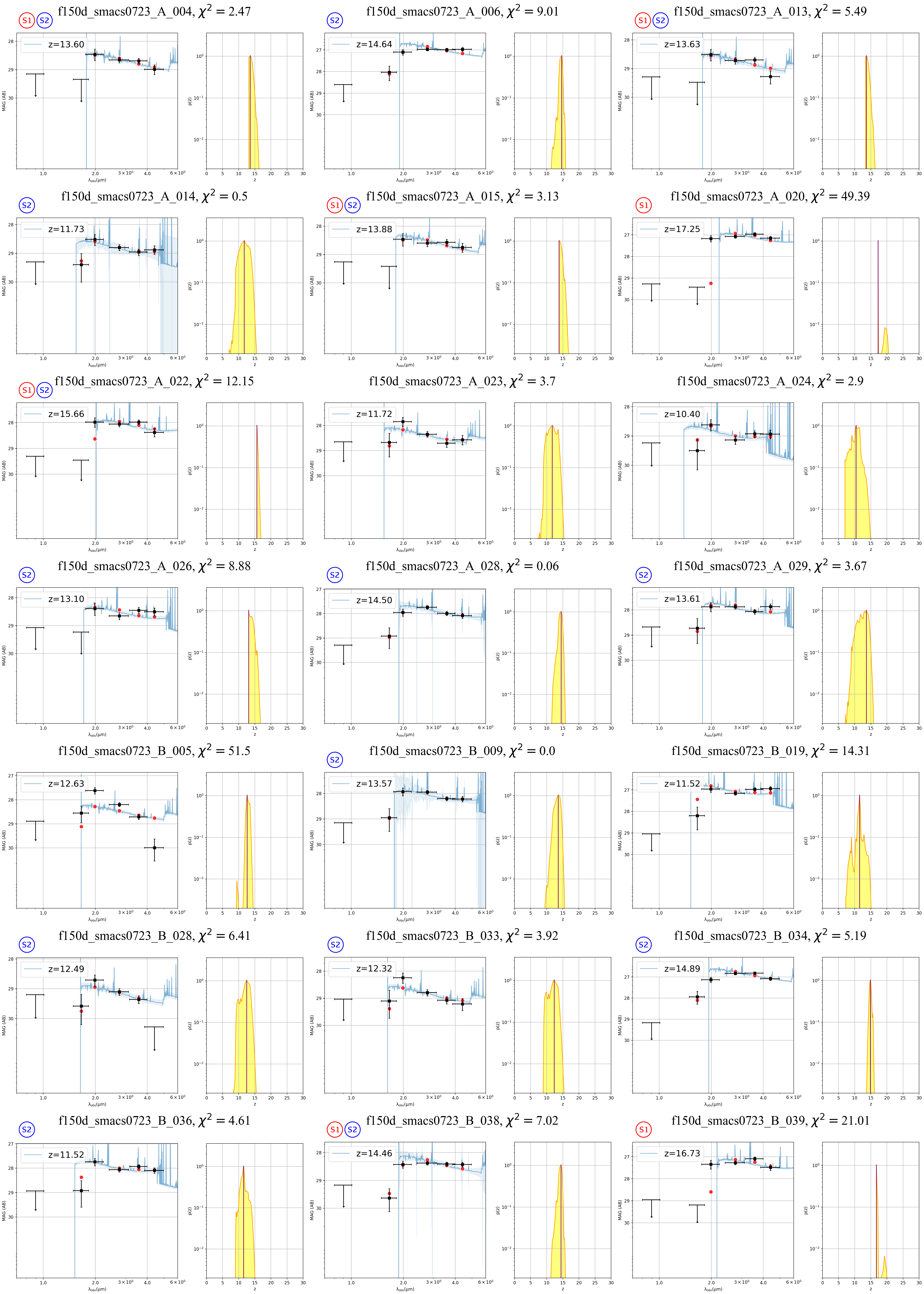}
    \caption{Same as Figure~\ref{fig:f150d_ez_uncover}; for the F150W
    dropouts in SMACS0723. 
   } 
    \label{fig:f150d_ez_smacs}
\end{figure*}


\begin{figure*}[htbp]
    \centering
    \includegraphics[width=0.85\textwidth]{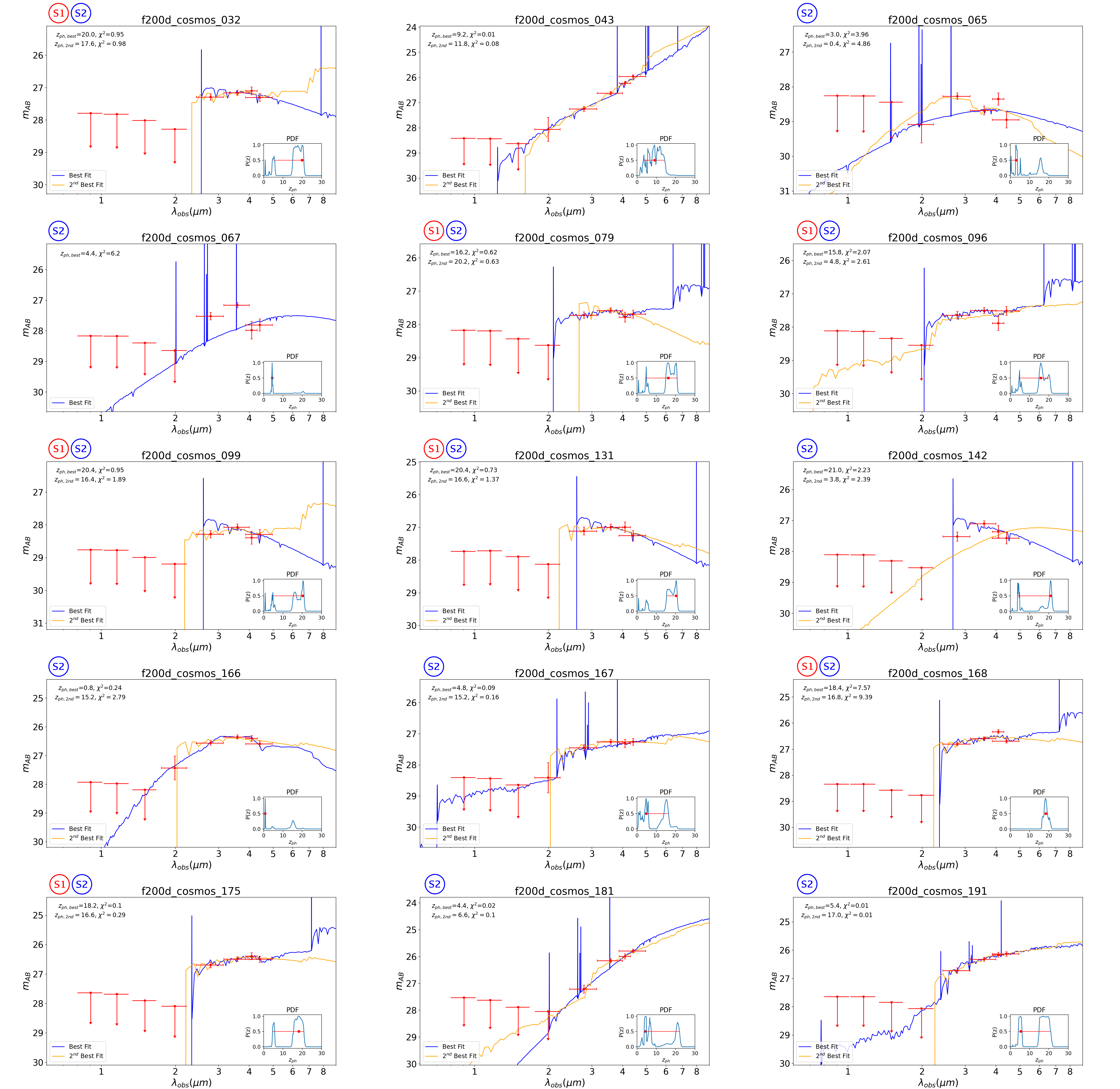}
    \includegraphics[width=0.85\textwidth]{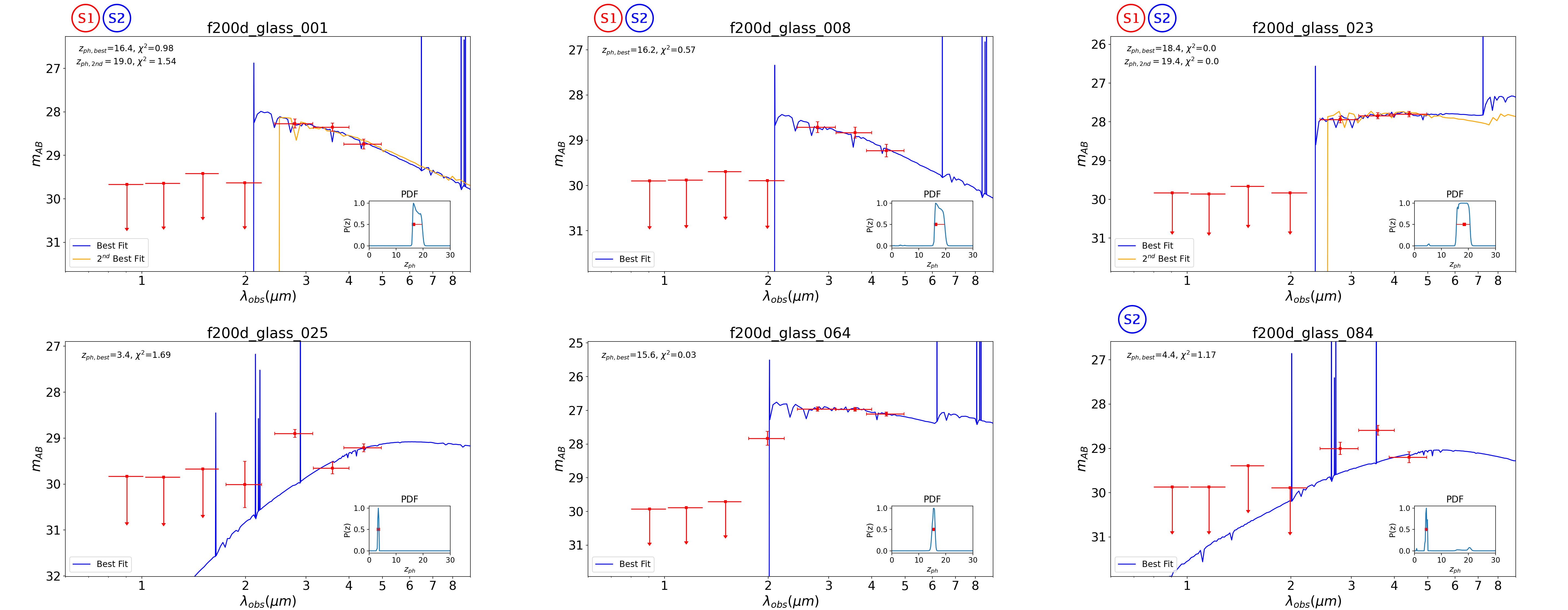}
    \caption{Same as Figure~\ref{fig:f150d_lp_uncover}; for the F200W
    dropouts in COSMOS and GLASS.
   }
    \label{fig:f200d_lp_cosmos_glass}
\end{figure*}

\begin{figure*}[htbp]
    \centering
    \includegraphics[width=0.85\textwidth]{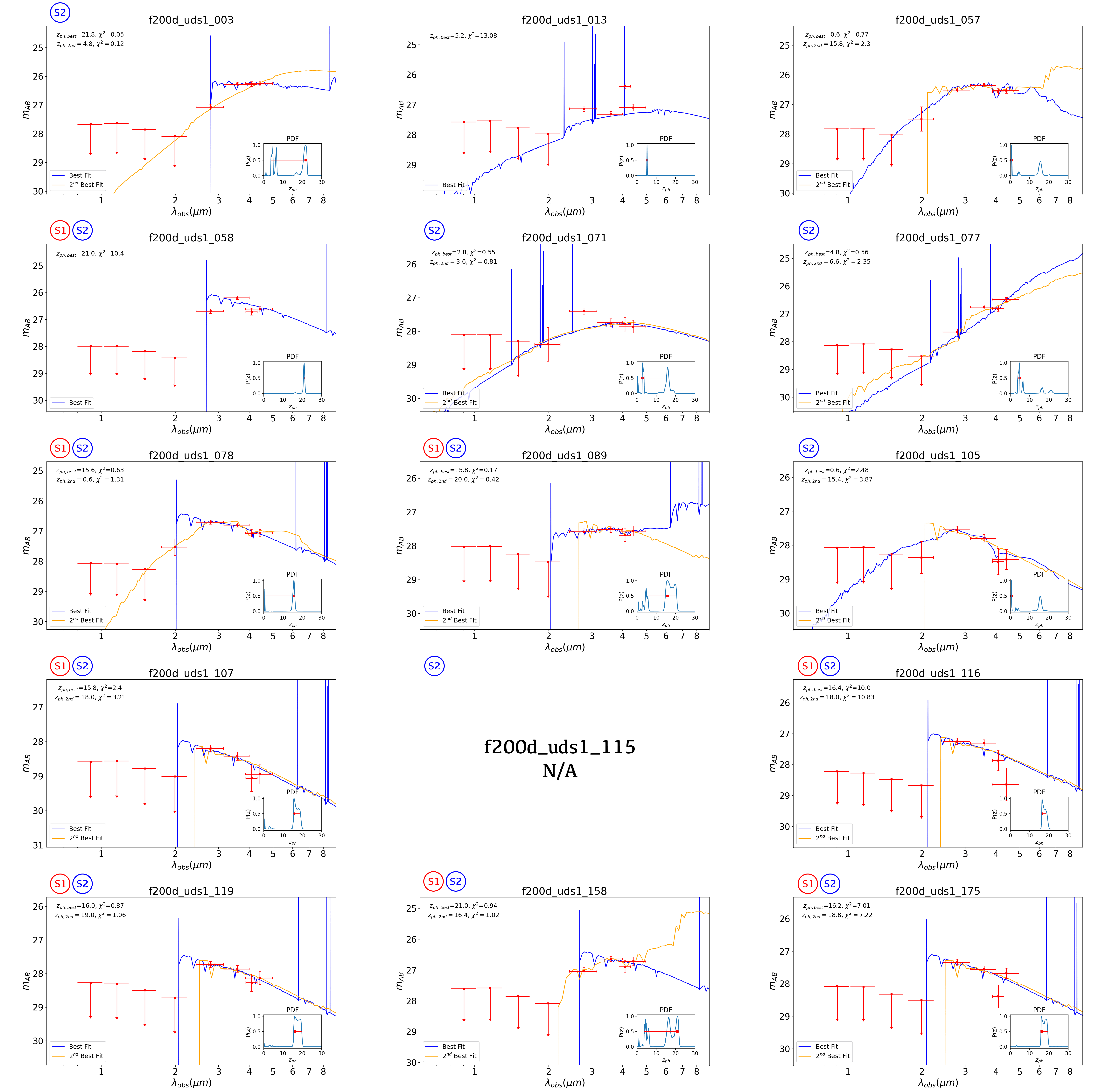}
    \includegraphics[width=0.85\textwidth]{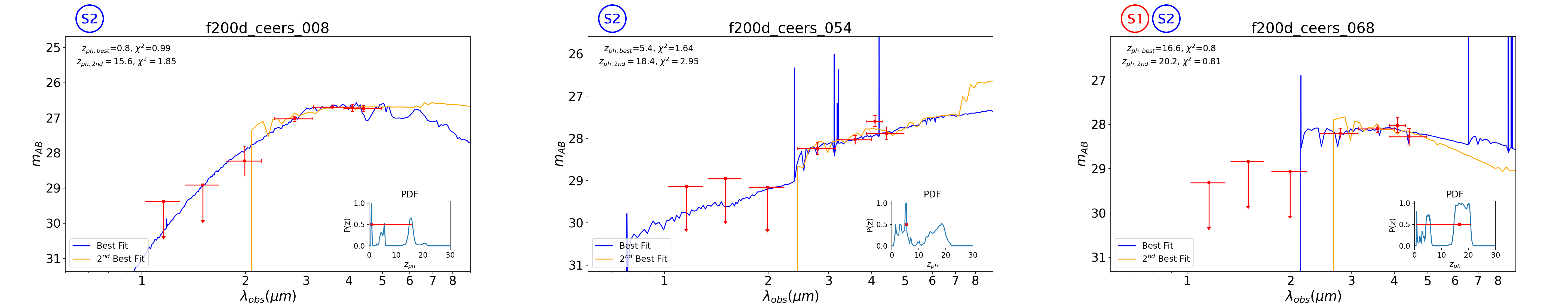}
    \includegraphics[width=0.28\textwidth]{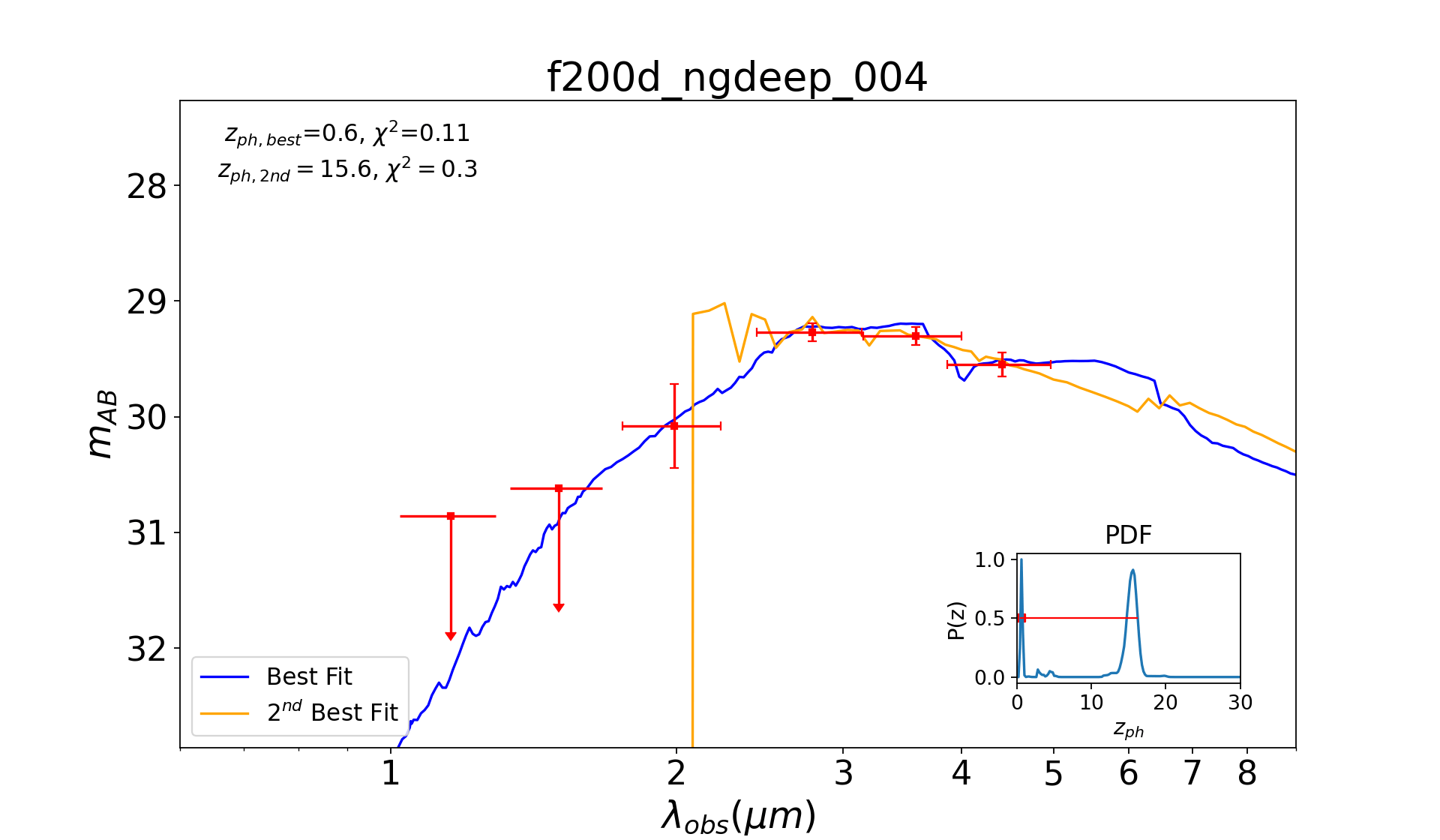}
    \caption{Same as Figure~\ref{fig:f150d_lp_uncover}; for the F200W
    dropouts in UDS1 (\texttt{f200d\_uds1\_115} not included due to the failed
    fit), CEERS and NGDEEP.
   }
    \label{fig:f200d_lp_uds1_ceers_ngdeep}
\end{figure*}

\begin{figure*}[htbp]
    \centering
    \includegraphics[width=0.85\textwidth]{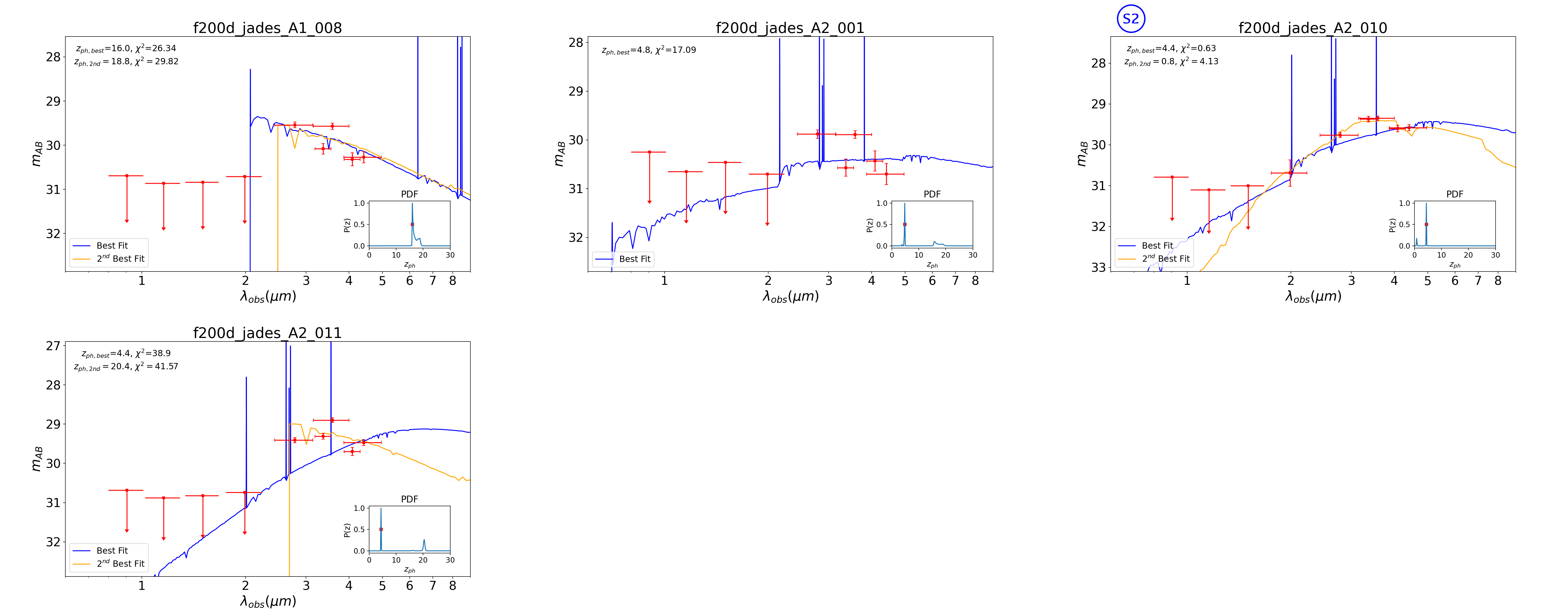}
    \includegraphics[width=0.85\textwidth]{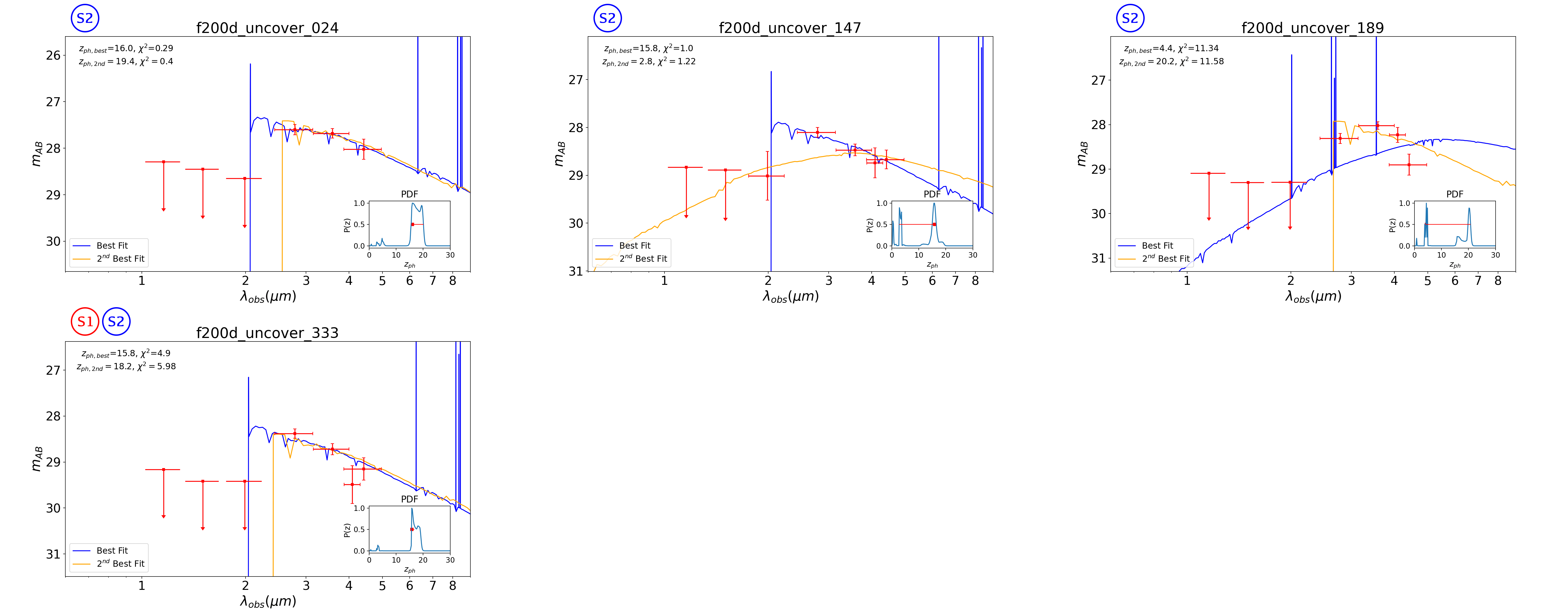}
    \includegraphics[width=0.85\textwidth]{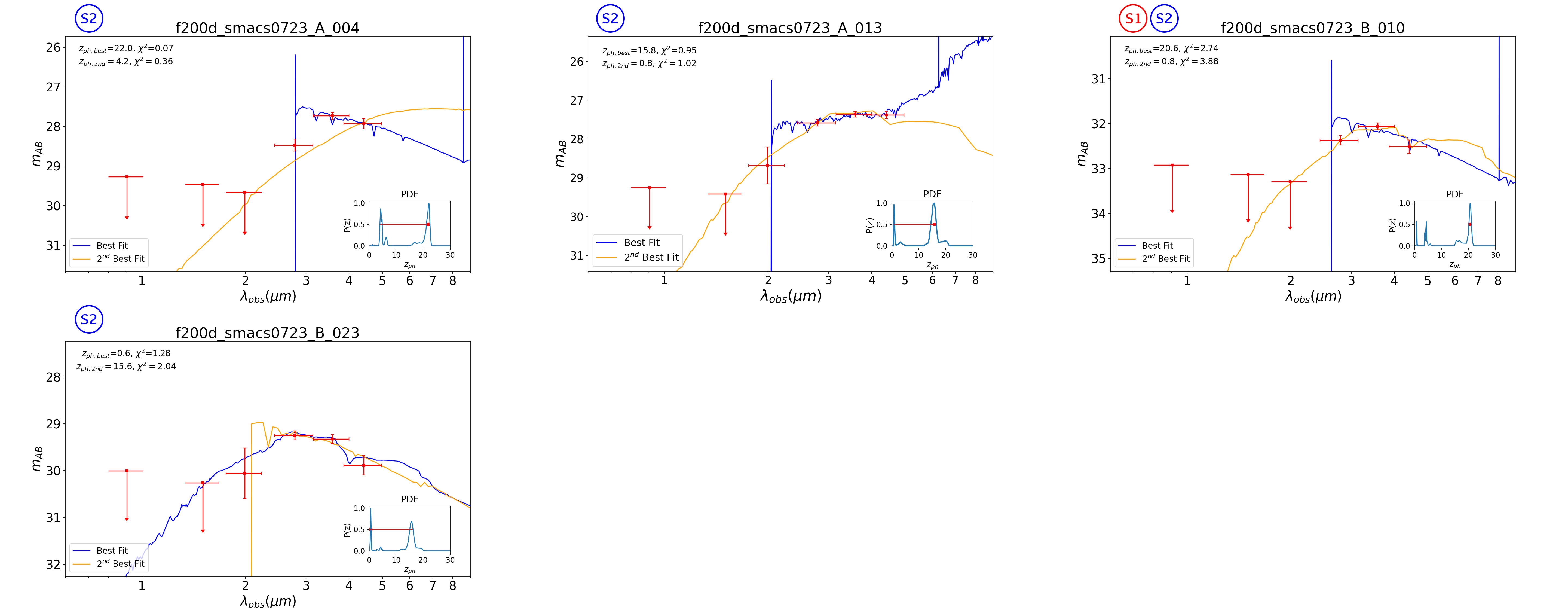}
    \caption{Same as Figure~\ref{fig:f150d_lp_uncover}; for the F200W
    dropouts in JADES GOODS-S, UNCOVER and SMACS0723.
   }
    \label{fig:f200d_lp_jades_uncover_smacs}
\end{figure*}


\begin{figure*}[htbp]
    \centering
    \includegraphics[width=0.85\textwidth]{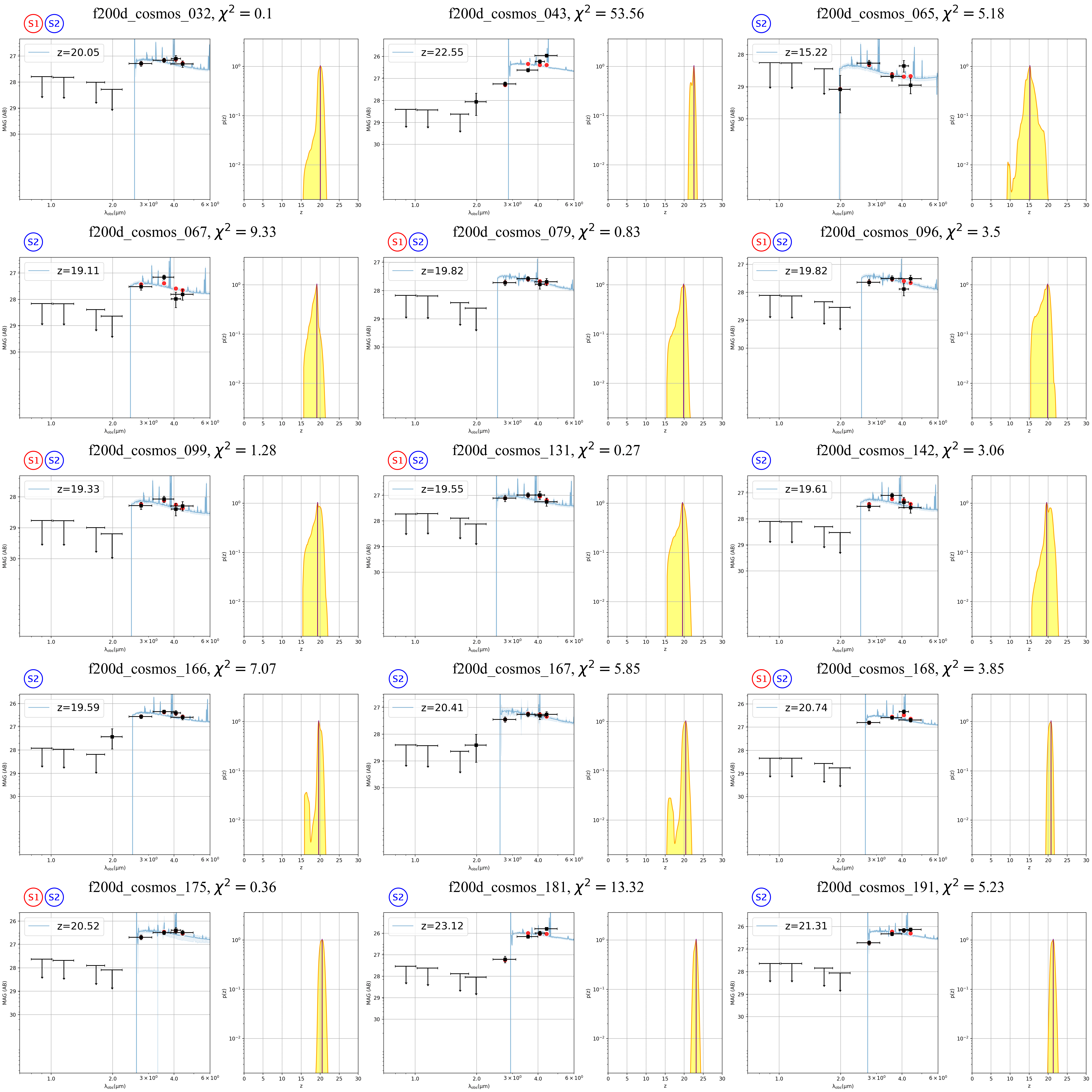}
    \includegraphics[width=0.85\textwidth]{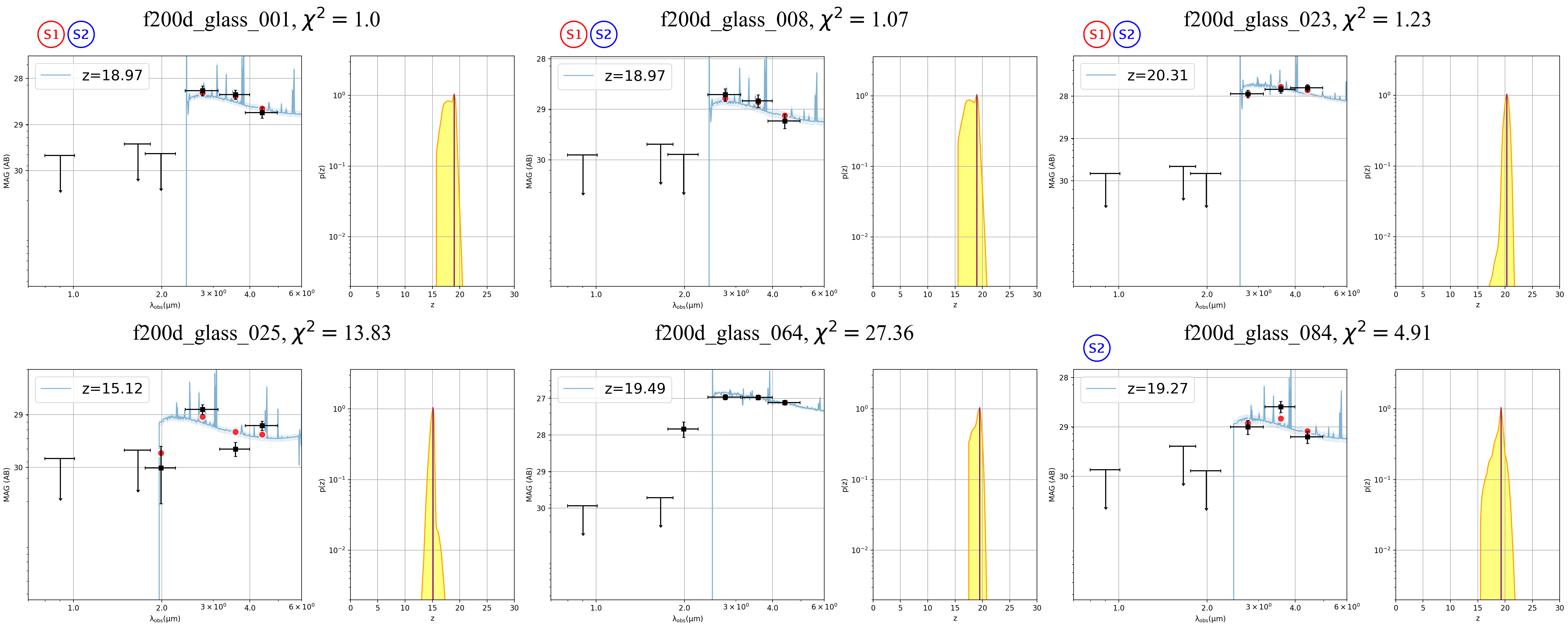}
    \caption{Same as Figure~\ref{fig:f150d_ez_uncover}; for the F200W
    dropouts in COSMOS and GLASS.
   }
    \label{fig:f200d_ez_cosmos_glass}
\end{figure*}

\begin{figure*}[htbp]
    \centering
    \includegraphics[width=0.85\textwidth]{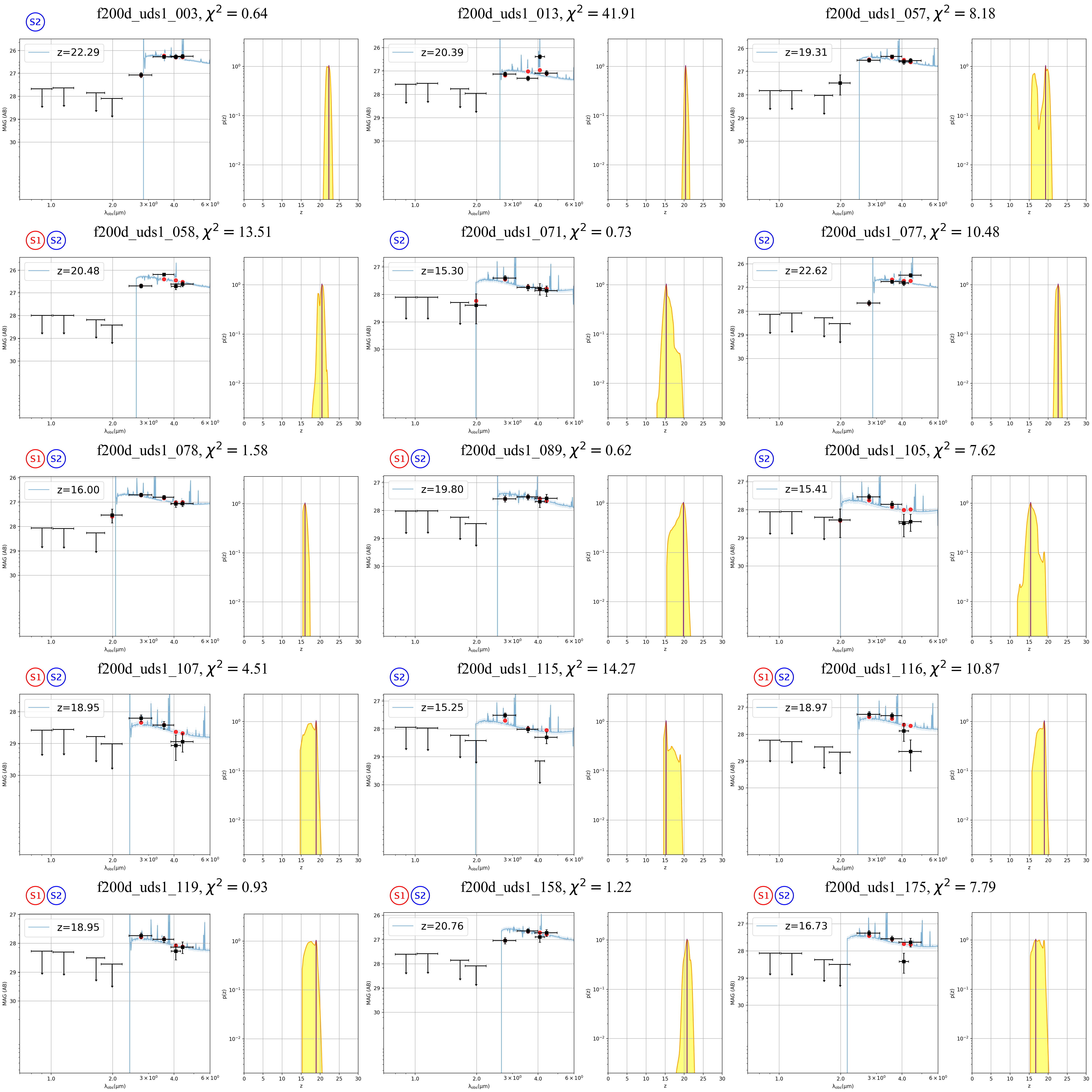}
    \includegraphics[width=0.85\textwidth]{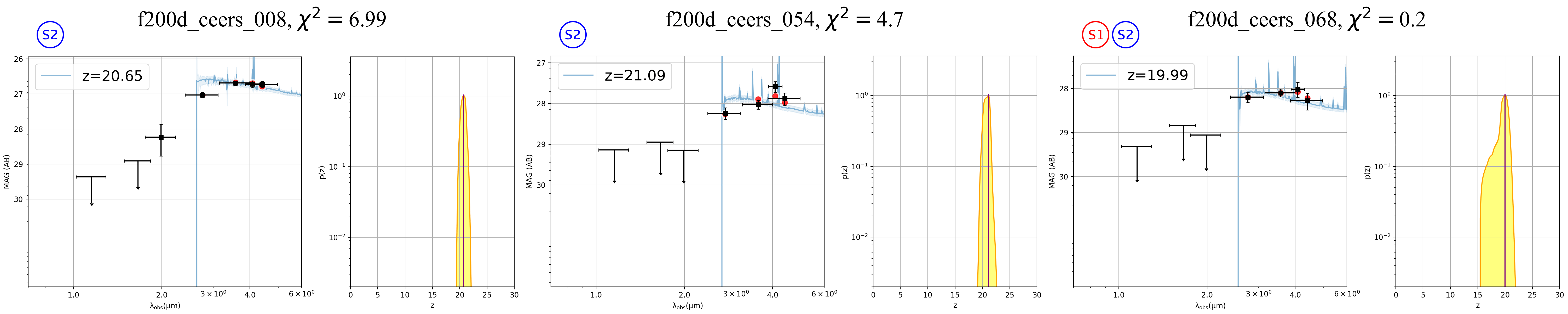}
    \includegraphics[width=0.28\textwidth]{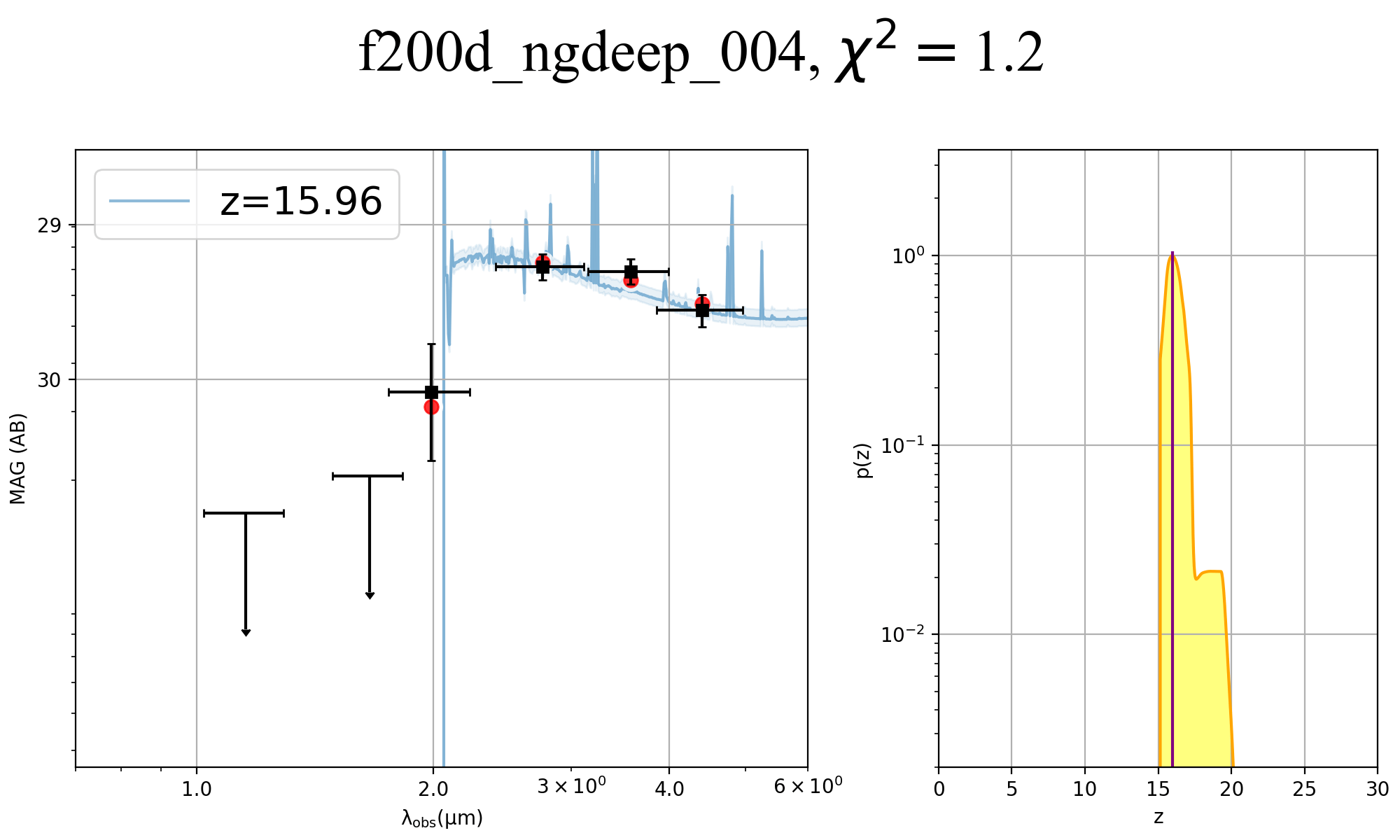}
    \caption{Same as Figure~\ref{fig:f150d_ez_uncover}; for the F200W
    dropouts in UDS1, CEERS and NGDEEP.
   }
    \label{fig:f200d_ez_uds1_ceers_ngdeep}
\end{figure*}

\begin{figure*}[htbp]
    \centering
    \includegraphics[width=0.85\textwidth]{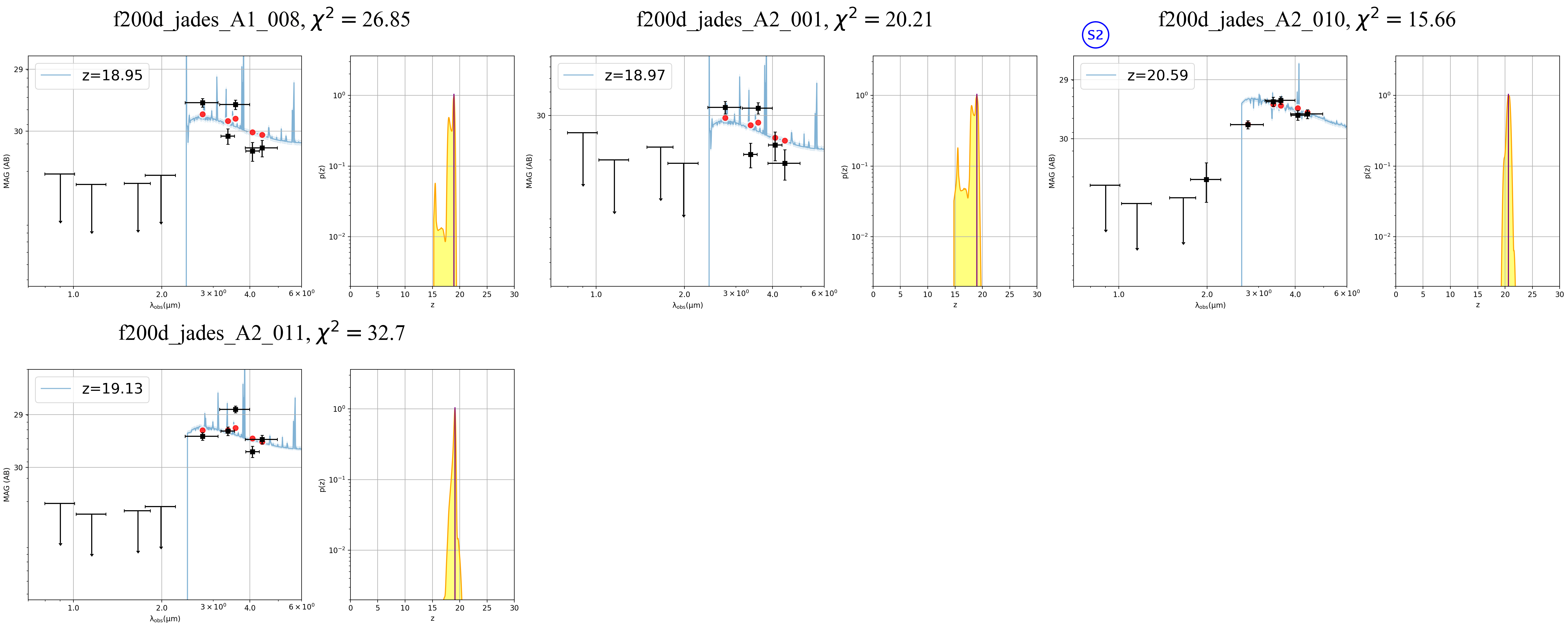}
    \includegraphics[width=0.85\textwidth]{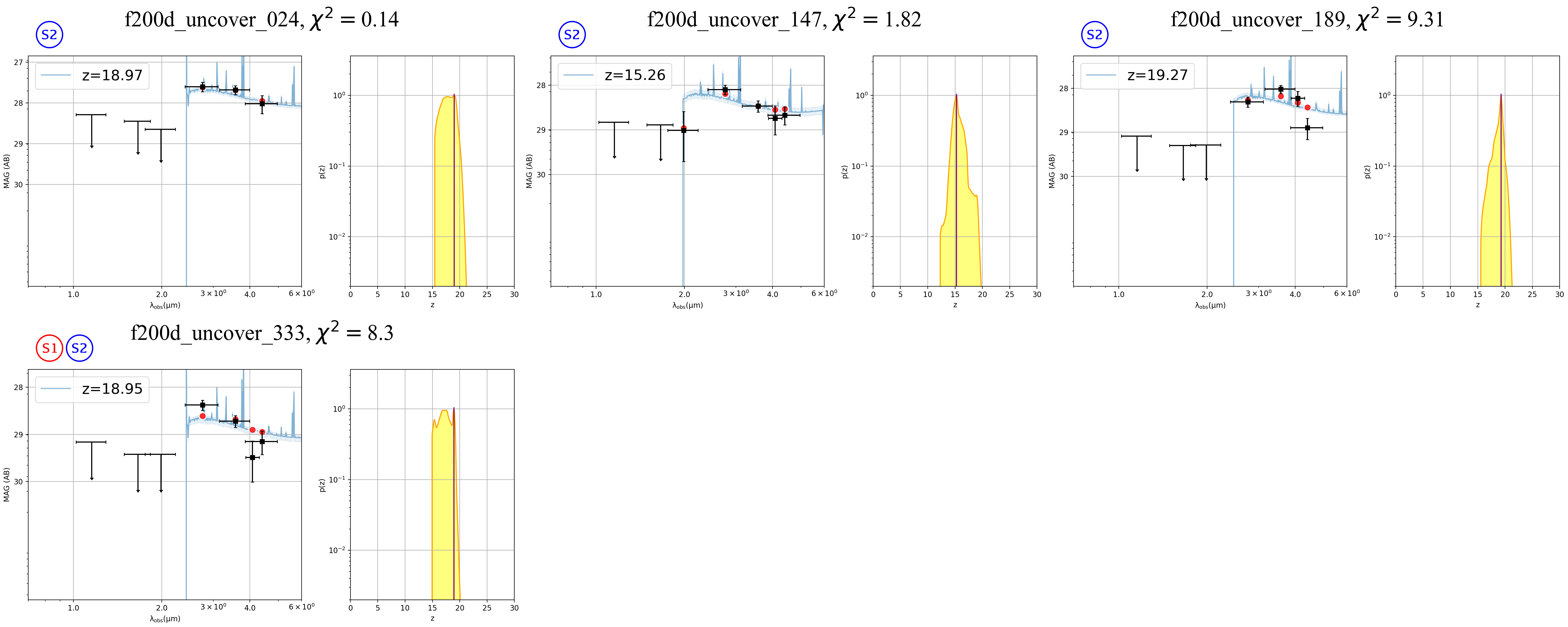}
    \includegraphics[width=0.85\textwidth]{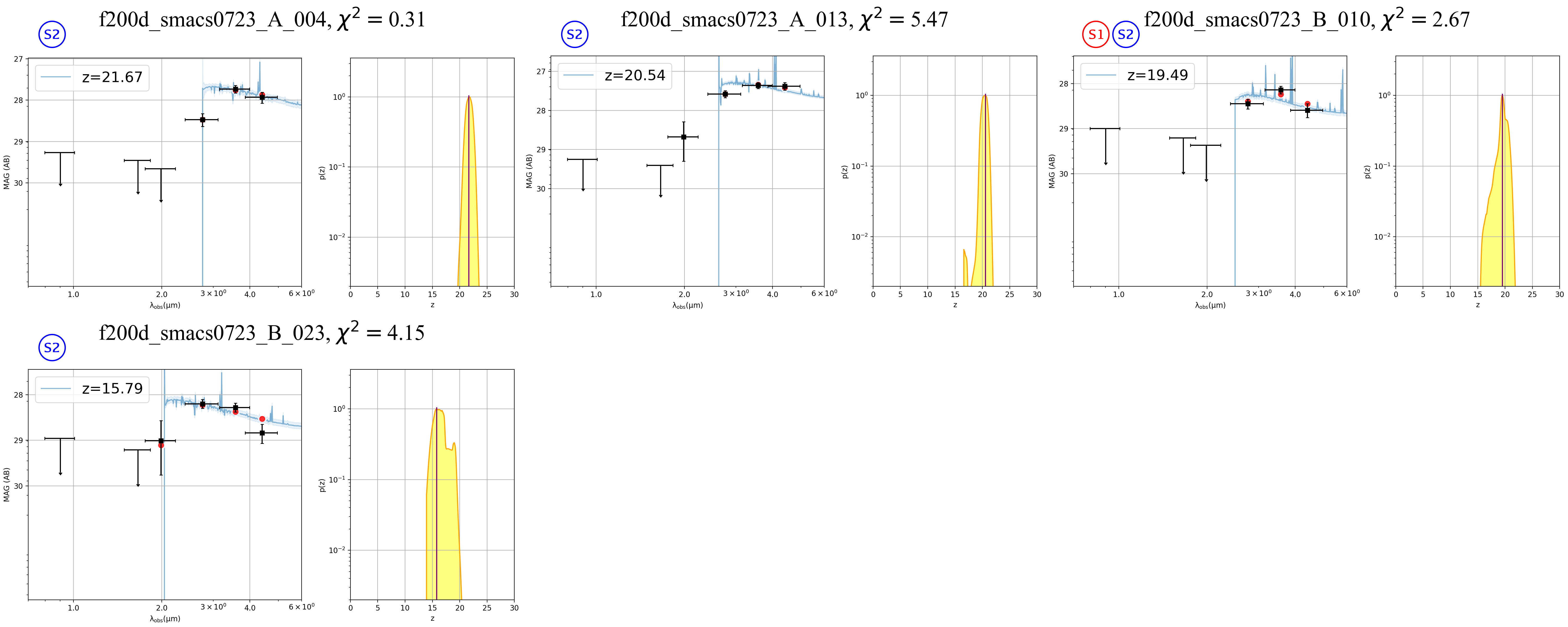}
    \caption{Same as Figure~\ref{fig:f150d_ez_uncover}; for the F200W
    dropouts in JADES GOODS-S, UNCOVER and SMACS0723.
   }
    \label{fig:f200d_ez_jades_uncover_smacs}
\end{figure*}

\section{``Depth Maps'' for Effective Survey Area Calculation}

   As explained in Section 6.1, we calculate the effective survey area,
$A_{\rm eff}$, for each of our dropouts individually. The total surface 
density of the dropouts found by a given survey is then the sum of the 
reciprocals of $A_{\rm eff}$. The implementation of the procedures outlined
in Section 6.1 is as follows.

   For a given survey, we first generate its 5~$\sigma$ and 2~$\sigma$ 
``depth maps'' in all relevant bands using their RMS maps. Such a depth map 
gives the depth at each pixel, where the depth is calculated within a circular
aperture of a certain radius. Using the depth maps, it is easy to find out
whether a pixel is above the specified depth thresholds and therefore whether
it contributes to the effective survey area of a given dropout. For 
illustration, let us use an F150W dropout with $m_{356}=29.25$~mag as an
example. If a pixel in the F356W 5~$\sigma$ depth map is 29.25~mag or deeper,
this pixel passes Step 1 in Section 6.1 for $A_1$ and goes to the next step.

   Strictly speaking, the depth at each pixel should be calculated using the
average of the pixels within the circular aperture centered at this pixel.
However, calculating such a moving average over all pixels extremely
time-consuming, and we simplify by assuming that the pixels within the 
aperture all have the same RMS value as the central pixel under question. The
size of the aperture is somewhat arbitrary. Considering that we use a
$r= $0\farcs2 aperture to calculate the 2~$\sigma$ limit when a candidate 
dropout is a non-detection in the drop-out band, here we adopt $r= $0\farcs2
as well for the depth map generation. A few smallest dropouts in our final
sample have \texttt{ISOAREA\_IMAGE}~$<34.9$~pixels in the F356W 60mas mosaics, 
which means that the $r= $0\farcs2 aperture (occupying $\sim$34.9 pixels at the
60 mas scale) would be too large for them when calculating the depths, which
would lead to an erroneously small effective area and thus an erroneously large
density. For these dropouts, we generate the depth maps individually for each
of them, with the aperture size tailored to occupy the same number of pixels
as their \texttt{ISOAREA\_IMAGE}.

\newpage

\bibliographystyle{aasjournal.bst}

\end{document}